\documentclass[aps,superscriptaddress,reprint]{revtex4-2}

\usepackage[utf8]{inputenc}
\usepackage{xcolor}
\usepackage{pst-node}
\usepackage{graphicx}
\usepackage{amsmath, amssymb, amsthm}
\usepackage{mathtools}
\usepackage{hyperref}
\usepackage[shortlabels]{enumitem}
\usepackage{dsfont}
\usepackage{amsthm}
\usepackage{amssymb}
\usepackage{amstext}
\usepackage{amsfonts}
\usepackage{nicefrac}
\usepackage{extarrows} 
\usepackage{graphicx}
\usepackage{relsize}
\usepackage{stmaryrd}
\usepackage{soul}

\newcommand{\ket}[1]{\vert #1 \rangle}

\newcommand{\calB}{{\mathcal B}}

\DeclareMathOperator{\Tr}{Tr}

\begin{document}
 
\title{Tensor network state methods and quantum information theory\\ for strongly correlated molecular systems}

\author{Mikl\'os Antal Werner}
\affiliation{%
Strongly Correlated Systems Lend\"ulet Research Group,
Wigner Research Centre for Physics, H-1525, Budapest, Hungary
}%

\author{Andor Menczer}
\affiliation{%
Strongly Correlated Systems Lend\"ulet Research Group,
Wigner Research Centre for Physics, H-1525, Budapest, Hungary
}%

\affiliation{%
Eötvös Loránd University, P\'azm\'any P\'eter S\'et\'any 1/C, 1117 Budapest, Hungary
}%

\author{\"Ors Legeza}
\email{legeza.ors@wigner.hu }
\affiliation{%
Strongly Correlated Systems Lend\"ulet Research Group,
Wigner Research Centre for Physics, H-1525, Budapest, Hungary
}%
\affiliation{
Institute for Advanced Study,Technical University of Munich, Germany, Lichtenbergstrasse 2a, 85748 Garching, Germany
}
\affiliation{Parmenides Stiftung, Hindenburgstr. 15, 82343, Pöcking, Germany}

\date{\today}

\begin{abstract}
\noindent 
A brief pedagogical overview of
recent advances in tensor network state methods are presented that have the potential to broaden their scope of application radically for strongly correlated molecular systems. These include global fermionic mode optimization, i.e., a general approach to find an optimal matrix product state (MPS) parametrization of a quantum many-body wave function
with the minimum number of parameters for a given error margin, 
the restricted active space DMRG-RAS-X method, multi-orbital correlations and entanglement, developments on hybrid CPU-multiGPU
parallelization, and an efficient treatment of non-Abelian symmetries on high-performance computing (HPC) infrastructures. Scaling analysis on NVIDIA DGX-A100 platform is also presented.
\end{abstract}

\maketitle

\section{Introduction}
\label{sec:intro}

In the past three decades,  
tensor network state (TNS) methods~\cite{Affleck-1987,Fannes-1992, White-1992b,White-1993,Nishino-1995, Ostlund-1995,Rommer-1997,Schollwock-2005, Hallberg-2006,Noack-2005,Legeza-2008, Chan-2009,Schollwock-2011,Chan-2011, Szalay-2015a,Orus-2019}, originating from the seminal work of S. R. White on the density matrix renormalization group method\cite{White-1992b},  have
become vital alternative approaches to treat strongly correlated, i.e., multireference problems in quantum chemistry \cite{White-1999,Legeza-2008,Chan-2008,Yanai-2009,Marti-2010c,Wouters-2014a,Legeza-2014,Szalay-2015a,Chan-2016,Baiardi-2020,Cheng-2022}.
Despite great successes in the past thirty years~\cite{Verstraete-2023}, TNS-based methods still witness important algorithmic and IT-technology-related developments broadening their scope of application radically by reducing computational time drastically.

A common feature of these methods is connected to singular value decomposition (SVD)~\cite{Roy-2014},  thus they provide an approximation of an eigenstate of the \emph{ab inito} Hamiltonian
as a product of low-rank matrices or tensors. Therefore, the computational demands are governed by the ranks of the component tensors,
also known as \emph{bond dimension}.
{During the optimization procedure, the bond dimension can be kept fixed, i.e., the eigenstate is then searched in a fixed submanifold of the full many-body state space. Alternatively, the tensor ranks can also be adapted dynamically by setting a fixed threshold in the so-called truncation error or discarded block entropy prior to the calculation\cite{Legeza-2003a,Holtz-2012a,Holtz-2012b}. In the latter case the bond dimension is not fixed within the TNS and its maximum, which also determines the required computational power, depends strongly on the network topology \cite{Legeza-2003b,Nakatani-2013,Murg-2015}, on the properties of the component tensors \cite{Gunst-2018}, and, in systems of identical particles (bosons/fermions), also on the chosen set of single particle modes. 
Finding an optimal \emph{set of the modes} (i.e., orbitals in quantum chemistry) dramatically influences the efficiency of the TNS methods \cite{Rissler-2006,Murg-2010a,Stein-2016},
because entanglement, a key ingredient in TNS factorization, may gets strongly localized if the optimal modes are used, therefore correlation effects can get highly suppressed in the system \cite{Fertitta-2014} which leads to a drastic reduction of the bond dimension \cite{Krumnow-2016,Krumnow-2021}.
Therefore, a joint optimization protocol in which the tensors are optimized parallel to the underlying modes can enable a black box application of TNS methods in strongly correlated problems, where one does not need heuristic and uncontrolled ways to define the used orbitals. Such an \emph{in situ} optimization of orbitals is expected to outperform standard methods in systems with strong multireference character, where the optimal set of modes is usually far away from the initial basis set usually provided by Hartree-Fock theory.

Optimization of the orbitals in quantum chemistry has been the research focus for many decades. 
\emph{Localized molecular orbitals} (LMOs), for example, have been introduced to form chemically intuitive orbitals that characterize more transparently the electronic structure of molecules while LMOs have also proven to be useful in developing computationally more tractable methods for strongly correlated quantum chemical systems.
The original, seminal methods search specially constructed unitary operators that are usually optimized through the expectation value of a heuristically introduced operator.
Among many others, we refer to
Foster--Boys localization \cite{Foster-1960,Boys-1960},
which minimizes the radial extent of the localized orbitals,
or Pipek--Mezey localization \cite{Pipek-1989},
which is based on maximizing the sum of the partial atomic charges of each orbital.

Tensor network states naturally call for an orbital optimization method that results in orbitals that allow for a more compressed TNS description of the targeted eigenstate. Applying such \emph{orbital optimization} protocol for quantum chemical systems demonstrated its potential to compress the multireference character of the wave functions by finding optimal MOs based on entanglement localization~\cite{Mate-2022,Petrov-2023}. Here, we go further by using the recently proposed global mode optimization protocol of Ref.~\onlinecite{Friesecke-2024} for quantum chemical problems which protocol is an improved version of methods in Refs.~\onlinecite{Krumnow-2016,Krumnow-2021}.
Numerical simulations have been performed for a simple two-dimensional lattice model, for the $F_2$ dimer, for the $N_2$ dimer both in equilibrium and stretched geometries, and for the computationally more challenging the chromium dimer.

These algorithmic developments have been further boosted drastically by technical solutions exploiting the tremendous computational power offered by modern high-performance computing (HPC) centers~\cite{Hager-2004,Stoudenmire-2013,Nemes-2014,Ganahl-2019,Milsted-2019,Brabec-2021,Zhai-2021,Gray-2021,Unfried-2023,Ganahl-2023,Menczer-2023a,Menczer-2024d,Menczer-2024a,Menczer-2024c,Xiang-2024}.
Quite recently, novel implementations of ab initio DMRG on hybrid CPU-GPU based architectures have been introduced~\cite{Menczer-2023a,Xiang-2024} that are even capable of utilizing AI-accelerators via NVIDIA tensor core units \cite{Menczer-2024d, Menczer-2024a}
reaching a quarter petaflops performance on a single DGX-H100 node~\cite{Menczer-2023c}.
In addition, simple heterogeneous multiNode-multiGPU solutions by
exploiting the different bandwidths of
the different communication channels at different layers of the algorithm resulted in a state-of-the-art implementation of the quantum chemical version of DMRG~\cite{Menczer-2024b}.

In this work, we present a brief pedagogical overview of the main aspects of the \emph{global fermionic mode optimization} protocol in quantum chemical applications and recent advances on massively parallel spin adapted ab initio DMRG via AI accelerators. The setup of this work is as follows: in Section~\ref{sec:theory}, we briefly recall the basics of matrix product states (MPS, a special case of TNS)
and global orbital optimization.
In Section~\ref{sec:numerics}, we describe the applied numerical procedure on small systems in a pedagogical style. 
In Sec.~\ref{sec:postmode} we 
present performance analysis and scaling properties for large-scale \emph{post mode optimization} calculations via our massively parallel multiNode-multiGPU code, together with a brief discussion of post-DMRG approaches.
In Section~\ref{sec:conclusion}, we draw the conclusions.

\section{Theoretical background}
\label{sec:theory}
In this section, we first present a brief introduction to MPS representation of many-body wave functions, second the general procedure based on MPS and orbital optimizations is discussed, third we show how the $SU(2)$ spin symmetry can efficiently be exploited,
and finally, we overview highly scalable parallelization strategies for hybrid CPU-multiGPU systems.

\subsection{Wave function representations}
In the context of nonrelativistic quantum chemistry,
the Hamiltonian takes the second quantized form
\begin{eqnarray}
\label{eq:Hamiltonian}
H &=& \sum_{i,j=1}^d \sum_{\sigma\in\{\downarrow, \uparrow\}} 
t_{i,j}^\sigma c_{i,\sigma}^\dag c_{j,\sigma}
+ \nonumber \\
& & \sum_{i,j,k,l=1}^d \sum_{\sigma,\sigma'\in\{\downarrow, \uparrow\}}
v_{i,j,k,l}^{\sigma, \sigma'} c_{i,\sigma}^\dag c_{j,\sigma'}^\dag c_{l,\sigma'} c_{k,\sigma},
\end{eqnarray} 
where $i,j,k,l$ index electronic orbitals, $\sigma,\sigma' \in \lbrace \uparrow,\downarrow \rbrace$ are the spin indices,
while $c^\dag_{i,\sigma}$, and $c_{i,\sigma}$ are the fermionic creation and annihilation operators, respectively,
satisfying the canonical anti-commutation relations 
$\{c_{i,\sigma},c_{j,\sigma'}\}=0$ and 
$\{c_{i,\sigma}^\dag,c_{j,\sigma'}\}=\delta_{i,j}\delta_{\sigma,\sigma'}$. 
Due to the hermiticity of $H$ and the anti-commutation relations of fermion operators, there are symmetry relations among the elements of the spin-independent integrals $t_{i,j}$ and $v_{i,j,k,l}$. 

The Hilbert space of $N$ interacting electrons in $d$ orbitals
is the $N$-electron subspace of the fermionic Fock space $\mathcal{F}_d \cong \bigotimes_{i=1}^d \mathcal{H}_i$ that is spanned by the basis constituted by all Slater determinants 
$\{ \ket{\alpha_1,\dots,\alpha_d}=\bigotimes_{i=1}^d \ket{\alpha_i} \}$,
where the occupation indices $\alpha_i\in\{0,\downarrow,\uparrow,\downarrow\uparrow\}$ label
the basis in the occupation spaces $\mathcal{H}_i$ of orbitals $i\in\{1,2,\dots,d\}$.
The quantum many-electron wave function of the ground state or excited states can be written as a linear combination of all Slater determinants
\begin{equation}
\ket{\psi} = \sum_{\substack{\alpha_1,\ldots,\alpha_d\\\in\{0,\downarrow,\uparrow,\downarrow\uparrow\}}}
C_{\alpha_1,\ldots,\alpha_d}\ket{\alpha_1,\ldots,\alpha_d},
\label{eq:fulltensor}
\end{equation}
where the $d$-order complex-valued coefficient tensor $C_{\alpha_1,\ldots,\alpha_d}$ is determined
by the eigenvalue problem of the Hamiltonian given by Eq.~\eqref{eq:Hamiltonian}
in the $N$-electron subspace.
The tensor $C$ is a representation of the so-called full configuration interaction (full CI) wave function, and contains $4^d$ elements, which dimension gets numerically intractable for $d \gtrsim 20-30$, in which case the use of approximate methods is inevitable. For example, in the truncated CI approach, $\ket{\Psi}$ is
expressed as a linear combination of wave functions
corresponding to different excitation levels with respect to a reference determinant,
\begin{equation} 
\ket{\psi} = \sum_I \ket{\psi_I},
\label{eq:ci}
\end{equation}
where the $I=0$ term $\ket{\psi_0}$ refers to the reference determinant (most often the Hartree-Fock determinant), 
and terms $\ket{\psi_I}$ with $I=1,2,3\dots$ 
refer to the single, double, triple, etc. excitations, respectively~\cite{Helgaker-2000}. Keeping terms only up to some truncation threshold $I\le I_{max}$ leads to the corresponding truncated-CI theory. (CI-SD ($I_{\max}=2$), CI-SDT ($I_{\max}=3$), etc.) As we can see, this truncation scheme requires a good reference determinant, which usually exists only for the so-called single reference problems.

TNS methods follow a different truncation strategy that roots in quantum entanglement structure of the full-CI state $\ket{\Psi}$.  In the MPS representation, the wave function takes the form
\begin{equation}
\ket{\psi} =
\sum_{\substack{\alpha_1,\ldots,\alpha_d\\\in\{0,\downarrow,\uparrow,\downarrow\uparrow\}}} A^{\alpha_1}_{[1]}\cdots A^{\alpha_d}_{[d]}\ket{\alpha_1,\dots,\alpha_d},
\label{eq:DefinitionMPS}
\end{equation}
where the \emph{component tensors} are $A_{[i]}^{\alpha_i}\in \mathbb{C}^{D_{i-1}\times D_i}$,
with \emph{bond dimensions} $D_i$, and $D_0 = D_d = 1$.
It is easy to prove that every state vector of the Fock space $\mathcal{F}_d$ can be transformed to an MPS form
by applying consecutive SVDs \cite{Roy-2014,Vidal-2003b},
using sufficiently large bond dimensions.
However, in the generic case, this exact MPS representation needs exponentially large bond dimensions $D_i \lesssim 4^{d/2}$.
Restricting the bond dimensions to a fixed value of $D$, i.e. keeping only the $D$ largest singular values in SVD, can lead to a very good approximation, at least if the entanglement through the bond is small enough. Such fixed $D$ restricts possible many-body quantum states to a sub-manifold of the full state space. 
Eigenstates of the Hamiltonian \eqref{eq:Hamiltonian} can then be approximated
within this sub-manifold by use of
the \emph{density matrix renormalization group} (DMRG) algorithm~\cite{White-1992b, Schollwock-2005},
which, being an alternating least square method, optimizes the entries
of the MPS tensors $A_{[i]}$ iteratively 
\cite{Ostlund-1995,Verstraete-2004a,Verstraete-2004b, Legeza-2014,Szalay-2015a},
leading to a variational treatment of the eigenvalue problem of the Hamiltonian \eqref{eq:Hamiltonian}.

\subsection{Global Mode optimization}
Originally, DMRG was used for one-dimensional lattice models where the chain structure of the ansatz \eqref{eq:DefinitionMPS} also coincides with the spatial topology of the system. In quantum chemistry problems, finding an optimal ordering of modes for a fixed bond dimension $D$ is already a non-trivial problem. We note, however, that the order of modes is irrelevant if no truncation is employed.
\begin{figure}
    \centering
    \includegraphics[width=0.39\textwidth]{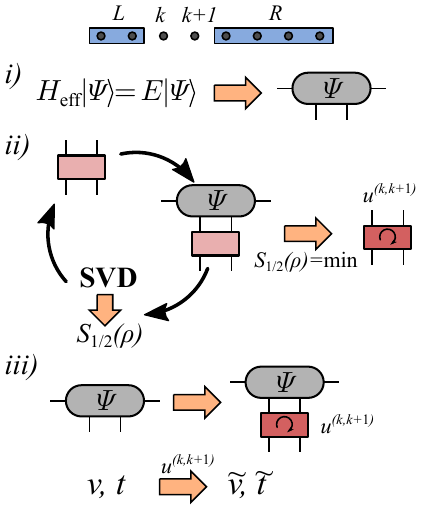}
    \caption{Steps of mode optimization micro iteration at bond position $(k,k+1)$. \textit{i)} The eigenstate $\Psi$ of the Hamiltonian in the effective Hilbert space $\mathcal{H}_\mathrm{eff}$ is numerically determined by an iterative eigensolver. \textit{ii)} The two site unitary $u^{(k,k+1)}$ that minimizes the half-Rényi block entropy $S_{1/2}(\rho)$ is searched iteratively. Note that the calculation of $S_{1/2}(\rho)$ requires an SVD in every cycle. \textit{iii)} The eigenstate wave function $\Psi$ is rotated by $u^{(k,k+1)}$ and the Hamiltonian parameters $t$ and $v$ are updated accordingly. 
    }    
    \label{fig:microiter}
\end{figure}
The ansatz \eqref{eq:DefinitionMPS} enables not only the re-ordering of modes but also a more flexible transformation: unitary rotations in the space of (single-)electron orbitals. Utilizing a unitary orbital-transformation $U \in \mathrm{U}(d)$, a linear transformation of the set of fermionic creation operators $\{c^\dag_{i,\sigma}\}$ to a new set $\{\tilde{c}^\dag_{i,\sigma}\}$ satisfying the canonical anti-commutation relations can be obtained, i.e., $\tilde{c}^\dag_{i,\sigma} = \sum_{j=1}^d U_{i,j,\sigma} c^\dag_{j,\sigma}$. A pleasant property of such transformation is that it preserves the structure of the Hamiltonian \eqref{eq:Hamiltonian}, one only has to replace the fermion operators $c \Rightarrow \tilde{c}$ and rotate the one- and two-body integrals $t \Rightarrow \tilde{t}$ and $v \Rightarrow \tilde{v}$ according to the transformation $U$ such that the Hamiltonian \eqref{eq:Hamiltonian} remains invariant. As we will see, such redefinition of modes can drastically reduce the entanglement of the state, especially in the case of multi-reference problems, where the Hartree-Fock state and orbitals are not expected to be optimal. The unitary matrix $U_{i,j,\sigma}$ can be restricted to being spin-independent but spin-dependent $U_{i,j,\uparrow} \ne U_{i,j,\downarrow}$ rotations are also possible. The latter situation arises naturally for open-shell problems, where the integrals $t$ and $v$ are spin-dependent within the unrestricted Hartree-Fock theory.

Optimization of $U$ needs a cost function whose (local) minimum is searched during the procedure. In Ref.~\onlinecite{Friesecke-2024} the block entropy area (BEA),
\begin{equation}
    \mathcal{B}_\alpha(C) = \sum_{k=1}^{d-1} S_\alpha(\rho_{\{1,2,\dots,k\}}) \; , \label{eq:bea_def}
\end{equation}
i.e. the sum of block Rényi-entropies of a suitably chosen parameter $\alpha$ (see below) along the bonds of the MPS,  has been introduced as a cost function. The block entropy $S_\alpha(\rho_{\{1,2,\dots,k\}})$ at bond position $k$ is related to the required bond dimension roughly as $D_k \sim e^{S_\alpha}$ at the corresponding bond for a given error margin and reducing the bond dimension results in reduced computational complexity of the DMRG algorithm~\cite{Menczer-2024a}. We remark that $\rho_{\{1,2,\dots,k\}}$ is the density operator of the first $k$ orbitals \cite{Szalay-2021,Boguslawski-2013}. Because block entropies are of central importance in the DMRG, their calculation does not involve any extra computational cost.

Finding an optimal $U$ that minimizes the cost function BEA is a hard numerical problem that cannot be solved directly. In the simplest approach, one attempts to build up $U_{i,j,\sigma}$ iteratively as a product of nearest-neighbor (two-site) rotations $u^{(k,k+1)}_{i,j,\sigma}$ which matrix is equivalent with the $d$-dimensional identity matrix except for the $2\times2$ block at rows and columns $k$ and $k+1$. It is easy to show that any unitary $U$ can be factorized as a product of such two-site rotations. Optimization of such nearest-neighbor rotations can be efficiently performed alongside the standard DMRG algorithm: after the approximate eigenstate (DMRG wave function) is determined at sweep position $(k,k+1)$, one also needs to find the $u^{(k,k+1)}$ that minimizes the entropy $S_\alpha(\rho_{\{1,2,\dots,k\}})$. In practice, here we use the half-R\'enyi block entropy
$S_{1/2}(\rho_{\{1,2,\dots,k\}}) = 2\ln(\Tr\sqrt{\rho_{\{1,2,\dots,k\}}})$, but other entropy definitions, like the more commonly used von Neumann entropy $S_1(\rho_{\{1,2,\dots,k\}}) = - \Tr \rho_{\{1,2,\dots,k\}} \ln \rho_{\{1,2,\dots,k\}}$, can also be used here. Using different entropies, however, may result in slightly different optimized modes at the end. As $S(\rho_{\{1,2,\dots,k\}})$ is the only block entropy that is varied by $u^{(k,k+1)}$ while block entropies at other bond positions are invariant, such a local optimization leads to the optimization of the cost function \eqref{eq:bea_def}. Once the optimal $u^{(k,k+1)}$ is found, it is applied to the DMRG wave function, and the Hamiltonian is also updated by rotation of the integrals $t$ and $v$. We call the DMRG iteration step and the optimization of $u^{(k,k+1)}$ together as a 'micro' iteration of the mode optimization procedure.

Unfortunately, performing only local mode optimization micro-iterations provides poor results. The fact that any global $U_{i,j,\sigma}$ can be factorized to a product of nearest neighbor two-site rotations does not guarantee that optimization can also be efficiently performed in the factorized form. The reason for the poor performance can be easily understood from the structure of the unitary group $U(d)$, which is generated by all the two-site (infinitesimal) rotations $u_{\delta}^{(k,l)}$, not only by nearest neighbor ones. Here the subscript $\delta$ indicates that the two-site rotation is close to the identity matrix. Finding a stationary point of the above proposed local mode optimization iteration, consequently, is not equivalent to finding a local minimum of $\mathcal{B}_\alpha$ in the manifold $U(d)$ of global unitaries, as zero derivatives are assured only for local (nearest neighbor) rotations which form only a small fraction of possible directions in the manifold~\cite{Friesecke-2024}. Introducing long-range rotations, however, needs serious algorithmic changes to the standard DMRG algorithm and further code development. However, we can partly circumvent the problem by re-ordering the modes after every 'macro' iteration, i.e. after several sweeps of local mode-optimization micro iterations. This re-ordering puts new mode pairs next to each other, consequently allowing for new optimization directions. In the original works~\cite{Murg-2010a,Krumnow-2016}, the new mode order was determined by the rather heuristic Fiedler-vector method from the two-mode mutual information that, therefore,  needed to be calculated for every mode pair at the end of every macro iteration. More recently, a general re-ordering protocol has been introduced~\cite{Friesecke-2024} that has the pleasant property that in $d/2$ or $(d+1)/2$ macro iteration steps for even/odd $d$ every mode-pair gets next to each other at least once. Moreover, in contrast to the Fiedler-vector technique, the MPS form of the wave function is preserved during the swap-gate-based re-ordering, i.e. there is no need to rebuild the MPS state by several DMRG sweeps after every macro iteration step. In this paper, this general protocol has been employed,  further details are discussed below. We must note that re-ordering the modes can lead to an increase in the cost function \eqref{eq:bea_def} both for the Fiedler-vector and for the swap-gate-based methods, the local mode-optimizations of the next macro-iteration are needed to push the cost function further down. As we will see, this non-monotonous behavior can lead to fluctuations or oscillations in the cost function, once the optimum is approached. 
\begin{figure}
    \centering
    \includegraphics[width=0.45\textwidth]{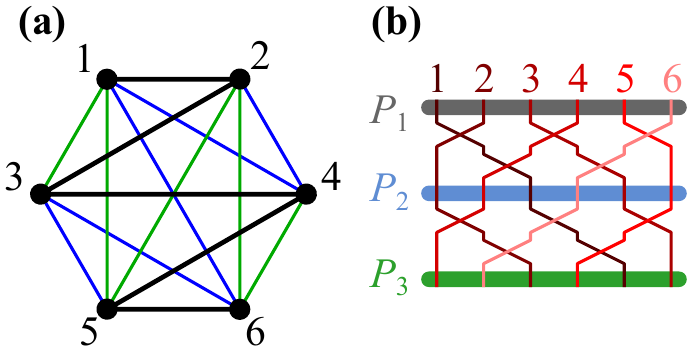}
    \caption{Generation of permutations by Walecki's construction for $d=6$. \textit{(a)} The modes are placed on the edges of a regular polygon (hexagon for $d=6$) such, that the original permutation $(1,2,\dots,d)$ follows the black zig-zag line. New permutations (blue and green zig-zag lines) are generated by repetitive clockwise rotation of the original zig-zag line by an angle $360^\circ/d$. \textit{(b)} Permutations $(P_1,P_2,P_3)$ can be generated consecutively from each other by applying two layers of swap gates in a brick-wall pattern. 
    }    
    \label{fig:walecki}
\end{figure}

\subsubsection{Global mode optimization with swap-gate generated permutations}
The basic idea behind the recently introduced protocol~\cite{Friesecke-2024} is that by generating appropriate permutations of modes all the mode pairs get neighbors at some macro iterations and, thus the global unitary $U$ is optimized in the direction of every generator of the unitary group $U(d)$. An optimal list of permutations for even $d$ can be generated by Walecki's method~\cite{Lucas-1882} that was introduced first in graph theory. In the method, the $d$ modes are first placed on the $d$ vertices of a regular polygon such that the original permutation forms a zig-zag line between the vertices (see Fig.~\ref{fig:walecki}a).  New permutations are then generated by clockwise rotation of the zig-zag line around the center of the polygon. After $d/2-1$ rotations, all the edges of the full graph of $d$ vertices are covered exactly once; consequently, all mode pairs get nearest neighbors exactly once within these $d/2$ permutations. For odd $d$ values, there does not exist such a completely optimal set of permutations. The method of Walecki still can be used: one has to add a 'dummy' auxiliary mode to the list, create the corresponding permutations for $d+1$ modes, and finally erase the auxiliary mode from the permutations. The resulting permutations cover all the mode pairs at least once, while some pairs get covered twice.

Besides its proven optimality, Walecki's construction has also the advantage that the consecutive permutations can be easily generated by two layers of nearest neighbor swap gates (see Fig.~\ref{fig:walecki}b). The application of these swap gates needs only a minimal change in the implementation of the local mode-optimization protocol, namely, at swap gate positions we fix $u^{(k,k+1)}$ to be the unitary that exchanges two modes while diagonalization is omitted in these steps. The two layers of swap gates are then executed in a forward and backward sweep using these modified and simplified micro iterations. We note that this simple protocol does not require the costly calculation of the two-mode mutual information and the MPS wave function is updated during the action of the swap gates, i.e. one does not need to re-calculate it from scratch. It is also important to note, that after the action of swap gates, if no truncation is employed, the bond dimension can increase by a factor of the dimension of the local site. One can therefore temporarily increase the maximal bond dimension at the swap gates by a factor denoted by $q$, and then compress the MPS back with a further sweep. However, as we numerically demonstrate below, such a bond expansion is usually unnecessary.

The bond dimension $D$ is usually kept small during the mode optimization procedure ($D\lesssim 256$) therefore the required computational power (memory and core number) remains moderate. However, one needs several tens or maybe hundred macro iteration steps that translate to several hundreds or maybe thousand usual DMRG sweeps with this low bond dimension. Once an optimal basis is found, one can perform a standard DMRG calculation with a high bond dimension ($D \gtrsim 10 000 - 20 000$) but without further mode optimization. To perform such high-$D$ calculation efficiently, we employ massive parallelization using a hybrid CPU-multiGPU kernel~\cite{Menczer-2023a,Menczer-2024d,Menczer-2024c}. If symmetries are also present, their use is also highly beneficial. However, it is important to exploit symmetries in such a simple way that does not lead to unacceptable overhead in the parallelization. In the next three subsections, we overview the basics of these technicalities.

\begin{figure*}
    \centering
    \includegraphics[width=0.8\textwidth]{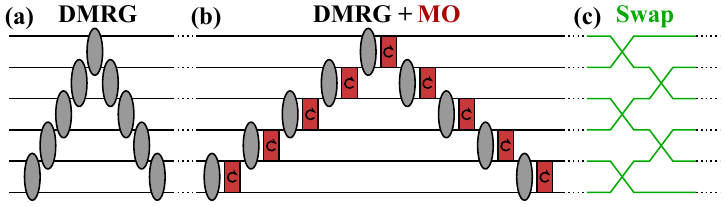}
    \caption{Steps of a mode optimization macro iteration. \textit{(a)} First several sweeps of standard DMRG iterations without the mode optimization steps are executed to reach an acceptable initial MPS state. \textit{(b)} Next, sweeps of local mode optimization micro iterations (see Fig.~\ref{fig:microiter}) are executed. The grey ellipses indicate that in the micro iterations, there are also diagonalization steps. \textit{(c)} The macro iteration is closed by two layers of swap gates. Note that the same coloring is used (grey-red-green) in the numerical figures about $E_{GS}$ and BEA to indicate the different types of steps.
    }    
    \label{fig:macroiter}
\end{figure*}
\subsection{Efficient exploitation of non-Abelian  $SU(2)$ spin symmetry}\label{subsec:sym}

In this subsection, we overview the use of symmetries in DMRG. We focus on the computationally most demanding step, the diagonalization of the effective Hamiltonian. In the standard two-site DMRG, at bond position $(k,k+1)$ the effective Hilbert space is $\mathcal{H}_\mathrm{eff} = \mathcal{H}_{L} \otimes \mathcal{H}_k \otimes \mathcal{H}_{k+1} \otimes \mathcal{H}_{R}$, where $L = \lbrace 1,2, \dots, k-1 \rbrace$ and $R = \lbrace k+2, k+3, \dots, d \rbrace $ stand for the left and right blocks of modes, whose state spaces $\mathcal{H}_L$ and $\mathcal{H}_R$ have been truncated by SVD in the previous renormalization steps~\cite{Schollwock-2005}. Wave functions in $\mathcal{H}_\mathrm{eff}$ are represented by their 4-index coefficient tensor $\Psi_{\alpha, \sigma_k,\sigma_{k+1},\beta}$, where the indices $\lbrace \alpha, \sigma_k,\sigma_{k+1},\beta \rbrace$ stand for the basis states of $ \lbrace \mathcal{H}_{L}, \mathcal{H}_k, \mathcal{H}_{k+1}, \mathcal{H}_{R} \rbrace$, respectively. 
The Hamiltonian \eqref{eq:Hamiltonian} is then transformed to the factorized form
\begin{equation}
    H = \sum_{\nu} O^{(\nu, L)} \otimes O^{(\nu, k)} \otimes O^{(\nu, k+1)} \otimes O^{(\nu, R)} \;,
\end{equation}
where $\nu$ enumerates the different terms in the Hamiltonian while the corresponding operators $O^{(\nu, i)}$ act in subsystems $i \in \lbrace L, k, k+1, R \rbrace$. The action of the Hamilton on state $\Psi$ translates to
\begin{eqnarray}
 & &\Psi'_{\alpha', \sigma_{k}', \sigma_{k+1}', \beta'} =  \sum_\nu \sum_{\alpha,\sigma_k,\sigma_{k+1},\beta} \nonumber \\ 
 & & O^{(\nu, L)}_{\alpha',\alpha} \, O^{(\nu, k)}_{\sigma_{k}',\sigma_{k}} \, O^{(\nu, k+1)}_{\sigma_{k+1}',\sigma_{k+1}} \, O^{(\nu, R)}_{\beta',\beta} \, \Psi_{\alpha, \sigma_{k}, \sigma_{k+1}, \beta} \label{eq:Hamilton_act_nonsym}
\end{eqnarray}

The Hamiltonian \eqref{eq:Hamiltonian} in its general form has two Abelian $U(1)$ symmetries, namely the total number $n_{\mathrm{tot}}$ of electrons and the $z$ component of the total spin $S^z_\mathrm{tot}$ are good quantum numbers. Exploiting these two $U(1)$ symmetries does not require serious changes in the algorithm: the quantum states get extra labels (their particle number $n$ and spin projection $s_z$) while the operators $O^{(\nu, i)}$ and the wave function tensor $\Psi_{\alpha, \sigma_k,\sigma_{k+1},\beta}$ have a block structure according to these labels, with many zero blocks due to quantum number conservation. The storage of zero blocks is avoided if block-sparse storage of matrices and tensors is used. For example, the operator $O^{(\nu, i)}_{a,b}$ is then stored as $O^{(\nu, i)}_{a,b}\lbrace p \rbrace$ where $p$ enumerates the blocks while the index ranges of $b$ and $a$ are restricted to the incoming and outgoing quantum number sectors of the block, respectively. Eq.~\eqref{eq:Hamilton_act_nonsym} is then further decomposed into a list of tasks, indexed by $\tau$, 
\begin{eqnarray}
 & &\Psi'_{\alpha', \sigma_{k}', \sigma_{k+1}', \beta'}(p'^{\tau}) =  \Psi'_{\alpha', \sigma_{k}', \sigma_{k+1}', \beta'}(p'^{\tau}) + \sum_{\alpha,\sigma_k,\sigma_{k+1},\beta} \nonumber \\
 & & O^{(\nu, L)}_{\alpha',\alpha} \lbrace p_L^\tau \rbrace \, O^{(\nu, k)}_{\sigma_{k}',\sigma_{k}}\lbrace p_k^\tau \rbrace \, O^{(\nu, k+1)}_{\sigma_{k+1}',\sigma_{k+1}}\lbrace p_{k+1}^\tau \rbrace \, O^{(\nu, R)}_{\beta',\beta}\lbrace p_R^\tau \rbrace \, \nonumber \\ 
 & & \phantom{aaaaaaaaaaaaaaaaaaaaaaaaaaaa} \Psi_{\alpha, \sigma_{k}, \sigma_{k+1}, \beta}\lbrace p^\tau \rbrace \; . \label{eq:Hamilton_act_U1sym}
\end{eqnarray}
Once the block indices $\lbrace p'^\tau, p^\tau, p^\tau_L, p^\tau_k, p^\tau_{k+1}, p^\tau_R \rbrace_\nu$ of the tasks $\tau$ for the Hamiltonian terms $\nu$ are collected, the whole list is provided to the numerical kernel that executes the tensor contractions. We note that this precalculated list of tasks allows for massive and scalable parallelization of the kernel~\cite{Nemes-2014}.

In many systems, not only the $z$ component of the total spin but also the total spin vector is conserved, therefore the spin symmetry is $SU(2)$ instead of $U(1)$. Using the $SU(2)$ symmetry allows for a large compression of the quantum states and operators because the $SU(2)$ group is non-Abelian and has $2S+1$ dimensional representations (multiplets) indexed by the total spin $S$. The DMRG algorithm can then be reformulated in a way, where indices $\lbrace \alpha, \sigma_k,\sigma_{k+1},\beta \rbrace$ stand for multiplets instead of individual states~\cite{McCulloch-2002a, Toth-2008,Sharma-2012a,Weichselbaum-2012,Keller-2016,Gunst-2019,Werner-2020}. One also has to group the operators $O^{(\nu, i)}$ into irreducible tensor operators also characterized by their operator spin $S^\mathrm{op}$. Then, the operators are stored by their reduced matrix elements $\mathbb{O}^{(\nu, i)}_{a,b}$ where, again, $a$ and $b$ are multiplet indices. These reduced matrix elements are connected to the standard matrix elements through the Wigner-Eckart theorem~\cite{Wigner-1959,Varshalovich-1988}.  Using multiplets instead of states results in a large compression of tensors.  However, one has to pay a small price for the transformation to reduced matrix elements: numerical prefactors $\mathcal{F}_\tau$ appear in the tensor multiplication tasks, dictated by the $SU(2)$ symmetry, and these prefactors depend on all the many quantum numbers and operator spins that appear in a given task,
\begin{eqnarray}
 & &\Psi'_{\alpha', \sigma_{k}', \sigma_{k+1}', \beta'}(p'^{\tau}) =  \Psi'_{\alpha', \sigma_{k}', \sigma_{k+1}', \beta'}(p'^{\tau}) + \mathcal{F}_\tau  \sum_{\alpha,\sigma_k,\sigma_{k+1},\beta} \nonumber \\
 & & \mathbb{O}^{(\nu, L)}_{\alpha',\alpha} \lbrace p_L^\tau \rbrace \, \mathbb{O}^{(\nu, k)}_{\sigma_{k}',\sigma_{k}}\lbrace p_k^\tau \rbrace \, \mathbb{O}^{(\nu, k+1)}_{\sigma_{k+1}',\sigma_{k+1}}\lbrace p_{k+1}^\tau \rbrace \, \mathbb{O}^{(\nu, R)}_{\beta',\beta}\lbrace p_R^\tau \rbrace \, \nonumber \\ 
 & & \phantom{aaaaaaaaaaaaaaaaaaaaaaaaaaaa} \Psi_{\alpha, \sigma_{k}, \sigma_{k+1}, \beta}\lbrace p^\tau \rbrace \; . \label{eq:Hamilton_act_su2}
\end{eqnarray}
Introduction of the numerical prefactor $\mathcal{F}_\tau$ in the task list is very simple, therefore the highly parallelized kernel can immediately used for $SU(2)$ symmetries too. However, one must access the values $\mathcal{F}_\tau$ rapidly during the preparation of tables to avoid an overhead~\cite{Menczer-2024d}.

The source of the prefactor $\mathcal{F}_\tau$ is in the product structure of the effective Hilbert space $\mathcal{H}_\mathrm{eff}$. In the case of $SU(2)$ multiplets, the product is better done hierarchically. First, we build $\mathcal{H}'_L = \mathcal{H}_L \otimes \mathcal{H}_k$ and $\mathcal{H}'_R = \mathcal{H}_{k+1} \otimes \mathcal{H}_R$, i.e. we generate the multiplets in the double products. The total effective Hilbert space is then formed again as a double product, namely, $\mathcal{H}_\mathrm{eff} = \mathcal{H}'_L \otimes \mathcal{H}'_R$. As the operators are represented by their reduced matrix elements, we must form tensor operators from the operator products and determine their reduced matrix elements. We briefly present the calculation for the left part. First, we form the multiplets of $\mathcal{H}'_L$ using the known spin addition rules,  $\ket{\alpha, S_\alpha} \otimes \ket{\sigma_k, S_k} \Rightarrow \ket{\gamma, S_\gamma}'$. Then we form tensor operators from the double products, which is done fully analogously to states using the spin addition rules,  $S^\mathrm{op}_L \otimes S^\mathrm{op}_k \Rightarrow {S^\mathrm{op}_L}'$. The reduced matrix element of the product is then given by
\begin{eqnarray}
\llbracket \mathbb{O}^{(\nu,L)} & & \otimes  \mathbb{O}^{(\nu,k)} \rrbracket_{\gamma', \gamma} = \mathbb{O}^{(\nu,L)} \mathbb{O}^{(\nu,k)} \nonumber \\ 
	& & F(S_\alpha,S_k,S_\gamma;S^\mathrm{op}_L,S^\mathrm{op}_k,{S^\mathrm{op}_L}';S_\alpha',S_k',S_\gamma') \, ,
\end{eqnarray}
where $F$ equals the famous Wigner-9j symbol~\cite{Wigner-1959} up to rescaling,
\begin{eqnarray}
	F(& & S_\alpha,S_k, S_\gamma;S^\mathrm{op}_L,S^\mathrm{op}_k,{S^\mathrm{op}_L}';S_\alpha',S_k',S_\gamma') = \nonumber \\
    & & \sqrt{  (2 S_L'+1)(2 S_k' + 1) (2 S_\gamma + 1)(2 S^\mathrm{op} + 1)} \nonumber \\
	& & \phantom{aa} W_{9j}(S_\alpha,S_k,S_\gamma;S^\mathrm{op}_L,S^\mathrm{op}_k,{S^\mathrm{op}_L}';S_\alpha',S_k',S_\gamma') \; .
\end{eqnarray}
We note, however, that the knowledge of $W_{9j}$ is not necessary, one can straightforwardly determine $F$ also from the contraction of six Clebsch-Gordan coefficient tensors, as it is shown in Fig.~\ref{fig:SU2FactorCell}. The total prefactor $\mathcal{F}_\tau$ is then a product of three elements of $F$ which correspond to the hierarchical product structure of $\mathcal{H}_\mathrm{eff}$ In practice, the nonzero elements of $F$ are tabulated in an efficiently searchable lookup table and used at the pre-calculation of the task list.
\begin{figure}
    \centering    
    \includegraphics[width=0.48\textwidth]{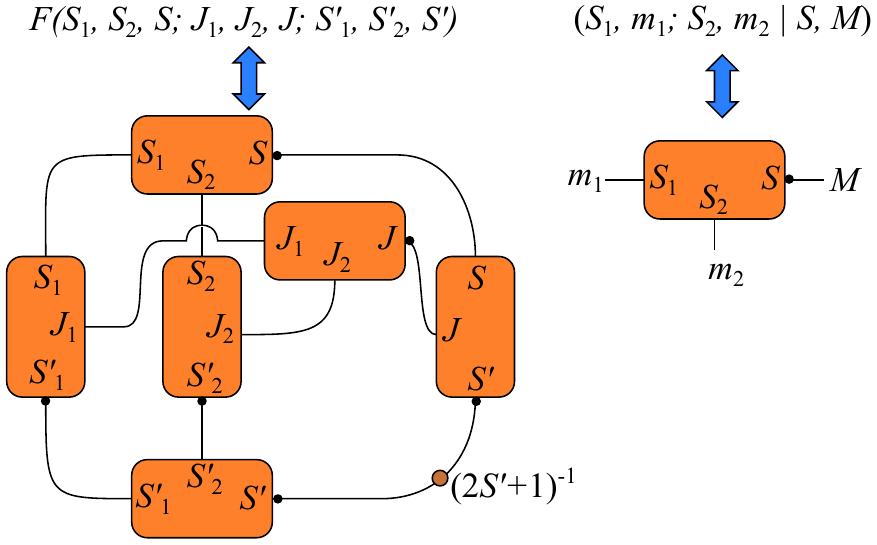}
    \caption{Generation of the factor $F$ as a contraction of six Clebsch-Gordan coefficients. The contracted lines indicate summation along the corresponding index. The coefficient $(S_1,m_1;S_2,m_2|S,M)$ is denoted by the tensor of three legs $\lbrace m_1,m_2,M \rbrace$. Outgoing leg $M$ is marked by a thick black dot, for the sake of clarity. Note the prefactor $(2S'+1)^{-1}$ indicated by the brown circle bottom right.   
    }    
    \label{fig:SU2FactorCell}
\end{figure}

\subsection{Parallelization strategies}

Besides the various algorithmic developments presented in the previous sections to reduce computational complexity, an efficient ab initio TNS/DMRG simulation also requires technical solutions to utilize the tremendous computational power offered by modern high-performance computing (HPC) centers. This is especially becoming a main issue in code developments nowadays
\cite{Hager-2004,Stoudenmire-2013,Nemes-2014,Ganahl-2019,Milsted-2019,Brabec-2021,Zhai-2021,Gray-2021,Unfried-2023,Ganahl-2023,Menczer-2023a,Menczer-2024d,Menczer-2024a,Menczer-2024c,Xiang-2024}
due to the enormous increase in accessible computational power by graphical processing units (GPUs) accelerated hardware~\cite{a100,gh200,mi300}.
In addition, considering the energy consumption of the TNS calculations, which is becoming one of the most important questions due
to high energy demands and costs, the power efficiency can be increased significantly via GPU accelerated hardware~\cite{Menczer-2023a}.

Quite recently, novel implementations of ab initio DMRG on hybrid CPU-GPU based architectures have been introduced~\cite{Menczer-2023a,Xiang-2024} being even capable of utilizing AI-accelerators via NVIDIA tensor core units \cite{Menczer-2024d, Menczer-2024a}
reaching a quarter petaflops performance on a single DGX-H100 node~\cite{Menczer-2023c}.
Further algorithmic developments based on
simple heterogeneous multiNode-multiGPU solutions by
exploiting the different bandwidths of
the different communication channels, like NVLink, PCIe, InfiniBand, and available storage media at different layers of the algorithm resulted in a state-of-the-art implementation of the quantum chemical version of DMRG~\cite{Menczer-2024b}.
\begin{figure}
    \centering    
    \includegraphics[width=0.48\textwidth]{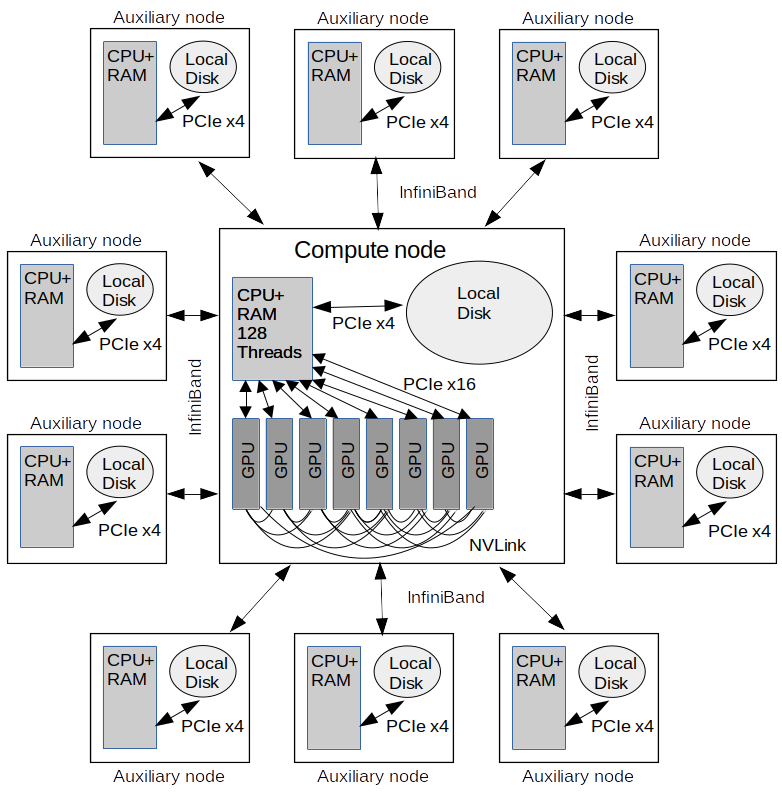}
    \caption{Schematic plot of hardware topology illustrating the various communication channels (arrows) between a powerful computational node, auxiliary nodes,
    GPU accelerators, and storage media. In practice, several powerful compute nodes can be utilized with strong GPU supports surrounded by many cheap auxiliary nodes with minimal computational capacity to handle IO operations. For further details, we guide interested readers to  Ref.~\cite{Menczer-2024b}.
    }    
    \label{fig:nodetopology}
\end{figure}

As discussed in Sec.~\ref{subsec:sym}, due to the conservation of quantum numbers, matrices and tensors are decomposed into smaller blocks in the DMRG framework, thus the underlying linear algebra is carried out via such block-sparse representation of the operators~\cite{Schollwock-2011,Nemes-2014}. Therefore, matrix and tensor operations, defined originally with full ranks in Eq.~\eqref{eq:Hamilton_act_nonsym}, e.g., are decomposed into several millions of independent operations (tasks) involving only certain subsets of blocks of operators (see Eqs.(\ref{eq:Hamilton_act_U1sym}- \ref{eq:Hamilton_act_su2})), which tasks are used to carry out the linear transformation of the effective Hamiltonian given by Eq.~\ref{eq:Hamiltonian}. As a consequence, DMRG poses an ideal case for massive parallelization; however, the reduction of the overhead due to communication among different computational units and the construction of efficient task schedulers based on various properties of heterogeneous hardware components are inevitable, and these pose a highly non-trivial piece of work.

Our high-speed DMRG code relies on various in-house built technical novelties.
These include \textit{(a)} a new mathematical framework for massive parallelization based on self-scheduled threading called contractor threads~\cite{Menczer-2024d}, \textit{(b)} a tree-graph based memory management model to reduce IO operations~\cite{Menczer-2023a}, \textit{(c)} a model for facilitating strided batched matrix operations for summation without the need for explicit sum reduction, \textit{(d)} an 
efficient handling of non-Abelian symmetries as discussed in Sec.~\ref{subsec:sym}, \textit{(e)} a massively parallel protocol to serialize and deserialize block sparse matrices and tensors, and \textit{(f)} asynchronous IO operation to minimize IO overhead, resulting in maximum performance rate for the main computing unit. For further details on such technical implementations, we refer interested readers to Refs.~\cite{Menczer-2023a,Menczer-2024d,Menczer-2024c}.

As a result, by exploiting all the algorithmic and technical developments, the computational complexity can be reduced drastically, and related computational time can be reduced by several orders of magnitudes~\cite{Menczer-2024a}. Therefore, those DMRG simulations that formerly took several weeks or even longer can nowadays be performed routinely on a daily basis,  opening new research directions to attack highly challenging complex problems in quantum chemistry and material science~\cite{Menczer-2025a}.

\section{Numerical approach}
\label{sec:numerics}

In this section, we present a detailed numerical analysis of concepts outlined in Sec.~\ref{sec:theory} for selected small system in a pedagogical style. The main purpose is to connect the theory and main steps of the numerical framework summarized in Fig.~\ref{fig:macroiter} to numerical simulations and results. We intend to highlight advantages obtained via mode optimization, but also show numerical issues related to convergence rate and stability.

\subsection{Spinless Hubbard model}

We start our discussion with a simple toy model, namely the two-dimensional spinless fermion model.
This is a simplified version of the general Hamiltonian given by Eq.~\ref{eq:Hamiltonian}
as it includes only nearest neighbor hopping and Coulomb interaction
\begin{equation}
H = \sum_{\langle i,j\rangle} \left( -t \,c^\dagger_ic_j + h.c \right) + \sum_{\langle i,j\rangle} V n_i n_j\,,
\label{eq:Spinless}
\end{equation}
where modes are spinless orbitals localized to sites of a square lattice. Spinless modes have only two states, they can be empty or occupied. The parameters $t$ and $V$ set the strength of hopping and interactions, respectively.  Despite the simple structure of the model, it poses a serious numerical challenge for DMRG when numerical simulations are performed on a two-dimensional quantum lattice with periodic boundary conditions in both spatial dimensions, i.e., on a torus topology.
In this case, local interactions become non-local after mapping the two-dimensional lattice to the one-dimensional, chain-like, topology of DMRG~\cite{Liang-1994, Noack-1994, Nishino-1995, White-2003, Menczer-2024a}.

In Fig.~\ref{fig:spinless_4x4_q2_BEA_swap} we show the block entropy area(BEA) and the ground state energy for model parameters $t=1$ and $V=1$ for a $4\times4$ lattice
using mode optimization with parameters presented in the caption of the figure.
In the thermodynamic limit, the model is characterized by a charge density wave ordered phase alternating partially occupied and almost empty orbitals on a checkerboard structure with two determinants having the same energy. 
The applied color coding of the individual DMRG sweeps of each macro iteration is according to Fig.~\ref{fig:macroiter}, i.e. black dots denote iterations with only DMRG, red dots mark micro iterations with local mode optimization while green dots mark iteration steps where only swap gates are applied.

As can be seen, in the figure both the BEA and the energy drop tremendously despite the very small value of $D_{\rm opt}=8$.
In this simulation, we used 6 DMRG sweeps (black), followed by 12 mode optimization sweeps (red) and a two swap gate layers (green) by increasing bond dimension by a factor of 2 in each layer, in order to preserve the energy value even after swapping the modes. This is visible as the last red data point and the two green dots correspond to the same energy values in each macro iteration step. After the swapgate layers, the first four DMRG sweeps are performed with such increased bond dimension, $D=32$, and a full forward and backward sweep is applied to cut back the bond dimension to $D_{\rm opt}=8$ before the subsequent mode optimization layer follows.
Therefore, the low energy values (black dots) below red data points are due to the increased bond dimension applied in the first four DMRG sweeps, while the 
small peaks in the BEA and energy profiles are due to the destruction of the MPS and the truncation of the bond dimension.
\begin{figure}
    \centering    
    \includegraphics[width=0.48\textwidth]{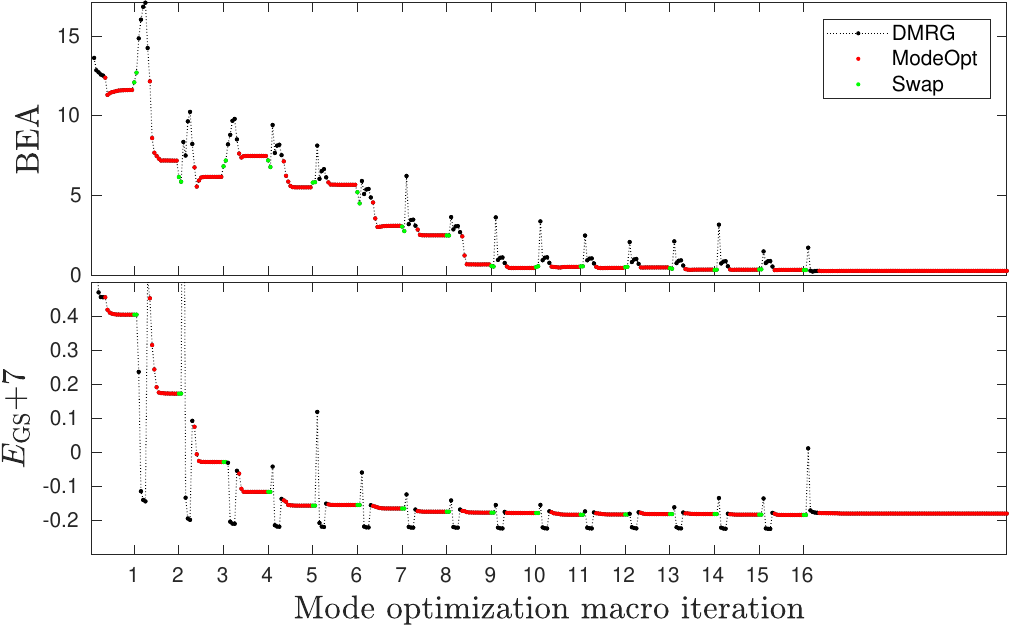}
    \caption{(a) Block entropy area $\calB_1(C)$ and (b) shifted ground state energy for the first 16 mode optimization macro iterations with $D_{\rm opt}=8$, $N_{\rm SW}^{\rm dmrg}=6$, $N_{\rm SW}^{\rm opt}=12$, and $N_{\rm SW}^{\rm opt,fin}=16$ for the $4\times 4$ spinless model.
    Black dots and dashed lines indicate data points obtained via DMRG, red dots refer to local mode optimization and green dots to the application of swap gates.
    The end of each mode optimization macro cycle corresponds to every subsequent green symbols.
    After 16 mode optimization macro iterations another 16
    mode optimization macro iterations are performed without utilizing swap gates for completeness.
    }    
    \label{fig:spinless_4x4_q2_BEA_swap}
\end{figure}
The big reduction in BEA becomes obvious by monitoring the block entropy profile shown in Fig.~\ref{fig:spinless_4x4_q2_S_swap} for a few selected mode optimization macro iteration steps. Note the two orders of magnitude reduction in the maximum of the block entropy. Since the computational complexity of DMRG is determined by the block entropy, i.e. by the level of entanglement encoded in the quantum many-body wave function, using optimized orbitals in large-scale DMRG calculations with large bond dimensions leads to far more accurate results than the original orbitals. 
The improvement of the basis  could also be monitored by the sum of the single mode entropies, called total correlation~\cite{Legeza-2004a,Legeza-2004b}, but BEA provides further information, i.e. the optimal ordering of the optimized modes as well.
For more details, we refer the interested readers to Refs.~\cite{Krumnow-2021,Menczer-2024a}. 
\begin{figure}
    \centering     
    \includegraphics[width=0.48\textwidth]{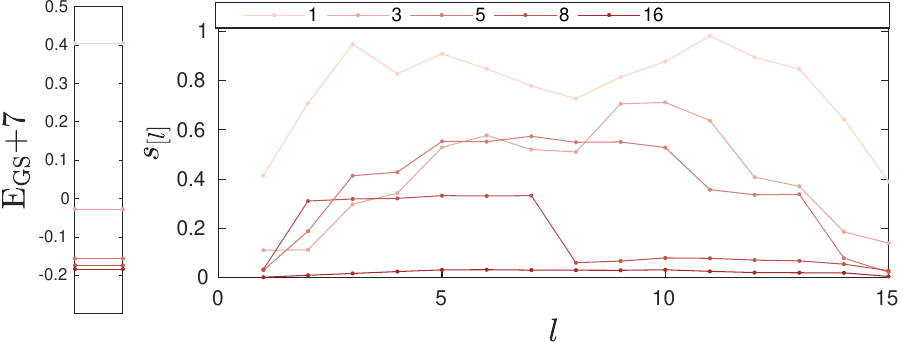} 
    \caption{
    Ground state energy (left) and block entropy, $s_{[l]}$, as a function of the left block size, $\ell$ (right) for some selected mode transformation macro iteration cycles 
    $1,3,5,8,16$, 
    with fixed bond dimension $D_{\rm mo}=8$ and 12 sweeps for the half-filled $4\times 4$ two-dimensional spinless fermionic model with model parameters given in the main text. The area under each curve corresponds to the block entropy area $\calB_1(C)$ shown in Fig.~\ref{fig:spinless_4x4_q2_BEA_swap}.
    Note that a full iteration to bring each mode to be neighbours at least once contains $N/2=16$ permutations, i.e., 16 mode transformation macro iterations.
    }  
    \label{fig:spinless_4x4_q2_S_swap}
\end{figure}

Next, we show results when the bond dimension is not increased for the swap gate layers,
i.e. keeping it fixed for the entire simulation, which makes the algorithm much faster, but we lose some exactness of the theory (see scattered energy data shown by the green dots). This is summarized in Fig.~\ref{fig:spinless_4x4_q1_BEA_swap}. 
Again a very stable and fast convergence is achieved, but larger peaks in the energy values appear. This, however, has no effect on the performance and very similar entropy profiles are obtained as has been shown in Fig.~\ref{fig:spinless_4x4_q2_S_swap}.
\begin{figure}
    \centering
    \includegraphics[width=0.48\textwidth]{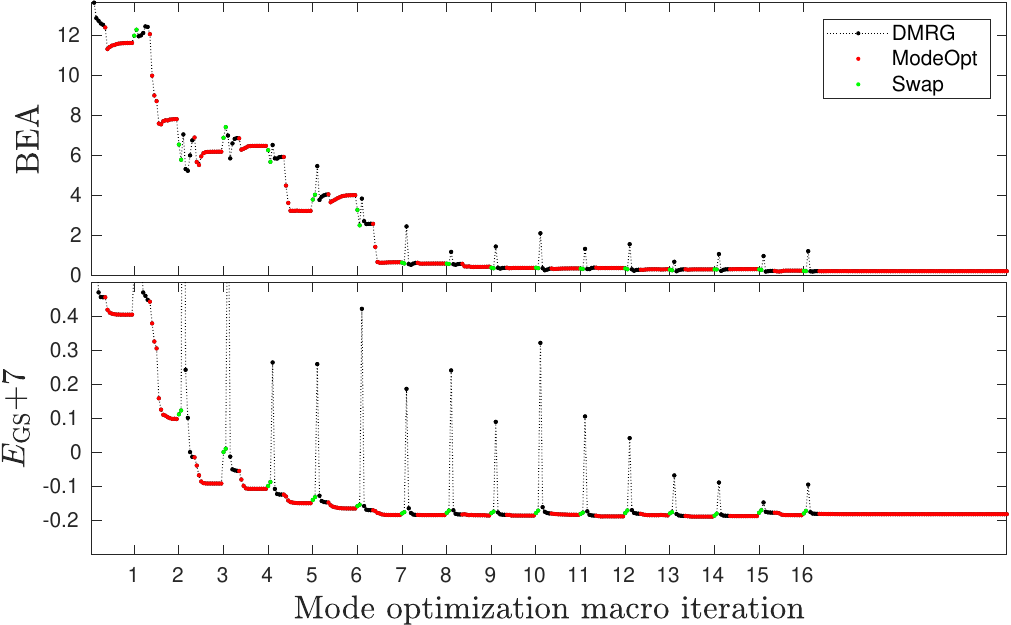}
    \caption{Similar to Fig.~\ref{fig:spinless_4x4_q2_BEA_swap}
    Similar to Fig.~\ref{fig:spinless_4x4_q2_BEA_swap}
    but bond dimension is not increased during the swap layer.}
    \label{fig:spinless_4x4_q1_BEA_swap}
\end{figure}

As a closing remark of our analysis, we discuss numerical stability. In Fig.~\ref{fig:spinless_4x4_m0080_q1_BEA_swap}we show the convergence profile using a higher bond dimension $D_{\rm opt}=80$ and keeping it fixed for the entire simulation. Here, we get much closer to the exact solution $E_\mathrm{GS}=-7.2295...$ that has been obtained by exact diagonalization of \eqref{eq:Spinless} (note the energy scale in the bottom panel). Interestingly, at some point DMRG can even loose the target state, that is a consequence of an almost degenerate first excited state with energy $E_1=-7.2242...$. Note that the energy difference is finite only for finite-sized systems and decreases to zero at the thermodynamic limit. As we can see, performing calculations at $D_{\rm opt}=80$ result an approximate ground state energy $\tilde{E}_\mathrm{GS} \approx -7.227$ that is already below $E_1$, i.e. the DMRG start to resolve the small gap between the two states. The algorithm, together with the scrambling caused by swap gates, provides an unsteady and fluctuating superposition of the two states during the iterations that is indicated by the increased BEA values. Note that $\tilde{E}_{\mathrm{GS}}$ remains almost constant meanwhile.
In practice this is not a problem as the global unitary, $U_{\rm tot}$, is saved after each macro iteration, thus optimized orbitals corresponding to the lowest BEA can be reconstructed and used for follow up calculations.
\begin{figure}
    \centering
    \includegraphics[width=0.48\textwidth]{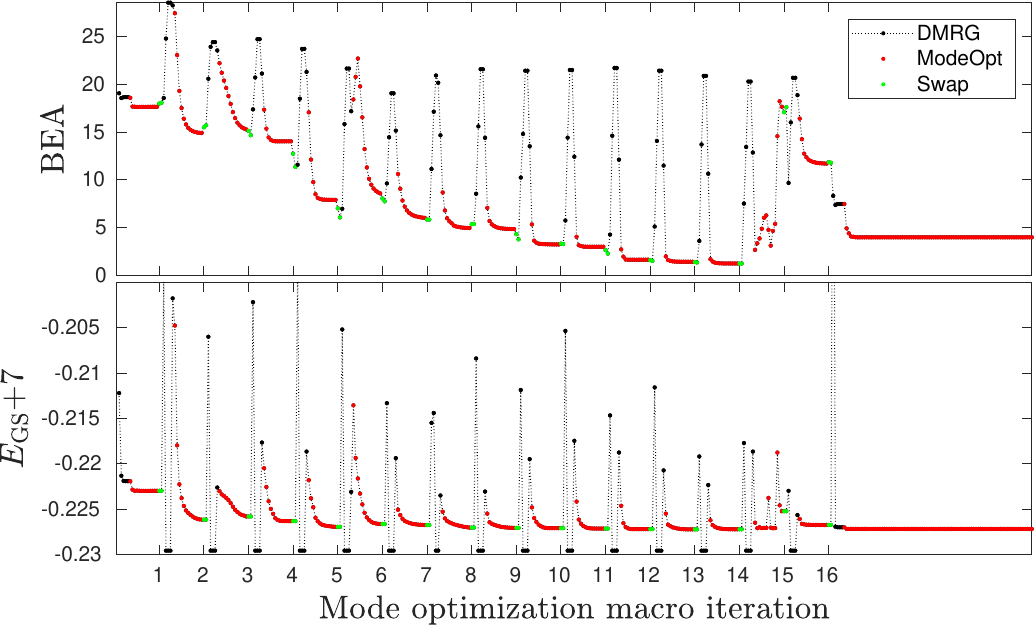}
    \caption{Similar to Fig.~\ref{fig:spinless_4x4_q2_BEA_swap}
        but $D_{\rm opt}=80$ has been used. Note that with $D=128$ the exact solution is recovered for the symmetric $k=8$ left-right block partitioning and, due to the $q=2$ bond extension parameter at the swap gate layers, this exact result is also displayed in the figure by the lowest energy black dots after following the swap gate layers.  
    }        
    \label{fig:spinless_4x4_m0080_q1_BEA_swap}
\end{figure}

\subsection{Quantum chemical systems}

In this section, we present our procedure for a few selected chemical systems. Again we do not intend to perform large-scale ab initio DMRG calculations, but rather discuss the numerical framework and its main features via examples.  

\subsubsection{F$_2$ dimer}

In Figs.~\ref{fig:f2_BEA_swap} and~\ref{fig:f2_q1_S_swap} we show results for the F$_2$ dimer at equilibrium bond length $r=2.68797a_0$, with $a_0$ being the Bohr radius, using split valence basis set~\cite{Schafer-1992} (for further details see Ref.~\cite{Legeza-2003a}) with $D_{\rm opt}=256$ starting the DMRG simulations by ordering orbitals energetically along the DMRG chain. Here we found that optimized rotation angle parameters of $u^{(k,k+1)}$ after each local mode optimization step were always very close to zero or $\pi/2$ reflecting the fact that the initial orbitals were already highly optimal and thus only their ordering gets optimized. 
Note that rotations with angles $\pi/2$ translate to swap gates.
The optimal initial basis is also apparent in the low initial BEA values.
In the current case, our procedure results in the optimal ordering of orbitals along the DMRG chain, as can be seen by the entropy profiles shown in \ref{fig:f2_q1_S_swap} for macro iterations with the lowest BEA values.
The two best configurations are practically the spatial reflection of each other.
Here, for tutorial purposes, we used $2\times d=36$ macro iteration steps to highlight the oscillation with a period $d/2$ of the permutations in Walecki's construction. Nevertheless, since the applied general unitary is much richer than the simpler permutation group used in the past, for example via the Fiedler-vector-based optimization protocols~\cite{Barcza-2011,Krumnow-2021}, 
our method can also be used as an automatized protocol for orbital ordering optimization as well. Finally, we remark on the relatively low value of the maximum of the block entropy which signals the weakly correlated, single reference character of the current problem. 

In Ref.~\onlinecite{Menczer-2024a} it has been demonstrated for the two-dimensional Hubbard model, that properties of the quantum many-body wave function determine whether a mode optimization by enforcing the same angles for the up and down spins is satisfactory or optimizing the two spin directions independently leads to significantly lower energies and BEA. Therefore, as a proof of principle, we have performed optimization by both options with $D_{\rm opt}=64$ and found the latter one resulting in only slightly lower energy values but very similar BEA profiles. However, mode optimization with different unitaries for the two spin directions results in spin-dependent integrals $t$ and $v$, therefore forbids the use of $SU(2)$ symmetry in the post mode optimization large-scale DMRG calculations~\cite{Menczer-2024d}. Consequently, enforcing spin-independent unitaries is more beneficial in the current case.
\begin{figure}
    \centering
    \includegraphics[width=0.48\textwidth]{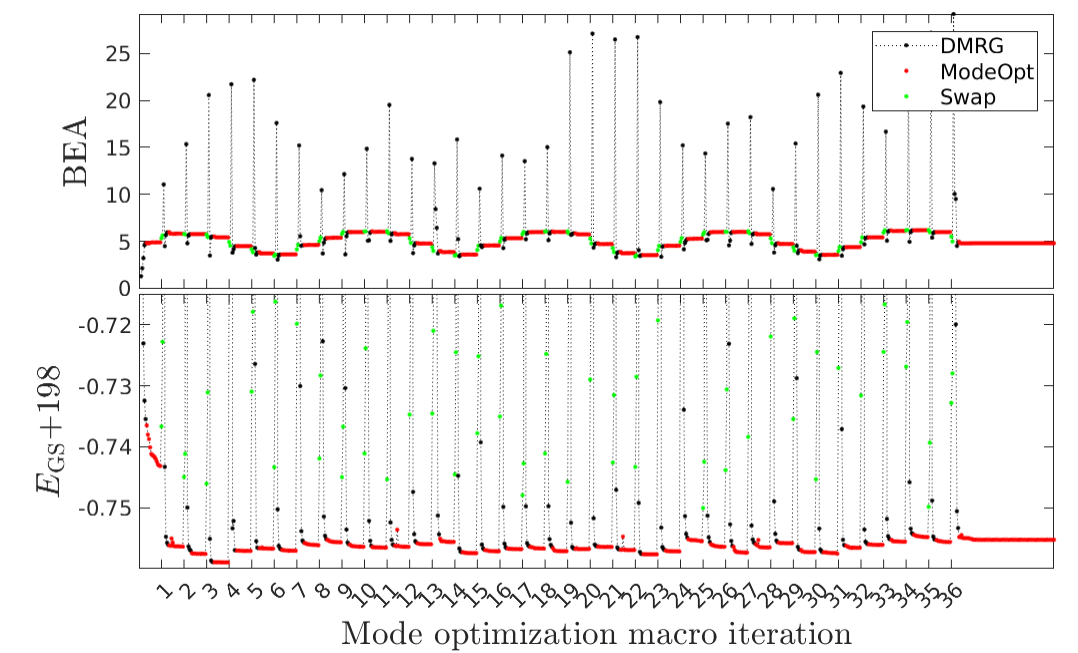}
    \caption{Similar to Fig.~\ref{fig:spinless_4x4_q1_BEA_swap}
    but for the F$_2$ molecule with $D_{\rm opt}=256$, $N_{\rm SW}^{\rm dmrg}=6$, $N_{\rm SW}^{\rm opt}=12$, and $N_{\rm SW}^{\rm opt,fin}=18$.
    }    
    \label{fig:f2_BEA_swap}
\end{figure}
\begin{figure}
    \centering     
    \includegraphics[width=0.48\textwidth]{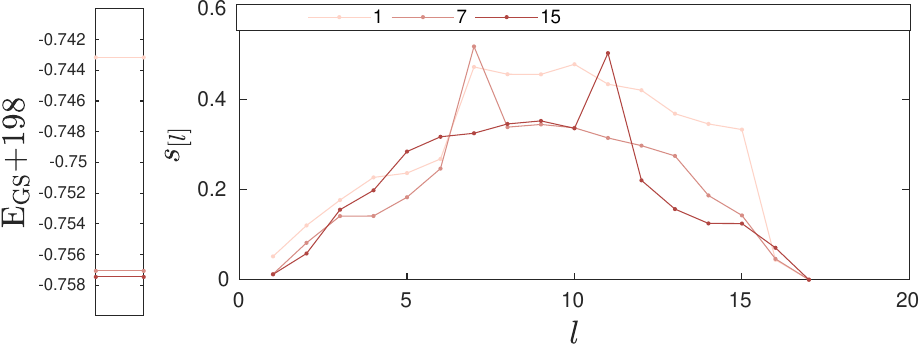}  
    \caption{
    Similar to Fig.~\ref{fig:spinless_4x4_q2_S_swap} but for the F$_2$ molecule obtained with parameters given in caption of Fig.~\ref{fig:f2_BEA_swap} for some selected mode transformation macro iteration cycles.
    }  
    \label{fig:f2_q1_S_swap}
\end{figure}

\subsubsection{Nitrogen dimer}

As a second example, we present an analysis for the nitrogen dimer in the cc-pVDZ basis~\cite{Dunning-1989}
for various bond lengths.
DMRG benchmark calculations on the full active space by correlating 14 electrons on 28 orbitals, CAS(14,28), of the stretched nitrogen dimer have been presented in various works of the past decades~\cite{Chan-2004b,Faulstich-2019b,Boguslawski-2013,Mate-2022}. 
Recently, it has also been shown how the multi-reference character of the wave function can be compressed by local mode optimization supplemented by Fiedler-vector-based reordering~\cite{Mate-2022,Petrov-2023}, in which works various entropic quantities and the decomposition of the wave function in terms of CI coefficients were monitored.
Here we repeat such analysis, but employing the global mode optimization protocol with swap gates, and present obtained convergence profiles for the equilibrium geometry $r=2.118 a_0$ and stretched ones at $r=3.6 a_0$ and $r=4.2a_0$.  
For the equilibrium geometry, our procure using $D_{\rm opt}=64$ leads again only to orbital optimization, and the maximum of the block entropy profile does not change significantly as summarized in Figs.~\ref{fig:n2_r2118_BEA_swap} and ~\ref{fig:n2_r2118_BEA_swap}. 
\begin{figure}
    \centering
    \includegraphics[width=0.48\textwidth]{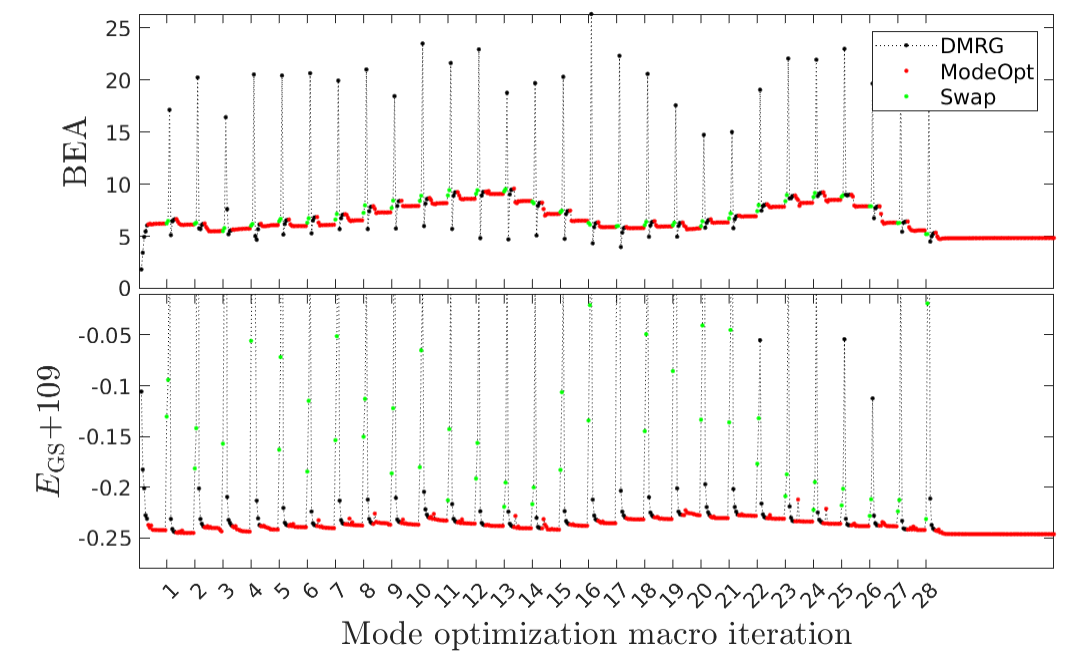}
    \caption{Similar to Fig.~\ref{fig:f2_BEA_swap}
    but for the N$_2$ dimer at the equilibrium geometry at $r=2.118a_0$ using $D_{\rm opt}=64$.
    } 
    \label{fig:n2_r2118_BEA_swap}
\end{figure}
\begin{figure}
    \centering     
    \includegraphics[width=0.48\textwidth]{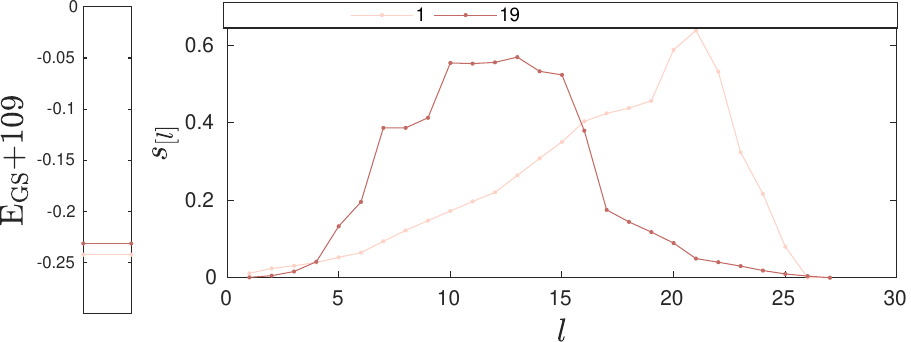}    
    \caption{Similar to Fig.~\ref{fig:f2_q1_S_swap}
    but for the N$_2$ dimer at the equilibrium geometry at $r=2.118a_0$ using $D_{\rm opt}=64$.
    }  
    \label{fig:n2_r2118_S_swap}
\end{figure}
In contrast to this, for $r=3.6a_0$ and $4.2a_0$
the optimized orbitals corresponding to macro iterations with the lowest BEA values display a significant reduction in the maximum of the block entropy profile\footnote{We remark that the initial slight increase of BEA in the first macro iteration is an artefact caused by a non-converged wave function.}. This is an important achievement since both the computational complexity and the memory demand depend polynomially on the bond dimension whose required value is an exponential function of  the block entropy  for a given pre-set numerical accuracy~\cite{Legeza-2003a,Legeza-2003b}.  
Therefore, even if the BEA is lowered only slightly, for example when the optimized block entropy profile is broadened compared to the initial profile, the new mode set is already more beneficial computationally if its maximum is lowered. Thanks to our massively parallel implementation for modern high-performance infrastructures, an increase in overall computational complexity can be well compensated ~\cite{Brabec-2021,Menczer-2023a,Menczer-2024d,Menczer-2024c}. However, the maximal system sizes and the achievable accuracy of the MPS wavefunction are roughly determined by the entropy peak due to the memory limits of the used hardware.

Finally, we note that the maximum of the block entropy increases as the molecule is stretched reflecting the stronger multi-reference character of the system when the initial basis is applied. In contrast to this, after mode optimization such multireference character is suppressed due to the big reduction of the maximum of the block entropy. In addition, while in our previous work using a protocol based on local mode optimization~\cite{Mate-2022} we could not obtain a well-converged data set for $r=3.6a_0$, now with the global mode optimization, we get well-converged profile also here due to the systematic permutation of modes, as it is apparent in Fig.~\ref{fig:n2_r3600_S_swap}.  
\begin{figure}
    \centering
    \includegraphics[width=0.48\textwidth]{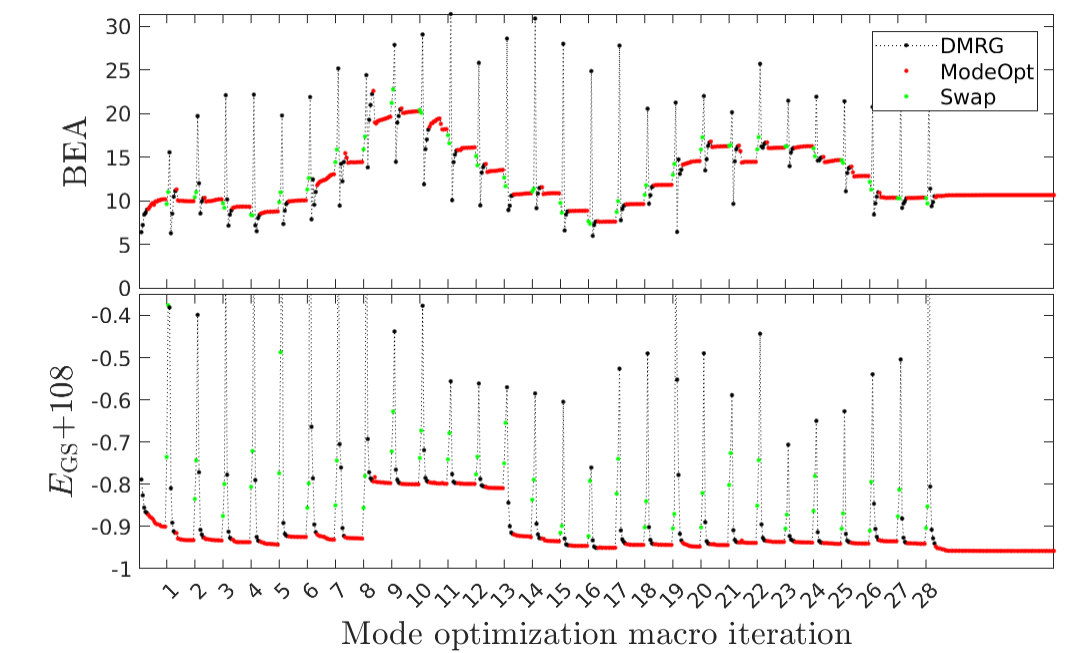}
    \caption{Similar to Fig.~\ref{fig:n2_r2118_BEA_swap} but for $r=3.600a_0$.
    }%
    \label{fig:n2_r3600_BEA_swap}
\end{figure}
\begin{figure}
    \centering     
    \includegraphics[width=0.48\textwidth]{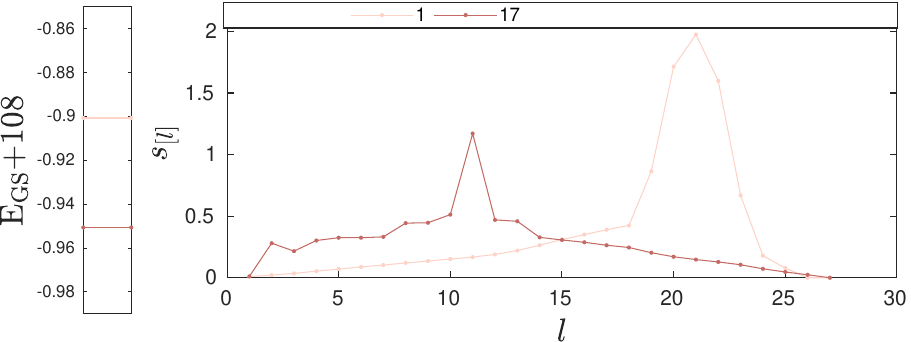}    
    \caption{Similar to Fig.~\ref{fig:n2_r2118_S_swap} but for $r=3.600a_0$.
    }  
    \label{fig:n2_r3600_S_swap}
\end{figure}

\begin{figure}
    \centering
    \includegraphics[width=0.48\textwidth]{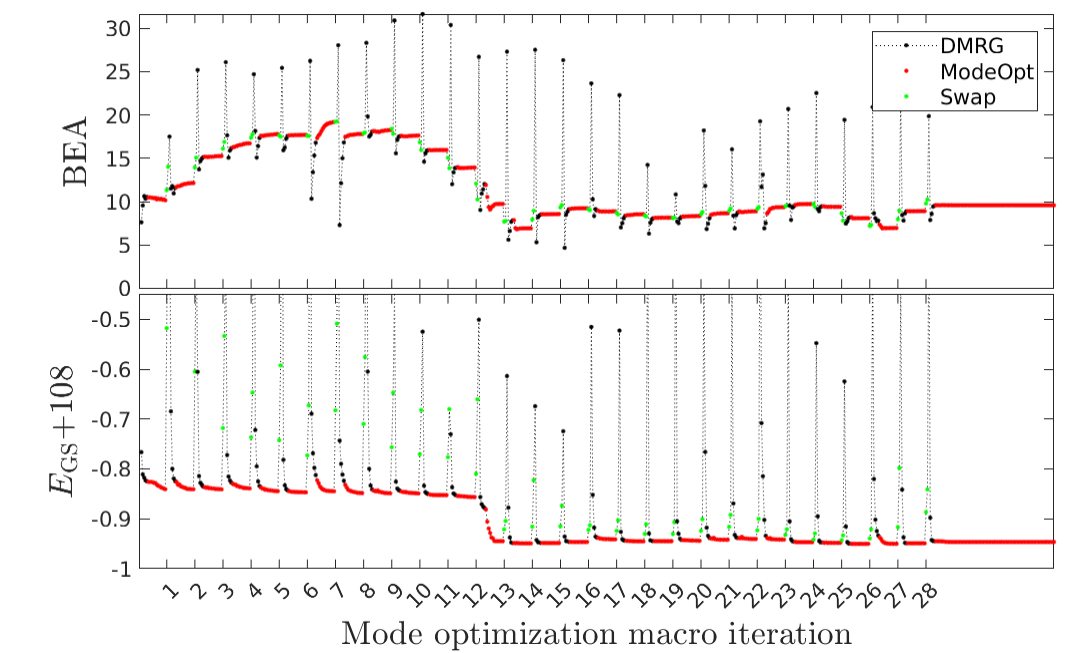}
    \caption{Similar to Fig.~\ref{fig:n2_r2118_BEA_swap} but for $r=4.200a_0$.
} 
    \label{fig:n2_r4200_BEA_swap}
\end{figure}
\begin{figure}
    \centering     
    \includegraphics[width=0.48\textwidth]{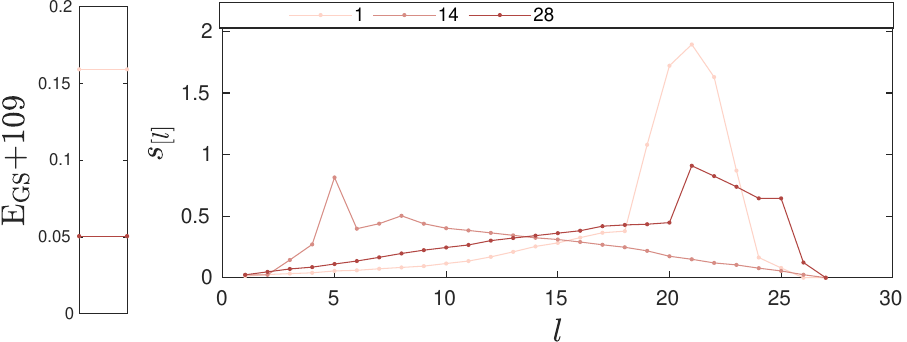}    
    \caption{Similar to Fig.~\ref{fig:n2_r2118_S_swap} but for $r=4.200a_0$
    }  
    \label{fig:n2_r4200_S_swap}
\end{figure}

\subsubsection{Chromium dimer}

For completeness, in this section we show results for
the notoriously strongly correlated chromium dimer which is subject to usual benchmark calculations even nowadays ~\cite{Kurashige-2011,Ma-2017,Sharma-2012,Veis-2016,Barcza-2022,Larsson-2022,Larsson-2021-arxiv}. 
Our result for
Cr$_2$ starting with a natural orbital basis obtained from the cc-pVDZ atomic basis (see Ref.~\cite{Barcza-2022}) at its equilibrium geometry, $d=1.6788 \AA$, corresponding to a full orbital space CAS(12,68) is shown in Figs.~\ref{fig:cr2_m0064_q1_sw7_19_pd2_ord0} and ~\ref{fig:cr2_S_swap}
\begin{figure}
    \centering
    \includegraphics[width=0.48\textwidth]{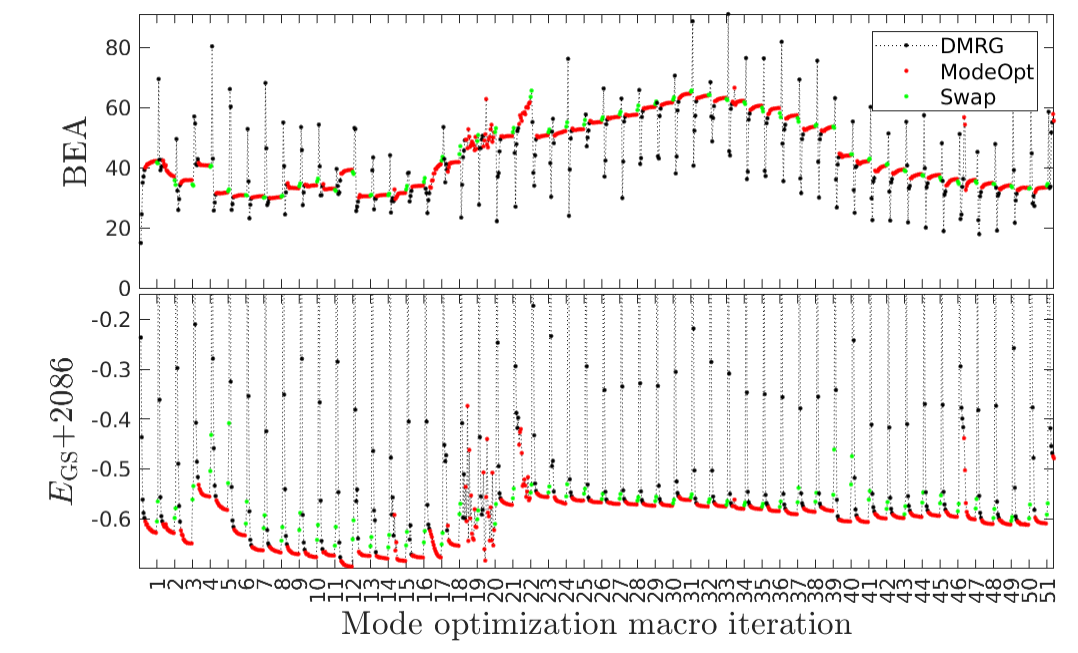}
    \caption{Similar to Fig.~\ref{fig:n2_r2118_BEA_swap} but for the Chromium dimer using $D_{\rm opt}=256$.
    }     \label{fig:cr2_m0064_q1_sw7_19_pd2_ord0}
\end{figure}
\begin{figure}
    \centering         \includegraphics[width=0.48\textwidth]{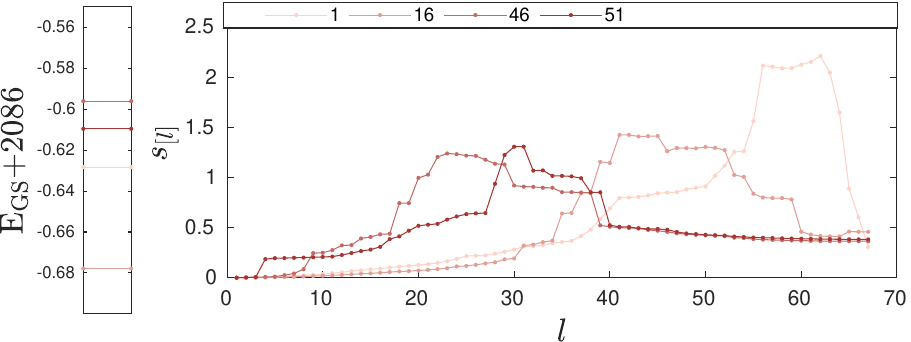}    
    \caption{Similar to Fig.~\ref{fig:n2_r2118_S_swap} but for the Chromium dimer
    using $D_{\rm opt}=256$.
    }  
    \label{fig:cr2_S_swap}
\end{figure}

We close this section by making a remark. In order to speed up mode optimization it is also possible to combine our former version based on the heuristic Fiedler-vector based approach to obtain a good ordering and starting mode configuration for the global mode optimization via swap-gates. Based on our numerical experiences on Hubbard-like model systems, a few macro iterations via Fiedler vector followed by swap gate based mode optimization lead to lower energy and BEA faster than pure application of global mode optimization protocol.

\section{Post mode optimization procedures}
\label{sec:postmode}

\subsection{Large scale multiNode-multiGPU DMRG simulations via AI accelerators}

Following mode optimization performed with low bond dimensions, large scale DMRG simulations are also performed in the optimized basis with large bond dimensions in order to determine electronic structures with high accuracy.
As an example, here we present such results for the Cr$_2$ dimer CAS(12,68) model space obtained via our massively parallel spin adapted DMRG code using a powerful compute node based on AMD EPYC 7702 CPUs with $2\times64$ cores supplied with eight NVIDIA A100-SXM4-80GB devices. Furthermore, to hide IO operations via asynchronous IO data management related to precontracted tensor network components, the compute node is supplied by ten auxiliary nodes according to the topology shown in Fig.~\ref{fig:nodetopology}.  
Using $D=4096$ $SU(2)$ multiplets, corresponding to largest $U(1)$ bond dimension values around 10800,
the decomposition of the total wall time as a function of DMRG iteration steps for a few sweeps  
is shown in Fig.~\ref{fig:N68_dmrgtime}.
Here, Diag$_{\rm H}$ stands for the diagonalization of the effective Hamiltonian,
Ren$_{\rm l}$ and Ren$_{\rm r}$ for the renormalization of the left and right blocks,
StVec for wave function transformation, i.e., for generation of the starting vector for the diagonalization step, 
SVD for singular vale decomposition,
Tables for generating meta data, like the task lists in Eq. \eqref{eq:Hamilton_act_su2}, for scheduling algorithms, and
IO for all read and write operations, respectively.
Functions already converted to our hybrid CPU-GPU kernel are indicated by asterisks.
For further details we guide interested readers to Ref.~\cite{Menczer-2024b}.
\begin{figure}
    \centering    
\includegraphics[width=0.48\textwidth]{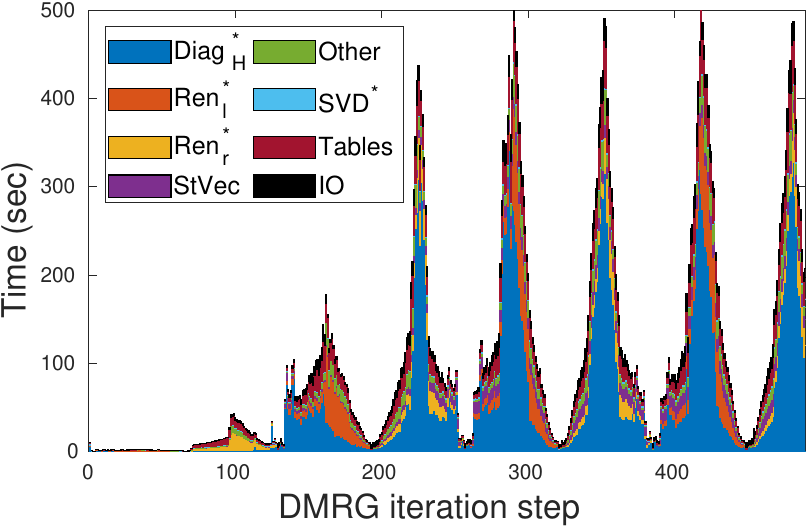}
    \caption{Decomposition of the total wall time 
    as a function of DMRG iteration steps for a few sweeps
    via asynchronous IO operations and 10 auxiliary nodes 
    for the Chromium dimer for the CAS(12,68) model space using $D=4096$ $SU(2)$ multiplets corresponding to largest $U(1)$ bond dimension values around 10800.
    The asterisks indicate functions converted to GPU already. The description of the legend is given in the main text. The first (warmup) sweep is performed with $D=512$.
    }
    \label{fig:N68_dmrgtime}
\end{figure}

Comparing our result to CCSD, CCSD(T), CCSDT, and CCSDTQ \cite{Barcza-2022}(E=-2086.7401, -2086.8785, -2086.8675, -2086.8689, respectively) we conclude that
DMRG variational energies converges to CCSDT with increasing bond dimension values, i.e., E=-2086.8079, -2086.8374, -2086.8580, -2086.8637 for D=1024, 2048, 3072, 4096, respectively. For more details on convergence properties we guide interested readers to Ref.~\cite{Szalay-2015b}.
Here we note that in practice the truncation error or the loss of quantum information entropy is kept fixed in large scale DMRG calculations and bond dimension is adjusted dynamically according to such pre-set error margin~\cite{Legeza-2003a,Legeza-2004b}.

Regarding performance and scaling properties we show the total diagonalization time together with IO overhead for seven sweeps in minutes as a function of $SU(2)$ bond dimension (multiplets) in Fig.~\ref{fig:cr2_perf}. Numbers next to the data points indicate the corresponding $U(1)$ bond dimension values.  
As can be seen for a broad range of $D_{SU(2)}$ bond dimension values an almost linear scaling is achieved with an exponent $p_1~\simeq 1.6$ while for larger bond dimension values, when saturation in performance is reached due to limited memory of GPUs, an exponent $p_2\simeq 2.9$ is recovered that is very close to the theoretical value $p_\mathrm{th}=3$. Note that for the current CAS(12,68) model space, slightly larger exponents are found compared to those reported in Refs.~\cite{Menczer-2023a,Menczer-2023b}, due to a significantly lower number of electrons leading to a lower number of sectors appearing in block sparse representation of the underlying matrices and tensors.
\begin{figure}
  \centering  \includegraphics[width=0.48\textwidth]{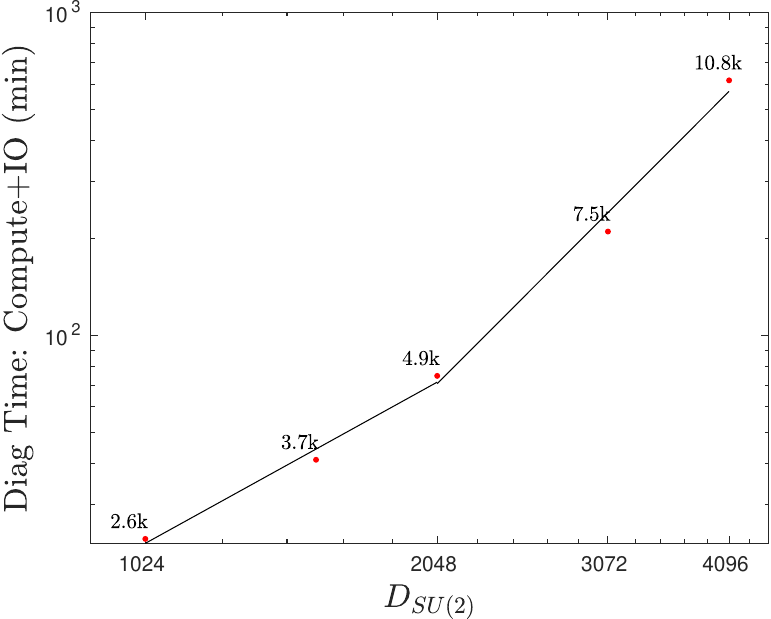}
  \caption{
  The corresponding total diagonalization time together with IO overhead for seven sweeps in minutes
  as a function of $SU(2)$ bond dimension (multiplets)
  for the Chromium dimer on a CAS(12,68) model space. 
  Numbers next to the data points indicate the corresponding $U(1)$ bond dimension values.  The solid lines are results of first-order polynomial fits leading to exponents $p_1=1.6$ and $p_2=2.9$.
}
 \label{fig:cr2_perf}
\end{figure}

\subsection{Post DMRG methods to capture both static and dynamic correlations}

Tensor network state methods in general are based on local optimization steps as discussed in Sec.~\ref{sec:theory}, thus the network must be swept through lot of times in order to capture also non-local correlations. Therefore, it is a natural step to combine DMRG with other conventional methods to exploit different scaling properties and benefits offered by them. In the past decades varios post-DMRG methods have been developed to capture both the so called static and dynamic correlations ~\cite{Piecuch-1996,Veis-2016,Faulstich-2019a,Faulstich-2019b,Leszczyk-2022,Kurashige-2011,Kurashige-2013,Sharma-2014b,Saitow-2013,Cheng-2022} which all provide corrections on top of the DMRG wave function.

Quite recently, however, 
building on the concept of dynamically extended active space(DEAS) procedure~\cite{Legeza-2003b},
the DMRG algorithm has been cross-fertilized with the concept of restricted active space (RAS) approach \cite{Barcza-2022,Larsson-2022,Cheng-2022} and even a rigorous mathematical analysis has lead to an efficient 
and stable extrapolation procedure to obtain the full-CI ground state energy within chemical accuracy using only limited CAS spaces~\cite{Friesecke-2022b}. 
Since the DMRG-RAS is an embedding method, in the sense that when orbitals are partitioned into two subspaces, CAS and EXT, 
the correlations between them are calculated self-consistently, in contrast to other post-DMRG approaches. Therefore, DMRG-RAS is
variational and the error exhibits monotone decay as the CAS-EXT split increases, unlike TCC where non-monotone behavior has been observed ~\cite{Faulstich-2019a,Faulstich-2019b}.
Connecting the efficiency of the DMRG-RAS-X to our discussions presented in previous sections, by performing DMRG calculations on 17 selected orbitals out of the 68 orbitals of the full model space used for Cr$_2$ we reported
an approximate, but variational energy $E^0(17,2)=-2086.8769$ in Ref.~\cite{Friesecke-2022b}. Note that this is already below the CCSDTQ by $8\times10^{-3}$. Moreover, an
extrapolated energy was found to be $E_{\rm RASX}=2086.891$.

Combination of this novel method, with our hybrid CPU-multiGPU kernel (see results for FeMoco in Ref.~\cite{Friesecke-2022b}) has the potential to raise DMRG to a routinely applied method on the daily basis to complex strongly correlated (multi reference) problems requiring large orbital spaces. In addition, sampling the RAS space via the GPU accelerated solution on-the-fly together with additional mathematical developments will easily boost DMRG to handle efficiently static and dynamic correlations for systems with several hundreds of orbitals.
This can be further boosted via DMRG-CASSCF \cite{Zgid-2008c} calculations where a combination of our DMRG code with ORCA~\cite{neese_orca_2020} already allowed us to attack large CAS(100,100) model spaces describing correlations among some 500 electrons on 1500 orbitals in the full model space~\cite{Menczer-2025a}.

In general, all the various concepts discussed in this tutorial review can be extended to networks with more general topologies than the chain-like topology of the DMRG. Promising extensions include ab initio tree-like tensor network state methods~\cite{Murg-2010a,Nakatani-2013,Murg-2015,Gunst-2018,Gunst-2019}.

\section{Conclusion}
\label{sec:conclusion}

In this brief pedagogical overview, we presented
recent advances in tensor network state methods that have the potential to broaden their scope of application radically for strongly correlated molecular systems. We discussed in detail the underlying theory behind global fermionic mode optimization, i.e., a general approach to finding an optimal matrix product state (MPS) parametrization of a quantum many-body wave function
with the minimum number of parameters for a given error margin. 

A short summary of main technical developments on parallelization strategies for hybrid CPU-GPU-based architectures together with
an efficient treatment of non-Abelian symmetries on high-performance computing (HPC) infrastructures was also discussed.

The numerical procedure is analyzed in a pedagogical style by connecting elements of the theory to results obtained via DMRG simulations on selected two-dimensional quantum lattice models and various molecular systems. 

Finally, large-scale DMRG simulations were presented for the Chromium dimer and future extensions based on  
restricted active space DMRG-RAS-X method, externally corrected tailored coupled cluster calculations and the DMRG-based self-consistent field framework are discussed which methods have the potential of capturing both static and dynamic correlations efficiently.

One natural extension of this work will be the direct optimization of long-range two-mode unitaries which is implemented only for neighboring modes in our present version. These ongoing developments require deeper modification of the original DMRG algorithm but they offer a more robust and mathematically precise way to finding the minimum of the cost function BEA. One can also avoid the swap layers that cause unintentional scrambling of the MPS which results in the undamped oscillations of the cost function.

\emph{Acknowledgements} -- This research was supported 
by the Hungarian National Research, Development and Innovation Office (NKFIH) through Grant Nos.~K134983 and TKP2021-NVA-04
by the Quantum Information National Laboratory of Hungary, 
and by the Hans Fischer Senior Fellowship programme funded by the Technical University
of Munich – Institute for Advanced Study. 
\"O.L. has also been supported
by the Center for Scalable and Predictive methods
for Excitation and Correlated phenomena (SPEC),
funded as part of the Computational Chemical Sciences Program FWP 70942, by the U.S. Department of Energy
(DOE), Office of Science, Office of Basic Energy Sciences, Division of Chemical Sciences, Geosciences, and Biosciences at Pacific Northwest National Laboratory. 
M.A.W. has also been supported by the Janos
Bolyai Research Scholarship of the Hungarian Academy of Sciences.
We thank computational support from the Wigner Scientific Computing Laboratory (WSCLAB) and the national supercomputer HPE Apollo Hawk at the High Performance Computing Center Stuttgart (HLRS) under the grant number MPTNS/44246.

\bibliographystyle{achemso}
\bibliography{main}{}

\begin{thebibliography}{112}%
\makeatletter
\providecommand \@ifxundefined [1]{%
 \@ifx{#1\undefined}
}%
\providecommand \@ifnum [1]{%
 \ifnum #1\expandafter \@firstoftwo
 \else \expandafter \@secondoftwo
 \fi
}%
\providecommand \@ifx [1]{%
 \ifx #1\expandafter \@firstoftwo
 \else \expandafter \@secondoftwo
 \fi
}%
\providecommand \natexlab [1]{#1}%
\providecommand \enquote  [1]{``#1''}%
\providecommand \bibnamefont  [1]{#1}%
\providecommand \bibfnamefont [1]{#1}%
\providecommand \citenamefont [1]{#1}%
\providecommand \href@noop [0]{\@secondoftwo}%
\providecommand \href [0]{\begingroup \@sanitize@url \@href}%
\providecommand \@href[1]{\@@startlink{#1}\@@href}%
\providecommand \@@href[1]{\endgroup#1\@@endlink}%
\providecommand \@sanitize@url [0]{\catcode `\\12\catcode `\$12\catcode
  `\&12\catcode `\#12\catcode `\^12\catcode `\_12\catcode `\%12\relax}%
\providecommand \@@startlink[1]{}%
\providecommand \@@endlink[0]{}%
\providecommand \url  [0]{\begingroup\@sanitize@url \@url }%
\providecommand \@url [1]{\endgroup\@href {#1}{\urlprefix }}%
\providecommand \urlprefix  [0]{URL }%
\providecommand \Eprint [0]{\href }%
\providecommand \doibase [0]{https://doi.org/}%
\providecommand \selectlanguage [0]{\@gobble}%
\providecommand \bibinfo  [0]{\@secondoftwo}%
\providecommand \bibfield  [0]{\@secondoftwo}%
\providecommand \translation [1]{[#1]}%
\providecommand \BibitemOpen [0]{}%
\providecommand \bibitemStop [0]{}%
\providecommand \bibitemNoStop [0]{.\EOS\space}%
\providecommand \EOS [0]{\spacefactor3000\relax}%
\providecommand \BibitemShut  [1]{\csname bibitem#1\endcsname}%
\let\auto@bib@innerbib\@empty
\bibitem [{\citenamefont {Affleck}\ \emph {et~al.}(1987)\citenamefont
  {Affleck}, \citenamefont {Kennedy}, \citenamefont {Lieb},\ and\ \citenamefont
  {Tasaki}}]{Affleck-1987}%
  \BibitemOpen
  \bibfield  {author} {\bibinfo {author} {\bibfnamefont {I.}~\bibnamefont
  {Affleck}}, \bibinfo {author} {\bibfnamefont {T.}~\bibnamefont {Kennedy}},
  \bibinfo {author} {\bibfnamefont {E.~H.}\ \bibnamefont {Lieb}},\ and\
  \bibinfo {author} {\bibfnamefont {H.}~\bibnamefont {Tasaki}},\ }\bibfield
  {title} {\bibinfo {title} {Rigorous results on valence-bond ground states in
  antiferromagnets},\ }\href {https://doi.org/10.1103/PhysRevLett.59.799}
  {\bibfield  {journal} {\bibinfo  {journal} {Phys. Rev. Lett.}\ }\textbf
  {\bibinfo {volume} {59}},\ \bibinfo {pages} {799} (\bibinfo {year}
  {1987})}\BibitemShut {NoStop}%
\bibitem [{\citenamefont {Fannes}\ \emph {et~al.}(1992)\citenamefont {Fannes},
  \citenamefont {Nachtergaele},\ and\ \citenamefont {Werner}}]{Fannes-1992}%
  \BibitemOpen
  \bibfield  {author} {\bibinfo {author} {\bibfnamefont {M.}~\bibnamefont
  {Fannes}}, \bibinfo {author} {\bibfnamefont {B.}~\bibnamefont
  {Nachtergaele}},\ and\ \bibinfo {author} {\bibfnamefont {R.~F.}\ \bibnamefont
  {Werner}},\ }\bibfield  {title} {\bibinfo {title} {Finitely correlated states
  on quantum spin chains},\ }\href {https://doi.org/10.1007/BF02099178}
  {\bibfield  {journal} {\bibinfo  {journal} {Communications in Mathematical
  Physics}\ }\textbf {\bibinfo {volume} {144}},\ \bibinfo {pages} {443}
  (\bibinfo {year} {1992})}\BibitemShut {NoStop}%
\bibitem [{\citenamefont {White}(1992)}]{White-1992b}%
  \BibitemOpen
  \bibfield  {author} {\bibinfo {author} {\bibfnamefont {S.~R.}\ \bibnamefont
  {White}},\ }\bibfield  {title} {\bibinfo {title} {Density matrix formulation
  for quantum renormalization groups},\ }\href
  {https://doi.org/10.1103/PhysRevLett.69.2863} {\bibfield  {journal} {\bibinfo
   {journal} {Phys. Rev. Lett.}\ }\textbf {\bibinfo {volume} {69}},\ \bibinfo
  {pages} {2863} (\bibinfo {year} {1992})}\BibitemShut {NoStop}%
\bibitem [{\citenamefont {White}(1993)}]{White-1993}%
  \BibitemOpen
  \bibfield  {author} {\bibinfo {author} {\bibfnamefont {S.~R.}\ \bibnamefont
  {White}},\ }\bibfield  {title} {\bibinfo {title} {Density-matrix algorithms
  for quantum renormalization groups},\ }\href
  {https://doi.org/10.1103/PhysRevB.48.10345} {\bibfield  {journal} {\bibinfo
  {journal} {Phys. Rev. B}\ }\textbf {\bibinfo {volume} {48}},\ \bibinfo
  {pages} {10345} (\bibinfo {year} {1993})}\BibitemShut {NoStop}%
\bibitem [{\citenamefont {Nishino}(1995)}]{Nishino-1995}%
  \BibitemOpen
  \bibfield  {author} {\bibinfo {author} {\bibfnamefont {T.}~\bibnamefont
  {Nishino}},\ }\bibfield  {title} {\bibinfo {title} {Density {Matrix}
  {Renormalization} {Group} {Method} for {2D} {Classical} {Models}},\ }\href
  {https://doi.org/10.1143/JPSJ.64.3598} {\bibfield  {journal} {\bibinfo
  {journal} {Journal of the Physical Society of Japan}\ }\textbf {\bibinfo
  {volume} {64}},\ \bibinfo {pages} {3598} (\bibinfo {year}
  {1995})}\BibitemShut {NoStop}%
\bibitem [{\citenamefont {\"Ostlund}\ and\ \citenamefont
  {Rommer}(1995)}]{Ostlund-1995}%
  \BibitemOpen
  \bibfield  {author} {\bibinfo {author} {\bibfnamefont {S.}~\bibnamefont
  {\"Ostlund}}\ and\ \bibinfo {author} {\bibfnamefont {S.}~\bibnamefont
  {Rommer}},\ }\bibfield  {title} {\bibinfo {title} {Thermodynamic limit of
  density matrix renormalization},\ }\href
  {https://doi.org/10.1103/PhysRevLett.75.3537} {\bibfield  {journal} {\bibinfo
   {journal} {Phys. Rev. Lett.}\ }\textbf {\bibinfo {volume} {75}},\ \bibinfo
  {pages} {3537} (\bibinfo {year} {1995})}\BibitemShut {NoStop}%
\bibitem [{\citenamefont {Rommer}\ and\ \citenamefont
  {\"Ostlund}(1997)}]{Rommer-1997}%
  \BibitemOpen
  \bibfield  {author} {\bibinfo {author} {\bibfnamefont {S.}~\bibnamefont
  {Rommer}}\ and\ \bibinfo {author} {\bibfnamefont {S.}~\bibnamefont
  {\"Ostlund}},\ }\bibfield  {title} {\bibinfo {title} {Class of ansatz wave
  functions for one-dimensional spin systems and their relation to the density
  matrix renormalization group},\ }\href
  {https://doi.org/10.1103/PhysRevB.55.2164} {\bibfield  {journal} {\bibinfo
  {journal} {Phys. Rev. B}\ }\textbf {\bibinfo {volume} {55}},\ \bibinfo
  {pages} {2164} (\bibinfo {year} {1997})}\BibitemShut {NoStop}%
\bibitem [{\citenamefont {Schollw\"ock}(2005)}]{Schollwock-2005}%
  \BibitemOpen
  \bibfield  {author} {\bibinfo {author} {\bibfnamefont {U.}~\bibnamefont
  {Schollw\"ock}},\ }\bibfield  {title} {\bibinfo {title} {The density-matrix
  renormalization group},\ }\href {https://doi.org/10.1103/RevModPhys.77.259}
  {\bibfield  {journal} {\bibinfo  {journal} {Rev. Mod. Phys.}\ }\textbf
  {\bibinfo {volume} {77}},\ \bibinfo {pages} {259} (\bibinfo {year}
  {2005})}\BibitemShut {NoStop}%
\bibitem [{\citenamefont {Hallberg}(2006)}]{Hallberg-2006}%
  \BibitemOpen
  \bibfield  {author} {\bibinfo {author} {\bibfnamefont {K.~A.}\ \bibnamefont
  {Hallberg}},\ }\bibfield  {title} {\bibinfo {title} {New trends in density
  matrix renormalization},\ }\href {https://doi.org/10.1080/00018730600766432}
  {\bibfield  {journal} {\bibinfo  {journal} {Advances in Physics}\ }\textbf
  {\bibinfo {volume} {55}},\ \bibinfo {pages} {477} (\bibinfo {year} {2006})},\
  \Eprint {https://arxiv.org/abs/http://dx.doi.org/10.1080/00018730600766432}
  {http://dx.doi.org/10.1080/00018730600766432} \BibitemShut {NoStop}%
\bibitem [{\citenamefont {Noack}\ and\ \citenamefont
  {Manmana}(2005)}]{Noack-2005}%
  \BibitemOpen
  \bibfield  {author} {\bibinfo {author} {\bibfnamefont {R.~M.}\ \bibnamefont
  {Noack}}\ and\ \bibinfo {author} {\bibfnamefont {S.~R.}\ \bibnamefont
  {Manmana}},\ }\bibfield  {title} {\bibinfo {title} {Diagonalization‐ and
  numerical renormalization‐group‐based methods for interacting quantum
  systems},\ }\href {https://doi.org/http://dx.doi.org/10.1063/1.2080349}
  {\bibfield  {journal} {\bibinfo  {journal} {AIP Conference Proceedings}\
  }\textbf {\bibinfo {volume} {789}},\ \bibinfo {pages} {93} (\bibinfo {year}
  {2005})}\BibitemShut {NoStop}%
\bibitem [{\citenamefont {Legeza}\ \emph {et~al.}(2008)\citenamefont {Legeza},
  \citenamefont {Noack}, \citenamefont {S\'olyom},\ and\ \citenamefont
  {Tincani}}]{Legeza-2008}%
  \BibitemOpen
  \bibfield  {author} {\bibinfo {author} {\bibfnamefont {{\"O}.}~\bibnamefont
  {Legeza}}, \bibinfo {author} {\bibfnamefont {R.}~\bibnamefont {Noack}},
  \bibinfo {author} {\bibfnamefont {J.}~\bibnamefont {S\'olyom}},\ and\
  \bibinfo {author} {\bibfnamefont {L.}~\bibnamefont {Tincani}},\ }\bibfield
  {title} {\bibinfo {title} {Applications of quantum information in the
  density-matrix renormalization group},\ }in\ \href
  {https://doi.org/10.1007/978-3-540-74686-7\_24} {\emph {\bibinfo {booktitle}
  {Computational Many-Particle Physics}}},\ \bibinfo {series} {Lecture Notes in
  Physics}, Vol.\ \bibinfo {volume} {739},\ \bibinfo {editor} {edited by\
  \bibinfo {editor} {\bibfnamefont {H.}~\bibnamefont {Fehske}}, \bibinfo
  {editor} {\bibfnamefont {R.}~\bibnamefont {Schneider}},\ and\ \bibinfo
  {editor} {\bibfnamefont {A.}~\bibnamefont {Weisse}}}\ (\bibinfo  {publisher}
  {Springer},\ \bibinfo {address} {Berlin, Heidelberg},\ \bibinfo {year}
  {2008})\ pp.\ \bibinfo {pages} {653--664}\BibitemShut {NoStop}%
\bibitem [{\citenamefont {Chan}\ and\ \citenamefont {Zgid}(2009)}]{Chan-2009}%
  \BibitemOpen
  \bibfield  {author} {\bibinfo {author} {\bibfnamefont {G.~K.-L.}\
  \bibnamefont {Chan}}\ and\ \bibinfo {author} {\bibfnamefont {D.}~\bibnamefont
  {Zgid}},\ }\bibfield  {title} {\bibinfo {title} {Chapter 7 the density matrix
  renormalization group in quantum chemistry}\ }(\bibinfo  {publisher}
  {Elsevier},\ \bibinfo {year} {2009})\ pp.\ \bibinfo {pages} {149 --
  162}\BibitemShut {NoStop}%
\bibitem [{\citenamefont {Schollw\"ock}(2011)}]{Schollwock-2011}%
  \BibitemOpen
  \bibfield  {author} {\bibinfo {author} {\bibfnamefont {U.}~\bibnamefont
  {Schollw\"ock}},\ }\bibfield  {title} {\bibinfo {title} {The density-matrix
  renormalization group in the age of matrix product states},\ }\href
  {https://doi.org/http://dx.doi.org/10.1016/j.aop.2010.09.012} {\bibfield
  {journal} {\bibinfo  {journal} {Annals of Physics}\ }\textbf {\bibinfo
  {volume} {326}},\ \bibinfo {pages} {96 } (\bibinfo {year} {2011})},\ \bibinfo
  {note} {january 2011 Special Issue}\BibitemShut {NoStop}%
\bibitem [{\citenamefont {Chan}\ and\ \citenamefont
  {Sharma}(2011)}]{Chan-2011}%
  \BibitemOpen
  \bibfield  {author} {\bibinfo {author} {\bibfnamefont {G.~K.-L.}\
  \bibnamefont {Chan}}\ and\ \bibinfo {author} {\bibfnamefont {S.}~\bibnamefont
  {Sharma}},\ }\bibfield  {title} {\bibinfo {title} {The density matrix
  renormalization group in quantum chemistry},\ }\href
  {https://doi.org/10.1146/annurev-physchem-032210-103338} {\bibfield
  {journal} {\bibinfo  {journal} {Annual Review of Physical Chemistry}\
  }\textbf {\bibinfo {volume} {62}},\ \bibinfo {pages} {465} (\bibinfo {year}
  {2011})},\ \bibinfo {note} {pMID: 21219144},\ \Eprint
  {https://arxiv.org/abs/http://dx.doi.org/10.1146/annurev-physchem-032210-103338}
  {http://dx.doi.org/10.1146/annurev-physchem-032210-103338} \BibitemShut
  {NoStop}%
\bibitem [{\citenamefont {{\relax Sz}alay}\ \emph {et~al.}(2015)\citenamefont
  {{\relax Sz}alay}, \citenamefont {Pfeffer}, \citenamefont {Murg},
  \citenamefont {Barcza}, \citenamefont {Verstraete}, \citenamefont
  {Schneider},\ and\ \citenamefont {Legeza}}]{Szalay-2015a}%
  \BibitemOpen
  \bibfield  {author} {\bibinfo {author} {\bibfnamefont {{\relax
  Sz}.}~\bibnamefont {{\relax Sz}alay}}, \bibinfo {author} {\bibfnamefont
  {M.}~\bibnamefont {Pfeffer}}, \bibinfo {author} {\bibfnamefont
  {V.}~\bibnamefont {Murg}}, \bibinfo {author} {\bibfnamefont {G.}~\bibnamefont
  {Barcza}}, \bibinfo {author} {\bibfnamefont {F.}~\bibnamefont {Verstraete}},
  \bibinfo {author} {\bibfnamefont {R.}~\bibnamefont {Schneider}},\ and\
  \bibinfo {author} {\bibfnamefont {{\"O}.}~\bibnamefont {Legeza}},\ }\bibfield
   {title} {\bibinfo {title} {Tensor product methods and entanglement
  optimization for ab initio quantum chemistry},\ }\href
  {https://doi.org/10.1002/qua.24898} {\bibfield  {journal} {\bibinfo
  {journal} {Int. J. Quantum Chem.}\ }\textbf {\bibinfo {volume} {115}},\
  \bibinfo {pages} {1342} (\bibinfo {year} {2015})}\BibitemShut {NoStop}%
\bibitem [{\citenamefont {Or{\'u}s}(2019)}]{Orus-2019}%
  \BibitemOpen
  \bibfield  {author} {\bibinfo {author} {\bibfnamefont {R.}~\bibnamefont
  {Or{\'u}s}},\ }\bibfield  {title} {\bibinfo {title} {Tensor networks for
  complex quantum systems},\ }\href {https://doi.org/10.1038/s42254-019-0086-7}
  {\bibfield  {journal} {\bibinfo  {journal} {Nature Reviews Physics}\ }\textbf
  {\bibinfo {volume} {1}},\ \bibinfo {pages} {538} (\bibinfo {year}
  {2019})}\BibitemShut {NoStop}%
\bibitem [{\citenamefont {White}\ and\ \citenamefont
  {Martin}(1999)}]{White-1999}%
  \BibitemOpen
  \bibfield  {author} {\bibinfo {author} {\bibfnamefont {S.~R.}\ \bibnamefont
  {White}}\ and\ \bibinfo {author} {\bibfnamefont {R.~L.}\ \bibnamefont
  {Martin}},\ }\bibfield  {title} {\bibinfo {title} {Ab initio quantum
  chemistry using the density matrix renormalization group},\ }\href
  {https://doi.org/http://dx.doi.org/10.1063/1.478295} {\bibfield  {journal}
  {\bibinfo  {journal} {The Journal of Chemical Physics}\ }\textbf {\bibinfo
  {volume} {110}},\ \bibinfo {pages} {4127} (\bibinfo {year}
  {1999})}\BibitemShut {NoStop}%
\bibitem [{\citenamefont {Chan}\ \emph {et~al.}(2008)\citenamefont {Chan},
  \citenamefont {Dorando}, \citenamefont {Ghosh}, \citenamefont {Hachmann},
  \citenamefont {Neuscamman}, \citenamefont {Wang},\ and\ \citenamefont
  {Yanai}}]{Chan-2008}%
  \BibitemOpen
  \bibfield  {author} {\bibinfo {author} {\bibfnamefont {G.~K.-L.}\
  \bibnamefont {Chan}}, \bibinfo {author} {\bibfnamefont {J.~J.}\ \bibnamefont
  {Dorando}}, \bibinfo {author} {\bibfnamefont {D.}~\bibnamefont {Ghosh}},
  \bibinfo {author} {\bibfnamefont {J.}~\bibnamefont {Hachmann}}, \bibinfo
  {author} {\bibfnamefont {E.}~\bibnamefont {Neuscamman}}, \bibinfo {author}
  {\bibfnamefont {H.}~\bibnamefont {Wang}},\ and\ \bibinfo {author}
  {\bibfnamefont {T.}~\bibnamefont {Yanai}},\ }\bibfield  {title} {\bibinfo
  {title} {An introduction to the density matrix renormalization group ansatz
  in quantum chemistry},\ }in\ \href
  {https://doi.org/10.1007/978-1-4020-8707-3} {\emph {\bibinfo {booktitle}
  {Frontiers in Quantum Systems in Chemistry and Physics}}},\ \bibinfo {series}
  {Progress in Theoretical Chemistry and Physics}, Vol.~\bibinfo {volume}
  {18},\ \bibinfo {editor} {edited by\ \bibinfo {editor} {\bibfnamefont
  {S.}~\bibnamefont {Wilson}}, \bibinfo {editor} {\bibfnamefont {P.~J.}\
  \bibnamefont {Grout}}, \bibinfo {editor} {\bibfnamefont {J.}~\bibnamefont
  {Maruani}}, \bibinfo {editor} {\bibfnamefont {G.}~\bibnamefont
  {Delgado-Barrio}},\ and\ \bibinfo {editor} {\bibfnamefont {P.}~\bibnamefont
  {Piecuch}}}\ (\bibinfo  {publisher} {Springer},\ \bibinfo {address}
  {Netherlands},\ \bibinfo {year} {2008})\BibitemShut {NoStop}%
\bibitem [{\citenamefont {Yanai}\ \emph {et~al.}(2009)\citenamefont {Yanai},
  \citenamefont {Kurashige}, \citenamefont {Ghosh},\ and\ \citenamefont
  {Chan}}]{Yanai-2009}%
  \BibitemOpen
  \bibfield  {author} {\bibinfo {author} {\bibfnamefont {T.}~\bibnamefont
  {Yanai}}, \bibinfo {author} {\bibfnamefont {Y.}~\bibnamefont {Kurashige}},
  \bibinfo {author} {\bibfnamefont {D.}~\bibnamefont {Ghosh}},\ and\ \bibinfo
  {author} {\bibfnamefont {G.~K.-L.}\ \bibnamefont {Chan}},\ }\bibfield
  {title} {\bibinfo {title} {Accelerating convergence in iterative solution for
  large-scale complete active space self-consistent-field calculations},\
  }\href {https://doi.org/10.1002/qua.22099} {\bibfield  {journal} {\bibinfo
  {journal} {International Journal of Quantum Chemistry}\ }\textbf {\bibinfo
  {volume} {109}},\ \bibinfo {pages} {2178} (\bibinfo {year}
  {2009})}\BibitemShut {NoStop}%
\bibitem [{\citenamefont {Marti}\ and\ \citenamefont
  {Reiher}(2010)}]{Marti-2010c}%
  \BibitemOpen
  \bibfield  {author} {\bibinfo {author} {\bibfnamefont {K.~H.}\ \bibnamefont
  {Marti}}\ and\ \bibinfo {author} {\bibfnamefont {M.}~\bibnamefont {Reiher}},\
  }\bibfield  {title} {\bibinfo {title} {The density matrix renormalization
  group algorithm in quantum chemistry},\ }\href
  {https://doi.org/10.1524/zpch.2010.6125} {\bibfield  {journal} {\bibinfo
  {journal} {Zeitschrift f\"ur Physikalische Chemie}\ }\textbf {\bibinfo
  {volume} {224}},\ \bibinfo {pages} {583} (\bibinfo {year}
  {2010})}\BibitemShut {NoStop}%
\bibitem [{\citenamefont {Wouters}\ \emph {et~al.}(2014)\citenamefont
  {Wouters}, \citenamefont {Poelmans}, \citenamefont {Ayers},\ and\
  \citenamefont {Neck}}]{Wouters-2014a}%
  \BibitemOpen
  \bibfield  {author} {\bibinfo {author} {\bibfnamefont {S.}~\bibnamefont
  {Wouters}}, \bibinfo {author} {\bibfnamefont {W.}~\bibnamefont {Poelmans}},
  \bibinfo {author} {\bibfnamefont {P.~W.}\ \bibnamefont {Ayers}},\ and\
  \bibinfo {author} {\bibfnamefont {D.~V.}\ \bibnamefont {Neck}},\ }\bibfield
  {title} {\bibinfo {title} {{CheMPS2}: A free open-source spin-adapted
  implementation of the density matrix renormalization group for ab initio
  quantum chemistry},\ }\href
  {https://doi.org/http://dx.doi.org/10.1016/j.cpc.2014.01.019} {\bibfield
  {journal} {\bibinfo  {journal} {Computer Physics Communications}\ }\textbf
  {\bibinfo {volume} {185}},\ \bibinfo {pages} {1501 } (\bibinfo {year}
  {2014})}\BibitemShut {NoStop}%
\bibitem [{\citenamefont {Legeza}\ \emph {et~al.}(2014)\citenamefont {Legeza},
  \citenamefont {Rohwedder}, \citenamefont {Schneider},\ and\ \citenamefont
  {{\relax Sz}alay}}]{Legeza-2014}%
  \BibitemOpen
  \bibfield  {author} {\bibinfo {author} {\bibfnamefont {{\"O}.}~\bibnamefont
  {Legeza}}, \bibinfo {author} {\bibfnamefont {T.}~\bibnamefont {Rohwedder}},
  \bibinfo {author} {\bibfnamefont {R.}~\bibnamefont {Schneider}},\ and\
  \bibinfo {author} {\bibfnamefont {{\relax Sz}.}~\bibnamefont {{\relax
  Sz}alay}},\ }\bibfield  {title} {\bibinfo {title} {Tensor product
  approximation ({DMRG}) and coupled cluster method in quantum chemistry},\
  }in\ \href {https://doi.org/10.1007/978-3-319-06379-9\_3} {\emph {\bibinfo
  {booktitle} {Many-Electron Approaches in Physics, Chemistry and
  Mathematics}}},\ \bibinfo {series and number} {Mathematical Physics
  Studies},\ \bibinfo {editor} {edited by\ \bibinfo {editor} {\bibfnamefont
  {V.}~\bibnamefont {Bach}}\ and\ \bibinfo {editor} {\bibfnamefont
  {L.}~\bibnamefont {Delle~Site}}}\ (\bibinfo  {publisher} {Springer
  International Publishing},\ \bibinfo {address} {Switzerland},\ \bibinfo
  {year} {2014})\ pp.\ \bibinfo {pages} {53--76}\BibitemShut {NoStop}%
\bibitem [{\citenamefont {Chan}\ \emph {et~al.}(2016)\citenamefont {Chan},
  \citenamefont {Keselman}, \citenamefont {Nakatani}, \citenamefont {Li},\ and\
  \citenamefont {White}}]{Chan-2016}%
  \BibitemOpen
  \bibfield  {author} {\bibinfo {author} {\bibfnamefont {G.~K.-L.}\
  \bibnamefont {Chan}}, \bibinfo {author} {\bibfnamefont {A.}~\bibnamefont
  {Keselman}}, \bibinfo {author} {\bibfnamefont {N.}~\bibnamefont {Nakatani}},
  \bibinfo {author} {\bibfnamefont {Z.}~\bibnamefont {Li}},\ and\ \bibinfo
  {author} {\bibfnamefont {S.~R.}\ \bibnamefont {White}},\ }\bibfield  {title}
  {\bibinfo {title} {Matrix product operators, matrix product states, and ab
  initio density matrix renormalization group algorithms},\ }\href
  {https://doi.org/10.1063/1.4955108} {\bibfield  {journal} {\bibinfo
  {journal} {The Journal of Chemical Physics}\ }\textbf {\bibinfo {volume}
  {145}},\ \bibinfo {pages} {014102} (\bibinfo {year} {2016})}\BibitemShut
  {NoStop}%
\bibitem [{\citenamefont {Baiardi}\ and\ \citenamefont
  {Reiher}(2020)}]{Baiardi-2020}%
  \BibitemOpen
  \bibfield  {author} {\bibinfo {author} {\bibfnamefont {A.}~\bibnamefont
  {Baiardi}}\ and\ \bibinfo {author} {\bibfnamefont {M.}~\bibnamefont
  {Reiher}},\ }\bibfield  {title} {\bibinfo {title} {The density matrix
  renormalization group in chemistry and molecular physics: Recent developments
  and new challenges},\ }\href {https://doi.org/10.1063/1.5129672} {\bibfield
  {journal} {\bibinfo  {journal} {The Journal of Chemical Physics}\ }\textbf
  {\bibinfo {volume} {152}},\ \bibinfo {pages} {040903} (\bibinfo {year}
  {2020})}\BibitemShut {NoStop}%
\bibitem [{\citenamefont {Cheng}\ \emph {et~al.}(2022)\citenamefont {Cheng},
  \citenamefont {Xie},\ and\ \citenamefont {Ma}}]{Cheng-2022}%
  \BibitemOpen
  \bibfield  {author} {\bibinfo {author} {\bibfnamefont {Y.}~\bibnamefont
  {Cheng}}, \bibinfo {author} {\bibfnamefont {Z.}~\bibnamefont {Xie}},\ and\
  \bibinfo {author} {\bibfnamefont {H.}~\bibnamefont {Ma}},\ }\bibfield
  {title} {\bibinfo {title} {Post-density matrix renormalization group methods
  for describing dynamic electron correlation with large active spaces},\
  }\href {https://doi.org/10.1021/acs.jpclett.1c04078} {\bibfield  {journal}
  {\bibinfo  {journal} {The Journal of Physical Chemistry Letters}\ }\textbf
  {\bibinfo {volume} {13}},\ \bibinfo {pages} {904} (\bibinfo {year}
  {2022})}\BibitemShut {NoStop}%
\bibitem [{\citenamefont {Verstraete}\ \emph {et~al.}(2023)\citenamefont
  {Verstraete}, \citenamefont {Nishino}, \citenamefont {Schollw{\"o}ck},
  \citenamefont {Ba{\~n}uls}, \citenamefont {Chan},\ and\ \citenamefont
  {Stoudenmire}}]{Verstraete-2023}%
  \BibitemOpen
  \bibfield  {author} {\bibinfo {author} {\bibfnamefont {F.}~\bibnamefont
  {Verstraete}}, \bibinfo {author} {\bibfnamefont {T.}~\bibnamefont {Nishino}},
  \bibinfo {author} {\bibfnamefont {U.}~\bibnamefont {Schollw{\"o}ck}},
  \bibinfo {author} {\bibfnamefont {M.~C.}\ \bibnamefont {Ba{\~n}uls}},
  \bibinfo {author} {\bibfnamefont {G.~K.}\ \bibnamefont {Chan}},\ and\
  \bibinfo {author} {\bibfnamefont {M.~E.}\ \bibnamefont {Stoudenmire}},\
  }\bibfield  {title} {\bibinfo {title} {Density matrix renormalization group,
  30 years on},\ }\href {https://doi.org/10.1038/s42254-023-00572-5} {\bibfield
   {journal} {\bibinfo  {journal} {Nature Reviews Physics}\ ,\ \bibinfo {pages}
  {1}} (\bibinfo {year} {2023})}\BibitemShut {NoStop}%
\bibitem [{\citenamefont {Roy}\ and\ \citenamefont
  {Banerjee}(2014)}]{Roy-2014}%
  \BibitemOpen
  \bibfield  {author} {\bibinfo {author} {\bibfnamefont {A.}~\bibnamefont
  {Roy}}\ and\ \bibinfo {author} {\bibfnamefont {S.}~\bibnamefont {Banerjee}},\
  }\href@noop {} {\emph {\bibinfo {title} {Linear algebra and matrix analysis
  for statistics}}}\ (\bibinfo  {publisher} {Chapman and Hall/CRC},\ \bibinfo
  {year} {2014})\BibitemShut {NoStop}%
\bibitem [{\citenamefont {Legeza}\ \emph {et~al.}(2003)\citenamefont {Legeza},
  \citenamefont {R\"oder},\ and\ \citenamefont {Hess}}]{Legeza-2003a}%
  \BibitemOpen
  \bibfield  {author} {\bibinfo {author} {\bibfnamefont {{\"O}.}~\bibnamefont
  {Legeza}}, \bibinfo {author} {\bibfnamefont {J.}~\bibnamefont {R\"oder}},\
  and\ \bibinfo {author} {\bibfnamefont {B.~A.}\ \bibnamefont {Hess}},\
  }\bibfield  {title} {\bibinfo {title} {Controlling the accuracy of the
  density-matrix renormalization-group method: The dynamical block state
  selection approach},\ }\href {https://doi.org/10.1103/PhysRevB.67.125114}
  {\bibfield  {journal} {\bibinfo  {journal} {Phys. Rev. B}\ }\textbf {\bibinfo
  {volume} {67}},\ \bibinfo {pages} {125114} (\bibinfo {year}
  {2003})}\BibitemShut {NoStop}%
\bibitem [{\citenamefont {Holtz}\ \emph
  {et~al.}(2012{\natexlab{a}})\citenamefont {Holtz}, \citenamefont
  {Rohwedder},\ and\ \citenamefont {Schneider}}]{Holtz-2012a}%
  \BibitemOpen
  \bibfield  {author} {\bibinfo {author} {\bibfnamefont {S.}~\bibnamefont
  {Holtz}}, \bibinfo {author} {\bibfnamefont {T.}~\bibnamefont {Rohwedder}},\
  and\ \bibinfo {author} {\bibfnamefont {R.}~\bibnamefont {Schneider}},\
  }\bibfield  {title} {\bibinfo {title} {On manifolds of tensors of fixed
  {TT}-rank},\ }\href {https://doi.org/10.1007/s00211-011-0419-7} {\bibfield
  {journal} {\bibinfo  {journal} {Numerische Mathematik}\ }\textbf {\bibinfo
  {volume} {120}},\ \bibinfo {pages} {701} (\bibinfo {year}
  {2012}{\natexlab{a}})}\BibitemShut {NoStop}%
\bibitem [{\citenamefont {Holtz}\ \emph
  {et~al.}(2012{\natexlab{b}})\citenamefont {Holtz}, \citenamefont
  {Rohwedder},\ and\ \citenamefont {Schneider}}]{Holtz-2012b}%
  \BibitemOpen
  \bibfield  {author} {\bibinfo {author} {\bibfnamefont {S.}~\bibnamefont
  {Holtz}}, \bibinfo {author} {\bibfnamefont {T.}~\bibnamefont {Rohwedder}},\
  and\ \bibinfo {author} {\bibfnamefont {R.}~\bibnamefont {Schneider}},\
  }\bibfield  {title} {\bibinfo {title} {The alternating linear scheme for
  tensor optimization in the tensor train format},\ }\href
  {https://doi.org/10.1137/100818893} {\bibfield  {journal} {\bibinfo
  {journal} {SIAM Journal on Scientific Computing}\ }\textbf {\bibinfo {volume}
  {34}},\ \bibinfo {pages} {A683} (\bibinfo {year}
  {2012}{\natexlab{b}})}\BibitemShut {NoStop}%
\bibitem [{\citenamefont {Legeza}\ and\ \citenamefont
  {S\'olyom}(2003)}]{Legeza-2003b}%
  \BibitemOpen
  \bibfield  {author} {\bibinfo {author} {\bibfnamefont {{\"O}.}~\bibnamefont
  {Legeza}}\ and\ \bibinfo {author} {\bibfnamefont {J.}~\bibnamefont
  {S\'olyom}},\ }\bibfield  {title} {\bibinfo {title} {Optimizing the
  density-matrix renormalization group method using quantum information
  entropy},\ }\href {https://doi.org/10.1103/PhysRevB.68.195116} {\bibfield
  {journal} {\bibinfo  {journal} {Phys. Rev. B}\ }\textbf {\bibinfo {volume}
  {68}},\ \bibinfo {pages} {195116} (\bibinfo {year} {2003})}\BibitemShut
  {NoStop}%
\bibitem [{\citenamefont {Nakatani}\ and\ \citenamefont
  {Chan}(2013)}]{Nakatani-2013}%
  \BibitemOpen
  \bibfield  {author} {\bibinfo {author} {\bibfnamefont {N.}~\bibnamefont
  {Nakatani}}\ and\ \bibinfo {author} {\bibfnamefont {G.~K.-L.}\ \bibnamefont
  {Chan}},\ }\bibfield  {title} {\bibinfo {title} {Efficient tree tensor
  network states ({TTNS}) for quantum chemistry: Generalizations of the density
  matrix renormalization group algorithm},\ }\href
  {https://doi.org/http://dx.doi.org/10.1063/1.4798639} {\bibfield  {journal}
  {\bibinfo  {journal} {The Journal of Chemical Physics}\ }\textbf {\bibinfo
  {volume} {138}},\ \bibinfo {eid} {134113} (\bibinfo {year}
  {2013})}\BibitemShut {NoStop}%
\bibitem [{\citenamefont {Murg}\ \emph {et~al.}(2015)\citenamefont {Murg},
  \citenamefont {Verstraete}, \citenamefont {Schneider}, \citenamefont {Nagy},\
  and\ \citenamefont {Legeza}}]{Murg-2015}%
  \BibitemOpen
  \bibfield  {author} {\bibinfo {author} {\bibfnamefont {V.}~\bibnamefont
  {Murg}}, \bibinfo {author} {\bibfnamefont {F.}~\bibnamefont {Verstraete}},
  \bibinfo {author} {\bibfnamefont {R.}~\bibnamefont {Schneider}}, \bibinfo
  {author} {\bibfnamefont {P.~R.}\ \bibnamefont {Nagy}},\ and\ \bibinfo
  {author} {\bibfnamefont {{\"O}.}~\bibnamefont {Legeza}},\ }\bibfield  {title}
  {\bibinfo {title} {Tree tensor network state with variable tensor order: An
  efficient multireference method for strongly correlated systems},\ }\href
  {https://doi.org/10.1021/ct501187j} {\bibfield  {journal} {\bibinfo
  {journal} {Journal of Chemical Theory and Computation}\ }\textbf {\bibinfo
  {volume} {11}},\ \bibinfo {pages} {1027} (\bibinfo {year}
  {2015})}\BibitemShut {NoStop}%
\bibitem [{\citenamefont {Gunst}\ \emph {et~al.}(2018)\citenamefont {Gunst},
  \citenamefont {Verstraete}, \citenamefont {Wouters}, \citenamefont {Legeza},\
  and\ \citenamefont {Van~Neck}}]{Gunst-2018}%
  \BibitemOpen
  \bibfield  {author} {\bibinfo {author} {\bibfnamefont {K.}~\bibnamefont
  {Gunst}}, \bibinfo {author} {\bibfnamefont {F.}~\bibnamefont {Verstraete}},
  \bibinfo {author} {\bibfnamefont {S.}~\bibnamefont {Wouters}}, \bibinfo
  {author} {\bibfnamefont {{\"O}.}~\bibnamefont {Legeza}},\ and\ \bibinfo
  {author} {\bibfnamefont {D.}~\bibnamefont {Van~Neck}},\ }\bibfield  {title}
  {\bibinfo {title} {{T3NS}: Three-legged tree tensor network states},\ }\href
  {https://doi.org/10.1021/acs.jctc.8b00098} {\bibfield  {journal} {\bibinfo
  {journal} {Journal of Chemical Theory and Computation}\ }\textbf {\bibinfo
  {volume} {14}},\ \bibinfo {pages} {2026} (\bibinfo {year}
  {2018})}\BibitemShut {NoStop}%
\bibitem [{\citenamefont {Rissler}\ \emph {et~al.}(2006)\citenamefont
  {Rissler}, \citenamefont {Noack},\ and\ \citenamefont
  {White}}]{Rissler-2006}%
  \BibitemOpen
  \bibfield  {author} {\bibinfo {author} {\bibfnamefont {J.}~\bibnamefont
  {Rissler}}, \bibinfo {author} {\bibfnamefont {R.~M.}\ \bibnamefont {Noack}},\
  and\ \bibinfo {author} {\bibfnamefont {S.~R.}\ \bibnamefont {White}},\
  }\bibfield  {title} {\bibinfo {title} {Measuring orbital interaction using
  quantum information theory},\ }\href
  {https://doi.org/http://dx.doi.org/10.1016/j.chemphys.2005.10.018} {\bibfield
   {journal} {\bibinfo  {journal} {Chemical Physics}\ }\textbf {\bibinfo
  {volume} {323}},\ \bibinfo {pages} {519 } (\bibinfo {year}
  {2006})}\BibitemShut {NoStop}%
\bibitem [{\citenamefont {Murg}\ \emph {et~al.}(2010)\citenamefont {Murg},
  \citenamefont {Verstraete}, \citenamefont {Legeza},\ and\ \citenamefont
  {Noack}}]{Murg-2010a}%
  \BibitemOpen
  \bibfield  {author} {\bibinfo {author} {\bibfnamefont {V.}~\bibnamefont
  {Murg}}, \bibinfo {author} {\bibfnamefont {F.}~\bibnamefont {Verstraete}},
  \bibinfo {author} {\bibfnamefont {{\"O}.}~\bibnamefont {Legeza}},\ and\
  \bibinfo {author} {\bibfnamefont {R.~M.}\ \bibnamefont {Noack}},\ }\bibfield
  {title} {\bibinfo {title} {Simulating strongly correlated quantum systems
  with tree tensor networks},\ }\href
  {https://doi.org/10.1103/PhysRevB.82.205105} {\bibfield  {journal} {\bibinfo
  {journal} {Phys. Rev. B}\ }\textbf {\bibinfo {volume} {82}},\ \bibinfo
  {pages} {205105} (\bibinfo {year} {2010})}\BibitemShut {NoStop}%
\bibitem [{\citenamefont {Stein}\ and\ \citenamefont
  {Reiher}(2016)}]{Stein-2016}%
  \BibitemOpen
  \bibfield  {author} {\bibinfo {author} {\bibfnamefont {C.~J.}\ \bibnamefont
  {Stein}}\ and\ \bibinfo {author} {\bibfnamefont {M.}~\bibnamefont {Reiher}},\
  }\bibfield  {title} {\bibinfo {title} {Automated selection of active orbital
  spaces},\ }\href {https://doi.org/10.1021/acs.jctc.6b00156} {\bibfield
  {journal} {\bibinfo  {journal} {Journal of Chemical Theory and Computation}\
  }\textbf {\bibinfo {volume} {12}},\ \bibinfo {pages} {1760} (\bibinfo {year}
  {2016})}\BibitemShut {NoStop}%
\bibitem [{\citenamefont {Fertitta}\ \emph {et~al.}(2014)\citenamefont
  {Fertitta}, \citenamefont {Paulus}, \citenamefont {Barcza},\ and\
  \citenamefont {Legeza}}]{Fertitta-2014}%
  \BibitemOpen
  \bibfield  {author} {\bibinfo {author} {\bibfnamefont {E.}~\bibnamefont
  {Fertitta}}, \bibinfo {author} {\bibfnamefont {B.}~\bibnamefont {Paulus}},
  \bibinfo {author} {\bibfnamefont {G.}~\bibnamefont {Barcza}},\ and\ \bibinfo
  {author} {\bibfnamefont {{\"O}.}~\bibnamefont {Legeza}},\ }\bibfield  {title}
  {\bibinfo {title} {Investigation of metal-insulator-like transition through
  the ab initio density matrix renormalization group approach},\ }\href
  {https://doi.org/10.1103/PhysRevB.90.245129} {\bibfield  {journal} {\bibinfo
  {journal} {Phys. Rev. B}\ }\textbf {\bibinfo {volume} {90}},\ \bibinfo
  {pages} {245129} (\bibinfo {year} {2014})}\BibitemShut {NoStop}%
\bibitem [{\citenamefont {Krumnow}\ \emph {et~al.}(2016)\citenamefont
  {Krumnow}, \citenamefont {Veis}, \citenamefont {Legeza},\ and\ \citenamefont
  {Eisert}}]{Krumnow-2016}%
  \BibitemOpen
  \bibfield  {author} {\bibinfo {author} {\bibfnamefont {C.}~\bibnamefont
  {Krumnow}}, \bibinfo {author} {\bibfnamefont {L.}~\bibnamefont {Veis}},
  \bibinfo {author} {\bibfnamefont {{\"O}.}~\bibnamefont {Legeza}},\ and\
  \bibinfo {author} {\bibfnamefont {J.}~\bibnamefont {Eisert}},\ }\bibfield
  {title} {\bibinfo {title} {Fermionic orbital optimization in tensor network
  states},\ }\href {https://doi.org/10.1103/PhysRevLett.117.210402} {\bibfield
  {journal} {\bibinfo  {journal} {Phys. Rev. Lett.}\ }\textbf {\bibinfo
  {volume} {117}},\ \bibinfo {pages} {210402} (\bibinfo {year}
  {2016})}\BibitemShut {NoStop}%
\bibitem [{\citenamefont {Krumnow}\ \emph {et~al.}(2021)\citenamefont
  {Krumnow}, \citenamefont {Veis}, \citenamefont {Eisert},\ and\ \citenamefont
  {Legeza}}]{Krumnow-2021}%
  \BibitemOpen
  \bibfield  {author} {\bibinfo {author} {\bibfnamefont {C.}~\bibnamefont
  {Krumnow}}, \bibinfo {author} {\bibfnamefont {L.}~\bibnamefont {Veis}},
  \bibinfo {author} {\bibfnamefont {J.}~\bibnamefont {Eisert}},\ and\ \bibinfo
  {author} {\bibfnamefont {{\"O}.}~\bibnamefont {Legeza}},\ }\bibfield  {title}
  {\bibinfo {title} {Effective dimension reduction with mode transformations:
  Simulating two-dimensional fermionic condensed matter systems with
  matrix-product states},\ }\href {https://doi.org/10.1103/PhysRevB.104.075137}
  {\bibfield  {journal} {\bibinfo  {journal} {Phys. Rev. B}\ }\textbf {\bibinfo
  {volume} {104}},\ \bibinfo {pages} {075137} (\bibinfo {year}
  {2021})}\BibitemShut {NoStop}%
\bibitem [{\citenamefont {Foster}\ and\ \citenamefont
  {Boys}(1960)}]{Foster-1960}%
  \BibitemOpen
  \bibfield  {author} {\bibinfo {author} {\bibfnamefont {J.~M.}\ \bibnamefont
  {Foster}}\ and\ \bibinfo {author} {\bibfnamefont {S.~F.}\ \bibnamefont
  {Boys}},\ }\bibfield  {title} {\bibinfo {title} {Canonical configurational
  interaction procedure},\ }\href {https://doi.org/10.1103/RevModPhys.32.300}
  {\bibfield  {journal} {\bibinfo  {journal} {Rev. Mod. Phys.}\ }\textbf
  {\bibinfo {volume} {32}},\ \bibinfo {pages} {300} (\bibinfo {year}
  {1960})}\BibitemShut {NoStop}%
\bibitem [{\citenamefont {Boys}(1960)}]{Boys-1960}%
  \BibitemOpen
  \bibfield  {author} {\bibinfo {author} {\bibfnamefont {S.~F.}\ \bibnamefont
  {Boys}},\ }\bibfield  {title} {\bibinfo {title} {Construction of some
  molecular orbitals to be approximately invariant for changes from one
  molecule to another},\ }\href {https://doi.org/10.1103/RevModPhys.32.296}
  {\bibfield  {journal} {\bibinfo  {journal} {Rev. Mod. Phys.}\ }\textbf
  {\bibinfo {volume} {32}},\ \bibinfo {pages} {296} (\bibinfo {year}
  {1960})}\BibitemShut {NoStop}%
\bibitem [{\citenamefont {Pipek}\ and\ \citenamefont
  {Mezey}(1989)}]{Pipek-1989}%
  \BibitemOpen
  \bibfield  {author} {\bibinfo {author} {\bibfnamefont {J.}~\bibnamefont
  {Pipek}}\ and\ \bibinfo {author} {\bibfnamefont {P.~G.}\ \bibnamefont
  {Mezey}},\ }\bibfield  {title} {\bibinfo {title} {A fast intrinsic
  localization procedure applicable for ab initio and semiempirical linear
  combination of atomic orbital wave functions},\ }\href
  {https://doi.org/http://dx.doi.org/10.1063/1.456588} {\bibfield  {journal}
  {\bibinfo  {journal} {The Journal of Chemical Physics}\ }\textbf {\bibinfo
  {volume} {90}},\ \bibinfo {pages} {4916} (\bibinfo {year}
  {1989})}\BibitemShut {NoStop}%
\bibitem [{\citenamefont {Máté}\ \emph {et~al.}(2023)\citenamefont {Máté},
  \citenamefont {Petrov}, \citenamefont {Szalay},\ and\ \citenamefont
  {Legeza}}]{Mate-2022}%
  \BibitemOpen
  \bibfield  {author} {\bibinfo {author} {\bibfnamefont {M.}~\bibnamefont
  {Máté}}, \bibinfo {author} {\bibfnamefont {K.}~\bibnamefont {Petrov}},
  \bibinfo {author} {\bibfnamefont {S.}~\bibnamefont {Szalay}},\ and\ \bibinfo
  {author} {\bibfnamefont {{\"O}.}~\bibnamefont {Legeza}},\ }\bibfield  {title}
  {\bibinfo {title} {Compressing multireference character of wave functions via
  fermionic mode optimization},\ }\href
  {https://doi.org/10.1007/s10910-022-01379-y} {\bibfield  {journal} {\bibinfo
  {journal} {Journal of Mathematical Chemistry}\ }\textbf {\bibinfo {volume}
  {61}},\ \bibinfo {pages} {362} (\bibinfo {year} {2023})}\BibitemShut
  {NoStop}%
\bibitem [{\citenamefont {Petrov}\ \emph {et~al.}(2024)\citenamefont {Petrov},
  \citenamefont {Ganyecz}, \citenamefont {Benedek}, \citenamefont {Olasz},
  \citenamefont {Barcza},\ and\ \citenamefont {Legeza}}]{Petrov-2023}%
  \BibitemOpen
  \bibfield  {author} {\bibinfo {author} {\bibfnamefont {K.}~\bibnamefont
  {Petrov}}, \bibinfo {author} {\bibfnamefont {A.}~\bibnamefont {Ganyecz}},
  \bibinfo {author} {\bibfnamefont {Z.}~\bibnamefont {Benedek}}, \bibinfo
  {author} {\bibfnamefont {A.}~\bibnamefont {Olasz}}, \bibinfo {author}
  {\bibfnamefont {G.}~\bibnamefont {Barcza}},\ and\ \bibinfo {author}
  {\bibfnamefont {{\"O}.}~\bibnamefont {Legeza}},\ }\bibfield  {title}
  {\bibinfo {title} {Low-cost generation of optimal molecular orbitals for
  multireference ci expansion: natural orbitals versus r´enyi entropy
  minimized orbitals provided by the density matrix renormalization group},\
  }in\ \href@noop {} {\emph {\bibinfo {booktitle} {Advances in Methods and
  Applications of Quantum Systems in Chemistry, Physics, and Biology Selected
  Proceed- ings of QSCP-XXV Conference (Torun, Poland, June 2022)}}},\ \bibinfo
  {editor} {edited by\ \bibinfo {editor} {\bibfnamefont {I.}~\bibnamefont
  {Grabowski}}, \bibinfo {editor} {\bibfnamefont {K.}~\bibnamefont
  {S\l{}owik}}, \bibinfo {editor} {\bibfnamefont {J.}~\bibnamefont {Maruani}},\
  and\ \bibinfo {editor} {\bibfnamefont {E.~J.}\ \bibnamefont {Br{\"a}ndas}}}\
  (\bibinfo  {publisher} {Springer, Cham},\ \bibinfo {year} {2024})\BibitemShut
  {NoStop}%
\bibitem [{\citenamefont {Friesecke}\ \emph {et~al.}(2024)\citenamefont
  {Friesecke}, \citenamefont {Werner}, \citenamefont {Kapás}, \citenamefont
  {Menczer},\ and\ \citenamefont {Örs Legeza}}]{Friesecke-2024}%
  \BibitemOpen
  \bibfield  {author} {\bibinfo {author} {\bibfnamefont {G.}~\bibnamefont
  {Friesecke}}, \bibinfo {author} {\bibfnamefont {M.~A.}\ \bibnamefont
  {Werner}}, \bibinfo {author} {\bibfnamefont {K.}~\bibnamefont {Kapás}},
  \bibinfo {author} {\bibfnamefont {A.}~\bibnamefont {Menczer}},\ and\ \bibinfo
  {author} {\bibnamefont {Örs Legeza}},\ }\href
  {https://arxiv.org/abs/2406.03449} {\bibinfo {title} {Global fermionic mode
  optimization via swap gates}} (\bibinfo {year} {2024}),\ \Eprint
  {https://arxiv.org/abs/2406.03449} {arXiv:2406.03449 [cond-mat.str-el]}
  \BibitemShut {NoStop}%
\bibitem [{\citenamefont {Hager}\ \emph {et~al.}(2004)\citenamefont {Hager},
  \citenamefont {Jeckelmann}, \citenamefont {Fehske},\ and\ \citenamefont
  {Wellein}}]{Hager-2004}%
  \BibitemOpen
  \bibfield  {author} {\bibinfo {author} {\bibfnamefont {G.}~\bibnamefont
  {Hager}}, \bibinfo {author} {\bibfnamefont {E.}~\bibnamefont {Jeckelmann}},
  \bibinfo {author} {\bibfnamefont {H.}~\bibnamefont {Fehske}},\ and\ \bibinfo
  {author} {\bibfnamefont {G.}~\bibnamefont {Wellein}},\ }\bibfield  {title}
  {\bibinfo {title} {Parallelization strategies for density matrix
  renormalization group algorithms on shared-memory systems},\ }\href
  {https://doi.org/http://dx.doi.org/10.1016/j.jcp.2003.09.018} {\bibfield
  {journal} {\bibinfo  {journal} {Journal of Computational Physics}\ }\textbf
  {\bibinfo {volume} {194}},\ \bibinfo {pages} {795 } (\bibinfo {year}
  {2004})}\BibitemShut {NoStop}%
\bibitem [{\citenamefont {Stoudenmire}\ and\ \citenamefont
  {White}(2013)}]{Stoudenmire-2013}%
  \BibitemOpen
  \bibfield  {author} {\bibinfo {author} {\bibfnamefont {E.~M.}\ \bibnamefont
  {Stoudenmire}}\ and\ \bibinfo {author} {\bibfnamefont {S.~R.}\ \bibnamefont
  {White}},\ }\bibfield  {title} {\bibinfo {title} {Real-space parallel density
  matrix renormalization group},\ }\href
  {https://doi.org/10.1103/PhysRevB.87.155137} {\bibfield  {journal} {\bibinfo
  {journal} {Phys. Rev. B}\ }\textbf {\bibinfo {volume} {87}},\ \bibinfo
  {pages} {155137} (\bibinfo {year} {2013})}\BibitemShut {NoStop}%
\bibitem [{\citenamefont {Nemes}\ \emph {et~al.}(2014)\citenamefont {Nemes},
  \citenamefont {Barcza}, \citenamefont {Nagy}, \citenamefont {Legeza},\ and\
  \citenamefont {Szolgay}}]{Nemes-2014}%
  \BibitemOpen
  \bibfield  {author} {\bibinfo {author} {\bibfnamefont {C.}~\bibnamefont
  {Nemes}}, \bibinfo {author} {\bibfnamefont {G.}~\bibnamefont {Barcza}},
  \bibinfo {author} {\bibfnamefont {Z.}~\bibnamefont {Nagy}}, \bibinfo {author}
  {\bibfnamefont {{\"O}.}~\bibnamefont {Legeza}},\ and\ \bibinfo {author}
  {\bibfnamefont {P.}~\bibnamefont {Szolgay}},\ }\bibfield  {title} {\bibinfo
  {title} {The density matrix renormalization group algorithm on kilo-processor
  architectures: Implementation and trade-offs},\ }\href
  {https://doi.org/http://dx.doi.org/10.1016/j.cpc.2014.02.021} {\bibfield
  {journal} {\bibinfo  {journal} {Computer Physics Communications}\ }\textbf
  {\bibinfo {volume} {185}},\ \bibinfo {pages} {1570 } (\bibinfo {year}
  {2014})}\BibitemShut {NoStop}%
\bibitem [{\citenamefont {Ganahl}\ \emph {et~al.}(2019)\citenamefont {Ganahl},
  \citenamefont {Milsted}, \citenamefont {Leichenauer}, \citenamefont
  {Hidary},\ and\ \citenamefont {Vidal}}]{Ganahl-2019}%
  \BibitemOpen
  \bibfield  {author} {\bibinfo {author} {\bibfnamefont {M.}~\bibnamefont
  {Ganahl}}, \bibinfo {author} {\bibfnamefont {A.}~\bibnamefont {Milsted}},
  \bibinfo {author} {\bibfnamefont {S.}~\bibnamefont {Leichenauer}}, \bibinfo
  {author} {\bibfnamefont {J.}~\bibnamefont {Hidary}},\ and\ \bibinfo {author}
  {\bibfnamefont {G.}~\bibnamefont {Vidal}},\ }\bibfield  {title} {\bibinfo
  {title} {Tensornetwork on tensorflow: Entanglement renormalization for
  quantum critical lattice models},\ }\href@noop {} {\bibfield  {journal}
  {\bibinfo  {journal} {arxiv:1906.1203}\ } (\bibinfo {year}
  {2019})}\BibitemShut {NoStop}%
\bibitem [{\citenamefont {Milsted}\ \emph {et~al.}(2019)\citenamefont
  {Milsted}, \citenamefont {Ganahl}, \citenamefont {Leichenauer}, \citenamefont
  {Hidary},\ and\ \citenamefont {Vidal}}]{Milsted-2019}%
  \BibitemOpen
  \bibfield  {author} {\bibinfo {author} {\bibfnamefont {A.}~\bibnamefont
  {Milsted}}, \bibinfo {author} {\bibfnamefont {M.}~\bibnamefont {Ganahl}},
  \bibinfo {author} {\bibfnamefont {S.}~\bibnamefont {Leichenauer}}, \bibinfo
  {author} {\bibfnamefont {J.}~\bibnamefont {Hidary}},\ and\ \bibinfo {author}
  {\bibfnamefont {G.}~\bibnamefont {Vidal}},\ }\bibfield  {title} {\bibinfo
  {title} {Tensornetwork on tensorflow: A spin chain application using tree
  tensor networks},\ }\href@noop {} {\bibfield  {journal} {\bibinfo  {journal}
  {arxiv:1905.01331}\ } (\bibinfo {year} {2019})}\BibitemShut {NoStop}%
\bibitem [{\citenamefont {Brabec}\ \emph {et~al.}(2021)\citenamefont {Brabec},
  \citenamefont {Brandejs}, \citenamefont {Kowalski}, \citenamefont {Xantheas},
  \citenamefont {Legeza},\ and\ \citenamefont {Veis}}]{Brabec-2021}%
  \BibitemOpen
  \bibfield  {author} {\bibinfo {author} {\bibfnamefont {J.}~\bibnamefont
  {Brabec}}, \bibinfo {author} {\bibfnamefont {J.}~\bibnamefont {Brandejs}},
  \bibinfo {author} {\bibfnamefont {K.}~\bibnamefont {Kowalski}}, \bibinfo
  {author} {\bibfnamefont {S.}~\bibnamefont {Xantheas}}, \bibinfo {author}
  {\bibfnamefont {{\"O}.}~\bibnamefont {Legeza}},\ and\ \bibinfo {author}
  {\bibfnamefont {L.}~\bibnamefont {Veis}},\ }\bibfield  {title} {\bibinfo
  {title} {Massively parallel quantum chemical density matrix renormalization
  group method},\ }\href {https://doi.org/https://doi.org/10.1002/jcc.26476}
  {\bibfield  {journal} {\bibinfo  {journal} {Journal of Computational
  Chemistry}\ }\textbf {\bibinfo {volume} {42}},\ \bibinfo {pages} {534}
  (\bibinfo {year} {2021})},\ \Eprint
  {https://arxiv.org/abs/https://onlinelibrary.wiley.com/doi/pdf/10.1002/jcc.26476}
  {https://onlinelibrary.wiley.com/doi/pdf/10.1002/jcc.26476} \BibitemShut
  {NoStop}%
\bibitem [{\citenamefont {Zhai}\ and\ \citenamefont {Chan}(2021)}]{Zhai-2021}%
  \BibitemOpen
  \bibfield  {author} {\bibinfo {author} {\bibfnamefont {H.}~\bibnamefont
  {Zhai}}\ and\ \bibinfo {author} {\bibfnamefont {G.~K.-L.}\ \bibnamefont
  {Chan}},\ }\bibfield  {title} {\bibinfo {title} {Low communication high
  performance ab initio density matrix renormalization group algorithms},\
  }\href@noop {} {\bibfield  {journal} {\bibinfo  {journal} {Journal of
  Chemical Physics}\ }\textbf {\bibinfo {volume} {154}},\ \bibinfo {pages}
  {0021} (\bibinfo {year} {2021})}\BibitemShut {NoStop}%
\bibitem [{\citenamefont {Gray}\ and\ \citenamefont
  {Kourtis}(2021)}]{Gray-2021}%
  \BibitemOpen
  \bibfield  {author} {\bibinfo {author} {\bibfnamefont {J.}~\bibnamefont
  {Gray}}\ and\ \bibinfo {author} {\bibfnamefont {S.}~\bibnamefont {Kourtis}},\
  }\bibfield  {title} {\bibinfo {title} {Hyper-optimized tensor network
  contraction},\ }\bibfield  {journal} {\bibinfo  {journal} {Quantum}\ }\textbf
  {\bibinfo {volume} {5}},\ \href {https://doi.org/Doi:
  https://doi.org/10.22331/q-2021-03-15-410} {Doi:
  https://doi.org/10.22331/q-2021-03-15-410} (\bibinfo {year}
  {2021})\BibitemShut {NoStop}%
\bibitem [{\citenamefont {Unfried}\ \emph {et~al.}(2023)\citenamefont
  {Unfried}, \citenamefont {Hauschild},\ and\ \citenamefont
  {Pollmann}}]{Unfried-2023}%
  \BibitemOpen
  \bibfield  {author} {\bibinfo {author} {\bibfnamefont {J.}~\bibnamefont
  {Unfried}}, \bibinfo {author} {\bibfnamefont {J.}~\bibnamefont {Hauschild}},\
  and\ \bibinfo {author} {\bibfnamefont {F.}~\bibnamefont {Pollmann}},\
  }\bibfield  {title} {\bibinfo {title} {Fast time evolution of matrix product
  states using the qr decomposition},\ }\href
  {https://doi.org/10.1103/PhysRevB.107.155133} {\bibfield  {journal} {\bibinfo
   {journal} {Phys. Rev. B}\ }\textbf {\bibinfo {volume} {107}},\ \bibinfo
  {pages} {155133} (\bibinfo {year} {2023})}\BibitemShut {NoStop}%
\bibitem [{\citenamefont {Ganahl}\ \emph {et~al.}(2023)\citenamefont {Ganahl},
  \citenamefont {Beall}, \citenamefont {Hauru}, \citenamefont {Lewis},
  \citenamefont {Wojno}, \citenamefont {Yoo}, \citenamefont {Zou},\ and\
  \citenamefont {Vidal}}]{Ganahl-2023}%
  \BibitemOpen
  \bibfield  {author} {\bibinfo {author} {\bibfnamefont {M.}~\bibnamefont
  {Ganahl}}, \bibinfo {author} {\bibfnamefont {J.}~\bibnamefont {Beall}},
  \bibinfo {author} {\bibfnamefont {M.}~\bibnamefont {Hauru}}, \bibinfo
  {author} {\bibfnamefont {A.~G.}\ \bibnamefont {Lewis}}, \bibinfo {author}
  {\bibfnamefont {T.}~\bibnamefont {Wojno}}, \bibinfo {author} {\bibfnamefont
  {J.~H.}\ \bibnamefont {Yoo}}, \bibinfo {author} {\bibfnamefont
  {Y.}~\bibnamefont {Zou}},\ and\ \bibinfo {author} {\bibfnamefont
  {G.}~\bibnamefont {Vidal}},\ }\bibfield  {title} {\bibinfo {title} {Density
  matrix renormalization group with tensor processing units},\ }\href@noop {}
  {\bibfield  {journal} {\bibinfo  {journal} {PRX Quantum}\ }\textbf {\bibinfo
  {volume} {4}},\ \bibinfo {pages} {010317} (\bibinfo {year}
  {2023})}\BibitemShut {NoStop}%
\bibitem [{\citenamefont {Menczer}\ and\ \citenamefont {Örs
  Legeza}(2023)}]{Menczer-2023a}%
  \BibitemOpen
  \bibfield  {author} {\bibinfo {author} {\bibfnamefont {A.}~\bibnamefont
  {Menczer}}\ and\ \bibinfo {author} {\bibnamefont {Örs Legeza}},\ }\bibfield
  {title} {\bibinfo {title} {Massively parallel tensor network state algorithms
  on hybrid cpu-gpu based architectures},\ }\href {https://arxiv.org
  abs/2305.05581} {\bibfield  {journal} {\bibinfo  {journal}
  {arXiv:2305.05581}\ } (\bibinfo {year} {2023})},\ \Eprint
  {https://arxiv.org/abs/2305.05581} {arXiv:2305.05581 [quant-ph]} \BibitemShut
  {NoStop}%
\bibitem [{\citenamefont {Menczer}\ and\ \citenamefont
  {Legeza}(2024)}]{Menczer-2024d}%
  \BibitemOpen
  \bibfield  {author} {\bibinfo {author} {\bibfnamefont {A.}~\bibnamefont
  {Menczer}}\ and\ \bibinfo {author} {\bibfnamefont {O.}~\bibnamefont
  {Legeza}},\ }\bibfield  {title} {\bibinfo {title} {Tensor network state
  algorithms on ai accelerators},\ }\href
  {https://pubs.acs.org/doi/full/10.1021/acs.jctc.4c00800} {\bibfield
  {journal} {\bibinfo  {journal} {Journal of Chemical Theory and Computation}\
  }\textbf {\bibinfo {volume} {20}},\ \bibinfo {pages} {8897} (\bibinfo {year}
  {2024})}\BibitemShut {NoStop}%
\bibitem [{\citenamefont {Menczer}\ \emph
  {et~al.}(2024{\natexlab{a}})\citenamefont {Menczer}, \citenamefont {Kap\'as},
  \citenamefont {Werner},\ and\ \citenamefont {Legeza}}]{Menczer-2024a}%
  \BibitemOpen
  \bibfield  {author} {\bibinfo {author} {\bibfnamefont {A.}~\bibnamefont
  {Menczer}}, \bibinfo {author} {\bibfnamefont {K.}~\bibnamefont {Kap\'as}},
  \bibinfo {author} {\bibfnamefont {M.~A.}\ \bibnamefont {Werner}},\ and\
  \bibinfo {author} {\bibfnamefont {{\"O}.}~\bibnamefont {Legeza}},\ }\bibfield
   {title} {\bibinfo {title} {Two-dimensional quantum lattice models via mode
  optimized hybrid cpu-gpu density matrix renormalization group method},\
  }\href {https://doi.org/10.1103/PhysRevB.109.195148} {\bibfield  {journal}
  {\bibinfo  {journal} {Phys. Rev. B}\ }\textbf {\bibinfo {volume} {109}},\
  \bibinfo {pages} {195148} (\bibinfo {year} {2024}{\natexlab{a}})}\BibitemShut
  {NoStop}%
\bibitem [{\citenamefont {Menczer}\ and\ \citenamefont {Örs
  Legeza}(2024)}]{Menczer-2024c}%
  \BibitemOpen
  \bibfield  {author} {\bibinfo {author} {\bibfnamefont {A.}~\bibnamefont
  {Menczer}}\ and\ \bibinfo {author} {\bibnamefont {Örs Legeza}},\ }\href
  {https://arxiv.org/abs/2412.04676} {\bibinfo {title} {Cost optimized ab
  initio tensor network state methods: industrial perspectives}} (\bibinfo
  {year} {2024}),\ \Eprint {https://arxiv.org/abs/2412.04676} {arXiv:2412.04676
  [physics.comp-ph]} \BibitemShut {NoStop}%
\bibitem [{\citenamefont {Xiang}\ \emph {et~al.}(2024)\citenamefont {Xiang},
  \citenamefont {Jia}, \citenamefont {Fang},\ and\ \citenamefont
  {Li}}]{Xiang-2024}%
  \BibitemOpen
  \bibfield  {author} {\bibinfo {author} {\bibfnamefont {C.}~\bibnamefont
  {Xiang}}, \bibinfo {author} {\bibfnamefont {W.}~\bibnamefont {Jia}}, \bibinfo
  {author} {\bibfnamefont {W.-H.}\ \bibnamefont {Fang}},\ and\ \bibinfo
  {author} {\bibfnamefont {Z.}~\bibnamefont {Li}},\ }\bibfield  {title}
  {\bibinfo {title} {A distributed multi-gpu ab initio density matrix
  renormalization group algorithm with applications to the p-cluster of
  nitrogenase},\ }\href
  {https://doi.org/https://doi.org/10.1021/acs.jctc.3c01228} {\bibfield
  {journal} {\bibinfo  {journal} {Journal of Chemical Theory and Computation}\
  }\textbf {\bibinfo {volume} {20}},\ \bibinfo {pages} {775–786} (\bibinfo
  {year} {2024})},\ \Eprint {https://arxiv.org/abs/2311.02854} {2311.02854}
  \BibitemShut {NoStop}%
\bibitem [{\citenamefont {Menczer}\ and\ \citenamefont
  {Legeza}(2023{\natexlab{a}})}]{Menczer-2023c}%
  \BibitemOpen
  \bibfield  {author} {\bibinfo {author} {\bibfnamefont {A.}~\bibnamefont
  {Menczer}}\ and\ \bibinfo {author} {\bibfnamefont {{\"O}.}~\bibnamefont
  {Legeza}},\ }\href@noop {} {\bibinfo {title} {Petaflops density matrix
  renormalization group method, unpublished}} (\bibinfo {year}
  {2023}{\natexlab{a}}),\ \Eprint {https://arxiv.org/abs/unpublished}
  {unpublished} \BibitemShut {NoStop}%
\bibitem [{\citenamefont {Menczer}\ \emph
  {et~al.}(2024{\natexlab{b}})\citenamefont {Menczer}, \citenamefont {van
  Damme}, \citenamefont {Rask}, \citenamefont {Huntington}, \citenamefont
  {Hammond}, \citenamefont {Xantheas}, \citenamefont {Ganahl},\ and\
  \citenamefont {Legeza}}]{Menczer-2024b}%
  \BibitemOpen
  \bibfield  {author} {\bibinfo {author} {\bibfnamefont {A.}~\bibnamefont
  {Menczer}}, \bibinfo {author} {\bibfnamefont {M.}~\bibnamefont {van Damme}},
  \bibinfo {author} {\bibfnamefont {A.}~\bibnamefont {Rask}}, \bibinfo {author}
  {\bibfnamefont {L.}~\bibnamefont {Huntington}}, \bibinfo {author}
  {\bibfnamefont {J.}~\bibnamefont {Hammond}}, \bibinfo {author} {\bibfnamefont
  {S.~S.}\ \bibnamefont {Xantheas}}, \bibinfo {author} {\bibfnamefont
  {M.}~\bibnamefont {Ganahl}},\ and\ \bibinfo {author} {\bibfnamefont
  {O.}~\bibnamefont {Legeza}},\ }\bibfield  {title} {\bibinfo {title} {Parallel
  implementation of the {D}ensity {M}atrix {R}enormalization {G}roup method
  achieving a quarter peta{FLOPS} performance on a single {DGX-H100 GPU}
  node},\ }\href {https://pubs.acs.org/doi/full/10.1021/acs.jctc.4c00903}
  {\bibfield  {journal} {\bibinfo  {journal} {Journal of Chemical Theory and
  Computation}\ }\textbf {\bibinfo {volume} {20}},\ \bibinfo {pages} {8397}
  (\bibinfo {year} {2024}{\natexlab{b}})}\BibitemShut {NoStop}%
\bibitem [{\citenamefont {Helgaker}\ \emph {et~al.}(2000)\citenamefont
  {Helgaker}, \citenamefont {Jorgensen},\ and\ \citenamefont
  {Olsen}}]{Helgaker-2000}%
  \BibitemOpen
  \bibfield  {author} {\bibinfo {author} {\bibfnamefont {T.}~\bibnamefont
  {Helgaker}}, \bibinfo {author} {\bibfnamefont {P.}~\bibnamefont
  {Jorgensen}},\ and\ \bibinfo {author} {\bibfnamefont {J.}~\bibnamefont
  {Olsen}},\ }\href
  {http://eu.wiley.com/WileyCDA/WileyTitle/productCd-0471967556.html} {\emph
  {\bibinfo {title} {Molecular electronic-structure theory}}}\ (\bibinfo
  {publisher} {Wiley New York},\ \bibinfo {year} {2000})\BibitemShut {NoStop}%
\bibitem [{\citenamefont {Vidal}(2003)}]{Vidal-2003b}%
  \BibitemOpen
  \bibfield  {author} {\bibinfo {author} {\bibfnamefont {G.}~\bibnamefont
  {Vidal}},\ }\bibfield  {title} {\bibinfo {title} {Efficient classical
  simulation of slightly entangled quantum computations},\ }\href
  {https://doi.org/10.1103/PhysRevLett.91.147902} {\bibfield  {journal}
  {\bibinfo  {journal} {Phys. Rev. Lett.}\ }\textbf {\bibinfo {volume} {91}},\
  \bibinfo {pages} {147902} (\bibinfo {year} {2003})}\BibitemShut {NoStop}%
\bibitem [{\citenamefont {Verstraete}\ and\ \citenamefont
  {Cirac}(2004)}]{Verstraete-2004a}%
  \BibitemOpen
  \bibfield  {author} {\bibinfo {author} {\bibfnamefont {F.}~\bibnamefont
  {Verstraete}}\ and\ \bibinfo {author} {\bibfnamefont {J.~I.}\ \bibnamefont
  {Cirac}},\ }\bibfield  {title} {\bibinfo {title} {Renormalization algorithms
  for quantum-many body systems in two and higher dimensions},\ }\href
  {http://arxiv.org/abs/cond-mat/0407066} {\bibfield  {journal} {\bibinfo
  {journal} {arXiv [cond-mat.str-el]}\ ,\ \bibinfo {eid} {cond-mat/0407066}}
  (\bibinfo {year} {2004})},\ \Eprint
  {https://arxiv.org/abs/http://arxiv.org/pdf/cond-mat/0407066}
  {http://arxiv.org/pdf/cond-mat/0407066} \BibitemShut {NoStop}%
\bibitem [{\citenamefont {Verstraete}\ \emph {et~al.}(2004)\citenamefont
  {Verstraete}, \citenamefont {Porras},\ and\ \citenamefont
  {Cirac}}]{Verstraete-2004b}%
  \BibitemOpen
  \bibfield  {author} {\bibinfo {author} {\bibfnamefont {F.}~\bibnamefont
  {Verstraete}}, \bibinfo {author} {\bibfnamefont {D.}~\bibnamefont {Porras}},\
  and\ \bibinfo {author} {\bibfnamefont {J.~I.}\ \bibnamefont {Cirac}},\
  }\bibfield  {title} {\bibinfo {title} {Density matrix renormalization group
  and periodic boundary conditions: A quantum information perspective},\ }\href
  {https://doi.org/10.1103/PhysRevLett.93.227205} {\bibfield  {journal}
  {\bibinfo  {journal} {Phys. Rev. Lett.}\ }\textbf {\bibinfo {volume} {93}},\
  \bibinfo {pages} {227205} (\bibinfo {year} {2004})}\BibitemShut {NoStop}%
\bibitem [{\citenamefont {{\relax Sz}alay}\ \emph {et~al.}(2021)\citenamefont
  {{\relax Sz}alay}, \citenamefont {Zimbor{\'a}s}, \citenamefont
  {M{\'a}t{\'e}}, \citenamefont {Barcza}, \citenamefont {Schilling},\ and\
  \citenamefont {Legeza}}]{Szalay-2021}%
  \BibitemOpen
  \bibfield  {author} {\bibinfo {author} {\bibfnamefont {{\relax
  Sz}.}~\bibnamefont {{\relax Sz}alay}}, \bibinfo {author} {\bibfnamefont
  {Z.}~\bibnamefont {Zimbor{\'a}s}}, \bibinfo {author} {\bibfnamefont
  {M.}~\bibnamefont {M{\'a}t{\'e}}}, \bibinfo {author} {\bibfnamefont
  {G.}~\bibnamefont {Barcza}}, \bibinfo {author} {\bibfnamefont
  {C.}~\bibnamefont {Schilling}},\ and\ \bibinfo {author} {\bibfnamefont
  {{\"O}.}~\bibnamefont {Legeza}},\ }\bibfield  {title} {\bibinfo {title}
  {Fermionic systems for quantum information people},\ }\href
  {https://doi.org/10.1088/1751-8121/ac0646} {\bibfield  {journal} {\bibinfo
  {journal} {Journal of Physics A: Mathematical and Theoretical}\ }\textbf
  {\bibinfo {volume} {54}},\ \bibinfo {pages} {393001} (\bibinfo {year}
  {2021})}\BibitemShut {NoStop}%
\bibitem [{\citenamefont {Boguslawski}\ \emph {et~al.}(2013)\citenamefont
  {Boguslawski}, \citenamefont {Tecmer}, \citenamefont {Barcza}, \citenamefont
  {Legeza},\ and\ \citenamefont {Reiher}}]{Boguslawski-2013}%
  \BibitemOpen
  \bibfield  {author} {\bibinfo {author} {\bibfnamefont {K.}~\bibnamefont
  {Boguslawski}}, \bibinfo {author} {\bibfnamefont {P.}~\bibnamefont {Tecmer}},
  \bibinfo {author} {\bibfnamefont {G.}~\bibnamefont {Barcza}}, \bibinfo
  {author} {\bibfnamefont {{\"O}.}~\bibnamefont {Legeza}},\ and\ \bibinfo
  {author} {\bibfnamefont {M.}~\bibnamefont {Reiher}},\ }\bibfield  {title}
  {\bibinfo {title} {Orbital entanglement in bond-formation processes},\ }\href
  {https://doi.org/10.1021/ct400247p} {\bibfield  {journal} {\bibinfo
  {journal} {Journal of Chemical Theory and Computation}\ }\textbf {\bibinfo
  {volume} {9}},\ \bibinfo {pages} {2959} (\bibinfo {year} {2013})}\BibitemShut
  {NoStop}%
\bibitem [{\citenamefont {Lucas}(1882)}]{Lucas-1882}%
  \BibitemOpen
  \bibfield  {author} {\bibinfo {author} {\bibfnamefont {{\'E}.}~\bibnamefont
  {Lucas}},\ }\href@noop {} {\emph {\bibinfo {title} {R{\'e}cr{\'e}ations
  Math{\'e}matiques}}},\ Vol.~\bibinfo {volume} {1}\ (\bibinfo  {publisher}
  {Gauthier-Villars},\ \bibinfo {year} {1882})\BibitemShut {NoStop}%
\bibitem [{\citenamefont {McCulloch}\ and\ \citenamefont
  {Gul\'acsi}(2002)}]{McCulloch-2002a}%
  \BibitemOpen
  \bibfield  {author} {\bibinfo {author} {\bibfnamefont {I.~P.}\ \bibnamefont
  {McCulloch}}\ and\ \bibinfo {author} {\bibfnamefont {M.}~\bibnamefont
  {Gul\'acsi}},\ }\bibfield  {title} {\bibinfo {title} {The non-abelian density
  matrix renormalization group algorithm},\ }\href
  {http://stacks.iop.org/0295-5075/57/i=6/a=852} {\bibfield  {journal}
  {\bibinfo  {journal} {EPL (Europhysics Letters)}\ }\textbf {\bibinfo {volume}
  {57}},\ \bibinfo {pages} {852} (\bibinfo {year} {2002})}\BibitemShut
  {NoStop}%
\bibitem [{\citenamefont {T\'oth}\ \emph {et~al.}(2008)\citenamefont {T\'oth},
  \citenamefont {Moca}, \citenamefont {Legeza},\ and\ \citenamefont
  {Zar\'and}}]{Toth-2008}%
  \BibitemOpen
  \bibfield  {author} {\bibinfo {author} {\bibfnamefont {A.~I.}\ \bibnamefont
  {T\'oth}}, \bibinfo {author} {\bibfnamefont {C.~P.}\ \bibnamefont {Moca}},
  \bibinfo {author} {\bibfnamefont {{\"O}.}~\bibnamefont {Legeza}},\ and\
  \bibinfo {author} {\bibfnamefont {G.}~\bibnamefont {Zar\'and}},\ }\bibfield
  {title} {\bibinfo {title} {Density matrix numerical renormalization group for
  non-abelian symmetries},\ }\href {https://doi.org/10.1103/PhysRevB.78.245109}
  {\bibfield  {journal} {\bibinfo  {journal} {Phys. Rev. B}\ }\textbf {\bibinfo
  {volume} {78}},\ \bibinfo {pages} {245109} (\bibinfo {year}
  {2008})}\BibitemShut {NoStop}%
\bibitem [{\citenamefont {Sharma}\ and\ \citenamefont
  {Chan}(2012{\natexlab{a}})}]{Sharma-2012a}%
  \BibitemOpen
  \bibfield  {author} {\bibinfo {author} {\bibfnamefont {S.}~\bibnamefont
  {Sharma}}\ and\ \bibinfo {author} {\bibfnamefont {G.~K.-L.}\ \bibnamefont
  {Chan}},\ }\bibfield  {title} {\bibinfo {title} {Spin-adapted density matrix
  renormalization group algorithms for quantum chemistry},\ }\href
  {https://doi.org/http://dx.doi.org/10.1063/1.3695642} {\bibfield  {journal}
  {\bibinfo  {journal} {The Journal of Chemical Physics}\ }\textbf {\bibinfo
  {volume} {136}},\ \bibinfo {eid} {124121} (\bibinfo {year}
  {2012}{\natexlab{a}})}\BibitemShut {NoStop}%
\bibitem [{\citenamefont {Weichselbaum}(2012)}]{Weichselbaum-2012}%
  \BibitemOpen
  \bibfield  {author} {\bibinfo {author} {\bibfnamefont {A.}~\bibnamefont
  {Weichselbaum}},\ }\bibfield  {title} {\bibinfo {title} {Non-abelian
  symmetries in tensor networks: A quantum symmetry space approach},\ }\href
  {https://doi.org/http://dx.doi.org/10.1016/j.aop.2012.07.009} {\bibfield
  {journal} {\bibinfo  {journal} {Annals of Physics}\ }\textbf {\bibinfo
  {volume} {327}},\ \bibinfo {pages} {2972 } (\bibinfo {year}
  {2012})}\BibitemShut {NoStop}%
\bibitem [{\citenamefont {Keller}\ and\ \citenamefont
  {Reiher}(2016)}]{Keller-2016}%
  \BibitemOpen
  \bibfield  {author} {\bibinfo {author} {\bibfnamefont {S.}~\bibnamefont
  {Keller}}\ and\ \bibinfo {author} {\bibfnamefont {M.}~\bibnamefont
  {Reiher}},\ }\bibfield  {title} {\bibinfo {title} {Spin-adapted matrix
  product states and operators},\ }\href {https://doi.org/10.1063/1.4944921}
  {\bibfield  {journal} {\bibinfo  {journal} {The Journal of Chemical Physics}\
  }\textbf {\bibinfo {volume} {144}},\ \bibinfo {pages} {134101} (\bibinfo
  {year} {2016})}\BibitemShut {NoStop}%
\bibitem [{\citenamefont {Gunst}\ \emph {et~al.}(2019)\citenamefont {Gunst},
  \citenamefont {Verstraete},\ and\ \citenamefont {Neck}}]{Gunst-2019}%
  \BibitemOpen
  \bibfield  {author} {\bibinfo {author} {\bibfnamefont {K.}~\bibnamefont
  {Gunst}}, \bibinfo {author} {\bibfnamefont {F.}~\bibnamefont {Verstraete}},\
  and\ \bibinfo {author} {\bibfnamefont {D.~V.}\ \bibnamefont {Neck}},\
  }\bibfield  {title} {\bibinfo {title} {Three-legged tree tensor networks with
  su(2) and molecular point group symmetry},\ }\href
  {https://doi.org/10.1021/acs.jctc.9b00071} {\bibfield  {journal} {\bibinfo
  {journal} {Journal of Chemical Theory and Computation}\ }\textbf {\bibinfo
  {volume} {15}},\ \bibinfo {pages} {2996} (\bibinfo {year}
  {2019})}\BibitemShut {NoStop}%
\bibitem [{\citenamefont {Werner}\ \emph {et~al.}(2020)\citenamefont {Werner},
  \citenamefont {Moca}, \citenamefont {Legeza},\ and\ \citenamefont
  {Zar{\'a}nd}}]{Werner-2020}%
  \BibitemOpen
  \bibfield  {author} {\bibinfo {author} {\bibfnamefont {M.~A.}\ \bibnamefont
  {Werner}}, \bibinfo {author} {\bibfnamefont {C.~P.}\ \bibnamefont {Moca}},
  \bibinfo {author} {\bibfnamefont {{\"O}.}~\bibnamefont {Legeza}},\ and\
  \bibinfo {author} {\bibfnamefont {G.}~\bibnamefont {Zar{\'a}nd}},\ }\bibfield
   {title} {\bibinfo {title} {Quantum quench and charge oscillations in the su
  (3) hubbard model: A test of time evolving block decimation with general
  non-abelian symmetries},\ }\href@noop {} {\bibfield  {journal} {\bibinfo
  {journal} {Physical Review B}\ }\textbf {\bibinfo {volume} {102}},\ \bibinfo
  {pages} {155108} (\bibinfo {year} {2020})}\BibitemShut {NoStop}%
\bibitem [{\citenamefont {Wigner}(1959)}]{Wigner-1959}%
  \BibitemOpen
  \bibfield  {author} {\bibinfo {author} {\bibfnamefont {E.~P.}\ \bibnamefont
  {Wigner}},\ }\href@noop {} {\emph {\bibinfo {title} {Group Theory and Its
  Application to the Quantum Mechanics of Atomic Spectra}}}\ (\bibinfo
  {publisher} {Academic Press},\ \bibinfo {year} {1959})\BibitemShut {NoStop}%
\bibitem [{\citenamefont {Varshalovich}\ \emph {et~al.}(1988)\citenamefont
  {Varshalovich}, \citenamefont {Moskalev},\ and\ \citenamefont
  {Khersonskii}}]{Varshalovich-1988}%
  \BibitemOpen
  \bibfield  {author} {\bibinfo {author} {\bibfnamefont {D.~A.}\ \bibnamefont
  {Varshalovich}}, \bibinfo {author} {\bibfnamefont {A.~N.}\ \bibnamefont
  {Moskalev}},\ and\ \bibinfo {author} {\bibfnamefont {V.~K.}\ \bibnamefont
  {Khersonskii}},\ }\href@noop {} {\emph {\bibinfo {title} {Quantum theory of
  angular momentum}}}\ (\bibinfo  {publisher} {World Scientific},\ \bibinfo
  {year} {1988})\BibitemShut {NoStop}%
\bibitem [{\citenamefont {NVIDIA}(2021)}]{a100}%
  \BibitemOpen
  \bibfield  {author} {\bibinfo {author} {\bibnamefont {NVIDIA}},\ }\bibfield
  {title} {\bibinfo {title} {Nvidia a100 tensor core gpu},\ }\href@noop {}
  {\bibfield  {journal} {\bibinfo  {journal}
  {\url{https://www.nvidia.com/content/dam/en-zz/Solutions/Data-Center/a100/pdf/nvidia-a100-datasheet-us-nvidia-1758950-r4-web.pdf}}\
  } (\bibinfo {year} {2021})}\BibitemShut {NoStop}%
\bibitem [{\citenamefont {NVIDIA}()}]{gh200}%
  \BibitemOpen
  \bibfield  {author} {\bibinfo {author} {\bibnamefont {NVIDIA}},\ }\href
  {https://resources.nvidia.com/en-us-dgx-systems/nvidia-dgx-gh200-datasheet-web-us}
  {\bibinfo {title} {{NVIDIA DGX GH200}}},\ \bibinfo {howpublished}
  {https://resources.nvidia.com/en-us-dgx-systems/nvidia-dgx-gh200-datasheet-web-us}\BibitemShut
  {NoStop}%
\bibitem [{\citenamefont {AMD}()}]{mi300}%
  \BibitemOpen
  \bibfield  {author} {\bibinfo {author} {\bibnamefont {AMD}},\ }\href
  {https://www.amd.com/content/dam/amd/en/documents/instinct-tech-docs/data-sheets/amd-instinct-mi300a-data-sheet.pdf}
  {\bibinfo {title} {{AMD INSTINCT MI300A APU}}},\ \bibinfo {howpublished}
  {https://www.amd.com/content/dam/amd/en/documents/instinct-tech-docs/data-sheets/amd-instinct-mi300a-data-sheet.pdf}\BibitemShut
  {NoStop}%
\bibitem [{\citenamefont {Menczer}\ \emph
  {et~al.}(2024{\natexlab{c}})\citenamefont {Menczer}, \citenamefont {Ganyecz},
  \citenamefont {Werner}, \citenamefont {Neese},\ and\ \citenamefont
  {Legeza}}]{Menczer-2025a}%
  \BibitemOpen
  \bibfield  {author} {\bibinfo {author} {\bibfnamefont {A.}~\bibnamefont
  {Menczer}}, \bibinfo {author} {\bibfnamefont {{\'A}.}~\bibnamefont
  {Ganyecz}}, \bibinfo {author} {\bibfnamefont {M.~A.}\ \bibnamefont {Werner}},
  \bibinfo {author} {\bibfnamefont {F.}~\bibnamefont {Neese}},\ and\ \bibinfo
  {author} {\bibfnamefont {{\"O}.}~\bibnamefont {Legeza}},\ }\bibfield  {title}
  {\bibinfo {title} {Orbital optimization of large active spaces via
  ai-accelerators},\ }\href@noop {} {\bibfield  {journal} {\bibinfo  {journal}
  {unpublished}\ } (\bibinfo {year} {2024}{\natexlab{c}})}\BibitemShut
  {NoStop}%
\bibitem [{\citenamefont {Liang}\ and\ \citenamefont
  {Pang}(1994)}]{Liang-1994}%
  \BibitemOpen
  \bibfield  {author} {\bibinfo {author} {\bibfnamefont {S.}~\bibnamefont
  {Liang}}\ and\ \bibinfo {author} {\bibfnamefont {H.}~\bibnamefont {Pang}},\
  }\bibfield  {title} {\bibinfo {title} {Approximate diagonalization using the
  density matrix renormalization-group method: A two-dimensional-systems
  perspective},\ }\href {https://doi.org/10.1103/PhysRevB.49.9214} {\bibfield
  {journal} {\bibinfo  {journal} {Physical Review B}\ }\textbf {\bibinfo
  {volume} {49}},\ \bibinfo {pages} {9214} (\bibinfo {year}
  {1994})}\BibitemShut {NoStop}%
\bibitem [{\citenamefont {Noack}\ \emph {et~al.}(1994)\citenamefont {Noack},
  \citenamefont {White},\ and\ \citenamefont {Scalapino}}]{Noack-1994}%
  \BibitemOpen
  \bibfield  {author} {\bibinfo {author} {\bibfnamefont {R.~M.}\ \bibnamefont
  {Noack}}, \bibinfo {author} {\bibfnamefont {S.~R.}\ \bibnamefont {White}},\
  and\ \bibinfo {author} {\bibfnamefont {D.~J.}\ \bibnamefont {Scalapino}},\
  }\bibfield  {title} {\bibinfo {title} {The density-matrix renormalization
  group for fermion systems},\ }in\ \href
  {https://link.springer.com/chapter/10.1007/978-3-642-79293-9_8} {\emph
  {\bibinfo {booktitle} {Computer Simulation Studies in Condensed-Matter
  Physics VII: Proceedings of the Seventh Workshop Athens, GA, USA, 28
  February--4 March 1994}}}\ (\bibinfo {organization} {Springer},\ \bibinfo
  {year} {1994})\ pp.\ \bibinfo {pages} {85--98}\BibitemShut {NoStop}%
\bibitem [{\citenamefont {White}\ and\ \citenamefont
  {Scalapino}(2003)}]{White-2003}%
  \BibitemOpen
  \bibfield  {author} {\bibinfo {author} {\bibfnamefont {S.~R.}\ \bibnamefont
  {White}}\ and\ \bibinfo {author} {\bibfnamefont {D.~J.}\ \bibnamefont
  {Scalapino}},\ }\bibfield  {title} {\bibinfo {title} {Stripes on a 6-leg
  hubbard ladder},\ }\href {https://doi.org/10.1103/PhysRevLett.91.136403}
  {\bibfield  {journal} {\bibinfo  {journal} {Physical Review Letters}\
  }\textbf {\bibinfo {volume} {91}},\ \bibinfo {pages} {136403} (\bibinfo
  {year} {2003})}\BibitemShut {NoStop}%
\bibitem [{\citenamefont {Legeza}\ and\ \citenamefont
  {S\'olyom}(2004{\natexlab{a}})}]{Legeza-2004a}%
  \BibitemOpen
  \bibfield  {author} {\bibinfo {author} {\bibfnamefont {{\"O}.}~\bibnamefont
  {Legeza}}\ and\ \bibinfo {author} {\bibfnamefont {J.}~\bibnamefont
  {S\'olyom}},\ }\bibfield  {title} {\bibinfo {title} {Optimizing
  density-matrix renormalization group method using quantum information
  entropy},\ }in\ \href
  {http://www.itp.uni-hannover.de/~jeckelm/dmrg/workshop/proceedings.html}
  {\emph {\bibinfo {booktitle} {International Workshop on Recent Progress and
  Prospects in Density-Matrix Renormalization}}}\ (\bibinfo  {publisher}
  {Lorentz Center, Leiden University, The Netherlands},\ \bibinfo {year}
  {2004})\BibitemShut {NoStop}%
\bibitem [{\citenamefont {Legeza}\ and\ \citenamefont
  {S\'olyom}(2004{\natexlab{b}})}]{Legeza-2004b}%
  \BibitemOpen
  \bibfield  {author} {\bibinfo {author} {\bibfnamefont {{\"O}.}~\bibnamefont
  {Legeza}}\ and\ \bibinfo {author} {\bibfnamefont {J.}~\bibnamefont
  {S\'olyom}},\ }\bibfield  {title} {\bibinfo {title} {Quantum data
  compression, quantum information generation, and the density-matrix
  renormalization-group method},\ }\href
  {https://doi.org/10.1103/PhysRevB.70.205118} {\bibfield  {journal} {\bibinfo
  {journal} {Phys. Rev. B}\ }\textbf {\bibinfo {volume} {70}},\ \bibinfo
  {pages} {205118} (\bibinfo {year} {2004}{\natexlab{b}})}\BibitemShut
  {NoStop}%
\bibitem [{\citenamefont {Sch\"afer}\ \emph {et~al.}(1992)\citenamefont
  {Sch\"afer}, \citenamefont {Horn},\ and\ \citenamefont
  {Ahlrichs}}]{Schafer-1992}%
  \BibitemOpen
  \bibfield  {author} {\bibinfo {author} {\bibfnamefont {A.}~\bibnamefont
  {Sch\"afer}}, \bibinfo {author} {\bibfnamefont {H.}~\bibnamefont {Horn}},\
  and\ \bibinfo {author} {\bibfnamefont {R.}~\bibnamefont {Ahlrichs}},\
  }\bibfield  {title} {\bibinfo {title} {Fully optimized contracted gaussian
  basis sets for atoms {L}i to {K}r},\ }\href@noop {} {\bibfield  {journal}
  {\bibinfo  {journal} {J. Chem. Phys.}\ }\textbf {\bibinfo {volume} {97}},\
  \bibinfo {pages} {2571} (\bibinfo {year} {1992})}\BibitemShut {NoStop}%
\bibitem [{\citenamefont {Barcza}\ \emph {et~al.}(2011)\citenamefont {Barcza},
  \citenamefont {Legeza}, \citenamefont {Marti},\ and\ \citenamefont
  {Reiher}}]{Barcza-2011}%
  \BibitemOpen
  \bibfield  {author} {\bibinfo {author} {\bibfnamefont {G.}~\bibnamefont
  {Barcza}}, \bibinfo {author} {\bibfnamefont {{\"O}.}~\bibnamefont {Legeza}},
  \bibinfo {author} {\bibfnamefont {K.~H.}\ \bibnamefont {Marti}},\ and\
  \bibinfo {author} {\bibfnamefont {M.}~\bibnamefont {Reiher}},\ }\bibfield
  {title} {\bibinfo {title} {Quantum-information analysis of electronic states
  of different molecular structures},\ }\href
  {https://doi.org/10.1103/PhysRevA.83.012508} {\bibfield  {journal} {\bibinfo
  {journal} {Phys. Rev. A}\ }\textbf {\bibinfo {volume} {83}},\ \bibinfo
  {pages} {012508} (\bibinfo {year} {2011})}\BibitemShut {NoStop}%
\bibitem [{\citenamefont {Dunning}(1989)}]{Dunning-1989}%
  \BibitemOpen
  \bibfield  {author} {\bibinfo {author} {\bibfnamefont {T.~H.}\ \bibnamefont
  {Dunning}},\ }\bibfield  {title} {\bibinfo {title} {Gaussian basis sets for
  use in correlated molecular calculations. {I}. {T}he atoms boron through neon
  and hydrogen},\ }\href {https://doi.org/10.1063/1.456153} {\bibfield
  {journal} {\bibinfo  {journal} {The Journal of Chemical Physics}\ }\textbf
  {\bibinfo {volume} {90}},\ \bibinfo {pages} {1007} (\bibinfo {year}
  {1989})}\BibitemShut {NoStop}%
\bibitem [{\citenamefont {Chan}\ \emph {et~al.}(2004)\citenamefont {Chan},
  \citenamefont {K\'allay},\ and\ \citenamefont {Gauss}}]{Chan-2004b}%
  \BibitemOpen
  \bibfield  {author} {\bibinfo {author} {\bibfnamefont {G.~K.-L.}\
  \bibnamefont {Chan}}, \bibinfo {author} {\bibfnamefont {M.}~\bibnamefont
  {K\'allay}},\ and\ \bibinfo {author} {\bibfnamefont {J.}~\bibnamefont
  {Gauss}},\ }\bibfield  {title} {\bibinfo {title} {State-of-the-art density
  matrix renormalization group and coupled cluster theory studies of the
  nitrogen binding curve},\ }\href
  {https://doi.org/http://dx.doi.org/10.1063/1.1783212} {\bibfield  {journal}
  {\bibinfo  {journal} {The Journal of Chemical Physics}\ }\textbf {\bibinfo
  {volume} {121}},\ \bibinfo {pages} {6110} (\bibinfo {year}
  {2004})}\BibitemShut {NoStop}%
\bibitem [{\citenamefont {Faulstich}\ \emph
  {et~al.}(2019{\natexlab{a}})\citenamefont {Faulstich}, \citenamefont {Mate},
  \citenamefont {Laestadius}, \citenamefont {Csirik}, \citenamefont {Veis},
  \citenamefont {Antalik}, \citenamefont {Brabec}, \citenamefont {Schneider},
  \citenamefont {Pittner}, \citenamefont {Kvaal},\ and\ \citenamefont
  {Legeza}}]{Faulstich-2019b}%
  \BibitemOpen
  \bibfield  {author} {\bibinfo {author} {\bibfnamefont {F.~M.}\ \bibnamefont
  {Faulstich}}, \bibinfo {author} {\bibfnamefont {M.}~\bibnamefont {Mate}},
  \bibinfo {author} {\bibfnamefont {A.}~\bibnamefont {Laestadius}}, \bibinfo
  {author} {\bibfnamefont {M.~A.}\ \bibnamefont {Csirik}}, \bibinfo {author}
  {\bibfnamefont {L.}~\bibnamefont {Veis}}, \bibinfo {author} {\bibfnamefont
  {A.}~\bibnamefont {Antalik}}, \bibinfo {author} {\bibfnamefont
  {J.}~\bibnamefont {Brabec}}, \bibinfo {author} {\bibfnamefont
  {R.}~\bibnamefont {Schneider}}, \bibinfo {author} {\bibfnamefont
  {J.}~\bibnamefont {Pittner}}, \bibinfo {author} {\bibfnamefont
  {S.}~\bibnamefont {Kvaal}},\ and\ \bibinfo {author} {\bibfnamefont
  {{\"O}.}~\bibnamefont {Legeza}},\ }\bibfield  {title} {\bibinfo {title}
  {Numerical and theoretical aspects of the dmrg-tcc method exemplified by the
  nitrogen dimer},\ }\href {https://doi.org/10.1021/acs.jctc.8b00960}
  {\bibfield  {journal} {\bibinfo  {journal} {Journal of Chemical Theory and
  Computation}\ }\textbf {\bibinfo {volume} {15}},\ \bibinfo {pages} {2206}
  (\bibinfo {year} {2019}{\natexlab{a}})},\ \Eprint
  {https://arxiv.org/abs/https://doi.org/10.1021/acs.jctc.8b00960}
  {https://doi.org/10.1021/acs.jctc.8b00960} \BibitemShut {NoStop}%
\bibitem [{Note1()}]{Note1}%
  \BibitemOpen
  \bibinfo {note} {We remark that the initial slight increase of BEA in the
  first macro iteration is an artefact caused by a non-converged wave
  function.}\BibitemShut {Stop}%
\bibitem [{\citenamefont {Kurashige}\ and\ \citenamefont
  {Yanai}(2011)}]{Kurashige-2011}%
  \BibitemOpen
  \bibfield  {author} {\bibinfo {author} {\bibfnamefont {Y.}~\bibnamefont
  {Kurashige}}\ and\ \bibinfo {author} {\bibfnamefont {T.}~\bibnamefont
  {Yanai}},\ }\bibfield  {title} {\bibinfo {title} {Second-order perturbation
  theory with a density matrix renormalization group self-consistent field
  reference function: Theory and application to the study of chromium dimer},\
  }\href {https://doi.org/http://dx.doi.org/10.1063/1.3629454} {\bibfield
  {journal} {\bibinfo  {journal} {The Journal of Chemical Physics}\ }\textbf
  {\bibinfo {volume} {135}},\ \bibinfo {eid} {094104} (\bibinfo {year}
  {2011})}\BibitemShut {NoStop}%
\bibitem [{\citenamefont {Ma}\ \emph {et~al.}(2017)\citenamefont {Ma},
  \citenamefont {Knecht}, \citenamefont {Keller},\ and\ \citenamefont
  {Reiher}}]{Ma-2017}%
  \BibitemOpen
  \bibfield  {author} {\bibinfo {author} {\bibfnamefont {Y.}~\bibnamefont
  {Ma}}, \bibinfo {author} {\bibfnamefont {S.}~\bibnamefont {Knecht}}, \bibinfo
  {author} {\bibfnamefont {S.}~\bibnamefont {Keller}},\ and\ \bibinfo {author}
  {\bibfnamefont {M.}~\bibnamefont {Reiher}},\ }\bibfield  {title} {\bibinfo
  {title} {Second-order self-consistent-field density-matrix renormalization
  group},\ }\href {https://doi.org/10.1021/acs.jctc.6b01118} {\bibfield
  {journal} {\bibinfo  {journal} {Journal of Chemical Theory and Computation}\
  }\textbf {\bibinfo {volume} {13}},\ \bibinfo {pages} {2533} (\bibinfo {year}
  {2017})},\ \bibinfo {note} {pMID: 28485978},\ \Eprint
  {https://arxiv.org/abs/https://doi.org/10.1021/acs.jctc.6b01118}
  {https://doi.org/10.1021/acs.jctc.6b01118} \BibitemShut {NoStop}%
\bibitem [{\citenamefont {Sharma}\ and\ \citenamefont
  {Chan}(2012{\natexlab{b}})}]{Sharma-2012}%
  \BibitemOpen
  \bibfield  {author} {\bibinfo {author} {\bibfnamefont {S.}~\bibnamefont
  {Sharma}}\ and\ \bibinfo {author} {\bibfnamefont {G.~K.-L.}\ \bibnamefont
  {Chan}},\ }\bibfield  {title} {\bibinfo {title} {Spin-adapted density matrix
  renormalization group algorithms for quantum chemistry},\ }\href
  {https://doi.org/10.1063/1.3695642} {\bibfield  {journal} {\bibinfo
  {journal} {The Journal of Chemical Physics}\ }\textbf {\bibinfo {volume}
  {136}},\ \bibinfo {pages} {124121} (\bibinfo {year} {2012}{\natexlab{b}})},\
  \Eprint {https://arxiv.org/abs/https://doi.org/10.1063/1.3695642}
  {https://doi.org/10.1063/1.3695642} \BibitemShut {NoStop}%
\bibitem [{\citenamefont {Veis}\ \emph {et~al.}(2016)\citenamefont {Veis},
  \citenamefont {Antalik}, \citenamefont {Brabec}, \citenamefont {Neese},
  \citenamefont {Legeza},\ and\ \citenamefont {Pittner}}]{Veis-2016}%
  \BibitemOpen
  \bibfield  {author} {\bibinfo {author} {\bibfnamefont {L.}~\bibnamefont
  {Veis}}, \bibinfo {author} {\bibfnamefont {A.}~\bibnamefont {Antalik}},
  \bibinfo {author} {\bibfnamefont {J.}~\bibnamefont {Brabec}}, \bibinfo
  {author} {\bibfnamefont {F.}~\bibnamefont {Neese}}, \bibinfo {author}
  {\bibfnamefont {{\"O}.}~\bibnamefont {Legeza}},\ and\ \bibinfo {author}
  {\bibfnamefont {J.}~\bibnamefont {Pittner}},\ }\bibfield  {title} {\bibinfo
  {title} {Coupled cluster method with single and double excitations tailored
  by matrix product state wave functions},\ }\href
  {https://doi.org/10.1021/acs.jpclett.6b01908} {\bibfield  {journal} {\bibinfo
   {journal} {The Journal of Physical Chemistry Letters}\ }\textbf {\bibinfo
  {volume} {7}},\ \bibinfo {pages} {4072} (\bibinfo {year} {2016})},\ \Eprint
  {https://arxiv.org/abs/https://doi.org/10.1021/acs.jpclett.6b01908}
  {https://doi.org/10.1021/acs.jpclett.6b01908} \BibitemShut {NoStop}%
\bibitem [{\citenamefont {Barcza}\ \emph {et~al.}(2022)\citenamefont {Barcza},
  \citenamefont {Werner}, \citenamefont {Zarand}, \citenamefont {Pershin},
  \citenamefont {Benedek}, \citenamefont {Legeza},\ and\ \citenamefont
  {Szilvasi}}]{Barcza-2022}%
  \BibitemOpen
  \bibfield  {author} {\bibinfo {author} {\bibfnamefont {G.}~\bibnamefont
  {Barcza}}, \bibinfo {author} {\bibfnamefont {M.~A.}\ \bibnamefont {Werner}},
  \bibinfo {author} {\bibfnamefont {G.}~\bibnamefont {Zarand}}, \bibinfo
  {author} {\bibfnamefont {A.}~\bibnamefont {Pershin}}, \bibinfo {author}
  {\bibfnamefont {Z.}~\bibnamefont {Benedek}}, \bibinfo {author} {\bibfnamefont
  {O.}~\bibnamefont {Legeza}},\ and\ \bibinfo {author} {\bibfnamefont
  {T.}~\bibnamefont {Szilvasi}},\ }\bibfield  {title} {\bibinfo {title} {Toward
  large-scale restricted active space calculations inspired by the schmidt
  decomposition},\ }\href {https://doi.org/10.1021/acs.jpca.2c05952} {\bibfield
   {journal} {\bibinfo  {journal} {The Journal of Physical Chemistry A}\
  }\textbf {\bibinfo {volume} {126}},\ \bibinfo {pages} {9709} (\bibinfo {year}
  {2022})},\ \Eprint
  {https://arxiv.org/abs/https://doi.org/10.1021/acs.jpca.2c05952}
  {https://doi.org/10.1021/acs.jpca.2c05952} \BibitemShut {NoStop}%
\bibitem [{\citenamefont {Larsson}\ \emph
  {et~al.}(2022{\natexlab{a}})\citenamefont {Larsson}, \citenamefont {Zhai},
  \citenamefont {Gunst},\ and\ \citenamefont {Chan}}]{Larsson-2022}%
  \BibitemOpen
  \bibfield  {author} {\bibinfo {author} {\bibfnamefont {H.~R.}\ \bibnamefont
  {Larsson}}, \bibinfo {author} {\bibfnamefont {H.}~\bibnamefont {Zhai}},
  \bibinfo {author} {\bibfnamefont {K.}~\bibnamefont {Gunst}},\ and\ \bibinfo
  {author} {\bibfnamefont {G.~K.-L.}\ \bibnamefont {Chan}},\ }\bibfield
  {title} {\bibinfo {title} {Matrix product states with large sites},\ }\href
  {https://doi.org/10.1021/acs.jctc.1c00957} {\bibfield  {journal} {\bibinfo
  {journal} {Journal of Chemical Theory and Computation}\ }\textbf {\bibinfo
  {volume} {18}},\ \bibinfo {pages} {749} (\bibinfo {year}
  {2022}{\natexlab{a}})},\ \bibinfo {note} {pMID: 35060382},\ \Eprint
  {https://arxiv.org/abs/https://doi.org/10.1021/acs.jctc.1c00957}
  {https://doi.org/10.1021/acs.jctc.1c00957} \BibitemShut {NoStop}%
\bibitem [{\citenamefont {Larsson}\ \emph
  {et~al.}(2022{\natexlab{b}})\citenamefont {Larsson}, \citenamefont {Zhai},
  \citenamefont {Umrigar},\ and\ \citenamefont {Chan}}]{Larsson-2021-arxiv}%
  \BibitemOpen
  \bibfield  {author} {\bibinfo {author} {\bibfnamefont {H.~R.}\ \bibnamefont
  {Larsson}}, \bibinfo {author} {\bibfnamefont {H.}~\bibnamefont {Zhai}},
  \bibinfo {author} {\bibfnamefont {C.~J.}\ \bibnamefont {Umrigar}},\ and\
  \bibinfo {author} {\bibfnamefont {G.~K.-L.}\ \bibnamefont {Chan}},\
  }\bibfield  {title} {\bibinfo {title} {The chromium dimer: Closing a chapter
  of quantum chemistry},\ }\href {https://doi.org/10.1021/jacs.2c06357}
  {\bibfield  {journal} {\bibinfo  {journal} {Journal of the American Chemical
  Society}\ }\textbf {\bibinfo {volume} {144}},\ \bibinfo {pages} {15932}
  (\bibinfo {year} {2022}{\natexlab{b}})},\ \bibinfo {note} {pMID: 36001866},\
  \Eprint {https://arxiv.org/abs/https://doi.org/10.1021/jacs.2c06357
  [doi.org]} {https://doi.org/10.1021/jacs.2c06357 [doi.org]} \BibitemShut
  {NoStop}%
\bibitem [{\citenamefont {{\relax Sz}alay}(2015)}]{Szalay-2015b}%
  \BibitemOpen
  \bibfield  {author} {\bibinfo {author} {\bibfnamefont {{\relax
  Sz}.}~\bibnamefont {{\relax Sz}alay}},\ }\bibfield  {title} {\bibinfo {title}
  {Multipartite entanglement measures},\ }\href
  {https://doi.org/10.1103/PhysRevA.92.042329} {\bibfield  {journal} {\bibinfo
  {journal} {Phys. Rev. A}\ }\textbf {\bibinfo {volume} {92}},\ \bibinfo
  {pages} {042329} (\bibinfo {year} {2015})}\BibitemShut {NoStop}%
\bibitem [{\citenamefont {Menczer}\ and\ \citenamefont
  {Legeza}(2023{\natexlab{b}})}]{Menczer-2023b}%
  \BibitemOpen
  \bibfield  {author} {\bibinfo {author} {\bibfnamefont {A.}~\bibnamefont
  {Menczer}}\ and\ \bibinfo {author} {\bibfnamefont {{\"O}.}~\bibnamefont
  {Legeza}},\ }\href {https://arxiv.org/abs/2309.16724} {\bibinfo {title}
  {Boosting the effective performance of massively parallel tensor network
  state algorithms on hybrid cpu-gpu based architectures via non-abelian
  symmetries}} (\bibinfo {year} {2023}{\natexlab{b}}),\ \Eprint
  {https://arxiv.org/abs/2309.16724} {arXiv:2309.16724 [physics.comp-ph]}
  \BibitemShut {NoStop}%
\bibitem [{\citenamefont {Piecuch}\ \emph {et~al.}(1996)\citenamefont
  {Piecuch}, \citenamefont {Tobol/a},\ and\ \citenamefont
  {Paldus}}]{Piecuch-1996}%
  \BibitemOpen
  \bibfield  {author} {\bibinfo {author} {\bibfnamefont {P.}~\bibnamefont
  {Piecuch}}, \bibinfo {author} {\bibfnamefont {R.}~\bibnamefont {Tobol/a}},\
  and\ \bibinfo {author} {\bibfnamefont {J.}~\bibnamefont {Paldus}},\
  }\bibfield  {title} {\bibinfo {title} {Approximate account of connected
  quadruply excited clusters in single-reference coupled-cluster theory via
  cluster analysis of the projected unrestricted hartree-fock wave function},\
  }\href {https://doi.org/10.1103/PhysRevA.54.1210} {\bibfield  {journal}
  {\bibinfo  {journal} {Phys. Rev. A}\ }\textbf {\bibinfo {volume} {54}},\
  \bibinfo {pages} {1210} (\bibinfo {year} {1996})}\BibitemShut {NoStop}%
\bibitem [{\citenamefont {Faulstich}\ \emph
  {et~al.}(2019{\natexlab{b}})\citenamefont {Faulstich}, \citenamefont
  {Laestadius}, \citenamefont {Legeza}, \citenamefont {Schneider},\ and\
  \citenamefont {Kvaal}}]{Faulstich-2019a}%
  \BibitemOpen
  \bibfield  {author} {\bibinfo {author} {\bibfnamefont {F.~M.}\ \bibnamefont
  {Faulstich}}, \bibinfo {author} {\bibfnamefont {A.}~\bibnamefont
  {Laestadius}}, \bibinfo {author} {\bibfnamefont {{\"O}.}~\bibnamefont
  {Legeza}}, \bibinfo {author} {\bibfnamefont {R.}~\bibnamefont {Schneider}},\
  and\ \bibinfo {author} {\bibfnamefont {S.}~\bibnamefont {Kvaal}},\ }\bibfield
   {title} {\bibinfo {title} {Analysis of the tailored coupled-cluster method
  in quantum chemistry},\ }\href {https://doi.org/10.1137/18M1171436}
  {\bibfield  {journal} {\bibinfo  {journal} {SIAM Journal on Numerical
  Analysis}\ }\textbf {\bibinfo {volume} {57}},\ \bibinfo {pages} {2579}
  (\bibinfo {year} {2019}{\natexlab{b}})},\ \Eprint
  {https://arxiv.org/abs/https://doi.org/10.1137/18M1171436}
  {https://doi.org/10.1137/18M1171436} \BibitemShut {NoStop}%
\bibitem [{\citenamefont {Leszczyk}\ \emph {et~al.}(2022)\citenamefont
  {Leszczyk}, \citenamefont {Mate}, \citenamefont {Legeza},\ and\ \citenamefont
  {Boguslawski}}]{Leszczyk-2022}%
  \BibitemOpen
  \bibfield  {author} {\bibinfo {author} {\bibfnamefont {A.}~\bibnamefont
  {Leszczyk}}, \bibinfo {author} {\bibfnamefont {M.}~\bibnamefont {Mate}},
  \bibinfo {author} {\bibfnamefont {O.}~\bibnamefont {Legeza}},\ and\ \bibinfo
  {author} {\bibfnamefont {K.}~\bibnamefont {Boguslawski}},\ }\bibfield
  {title} {\bibinfo {title} {Assessing the accuracy of tailored coupled cluster
  methods corrected by electronic wave functions of polynomial cost},\ }\href
  {https://doi.org/10.1021/acs.jctc.1c00284} {\bibfield  {journal} {\bibinfo
  {journal} {Journal of Chemical Theory and Computation}\ }\textbf {\bibinfo
  {volume} {18}},\ \bibinfo {pages} {96} (\bibinfo {year} {2022})},\ \Eprint
  {https://arxiv.org/abs/https://doi.org/10.1021/acs.jctc.1c00284}
  {https://doi.org/10.1021/acs.jctc.1c00284} \BibitemShut {NoStop}%
\bibitem [{\citenamefont {Kurashige}\ \emph {et~al.}(2013)\citenamefont
  {Kurashige}, \citenamefont {Chan},\ and\ \citenamefont
  {Yanai}}]{Kurashige-2013}%
  \BibitemOpen
  \bibfield  {author} {\bibinfo {author} {\bibfnamefont {Y.}~\bibnamefont
  {Kurashige}}, \bibinfo {author} {\bibfnamefont {G.~K.-L.}\ \bibnamefont
  {Chan}},\ and\ \bibinfo {author} {\bibfnamefont {T.}~\bibnamefont {Yanai}},\
  }\bibfield  {title} {\bibinfo {title} {Entangled quantum electronic
  wavefunctions of the {Mn$_4$CaO$_5$} cluster in photosystem {II}},\ }\href
  {https://doi.org/10.1038/nchem.1677} {\bibfield  {journal} {\bibinfo
  {journal} {Nature Chemistry}\ }\textbf {\bibinfo {volume} {5}},\ \bibinfo
  {pages} {660} (\bibinfo {year} {2013})}\BibitemShut {NoStop}%
\bibitem [{\citenamefont {Sharma}\ and\ \citenamefont
  {Chan}(2014)}]{Sharma-2014b}%
  \BibitemOpen
  \bibfield  {author} {\bibinfo {author} {\bibfnamefont {S.}~\bibnamefont
  {Sharma}}\ and\ \bibinfo {author} {\bibfnamefont {G.~K.-L.}\ \bibnamefont
  {Chan}},\ }\bibfield  {title} {\bibinfo {title} {Communication: A flexible
  multi-reference perturbation theory by minimizing the hylleraas functional
  with matrix product states},\ }\href {https://doi.org/10.1063/1.4895977}
  {\bibfield  {journal} {\bibinfo  {journal} {The Journal of Chemical Physics}\
  }\textbf {\bibinfo {volume} {141}},\ \bibinfo {pages} {111101} (\bibinfo
  {year} {2014})},\ \Eprint
  {https://arxiv.org/abs/https://doi.org/10.1063/1.4895977}
  {https://doi.org/10.1063/1.4895977} \BibitemShut {NoStop}%
\bibitem [{\citenamefont {Saitow}\ \emph {et~al.}(2013)\citenamefont {Saitow},
  \citenamefont {Kurashige},\ and\ \citenamefont {Yanai}}]{Saitow-2013}%
  \BibitemOpen
  \bibfield  {author} {\bibinfo {author} {\bibfnamefont {M.}~\bibnamefont
  {Saitow}}, \bibinfo {author} {\bibfnamefont {Y.}~\bibnamefont {Kurashige}},\
  and\ \bibinfo {author} {\bibfnamefont {T.}~\bibnamefont {Yanai}},\ }\bibfield
   {title} {\bibinfo {title} {Multireference configuration interaction theory
  using cumulant reconstruction with internal contraction of density matrix
  renormalization group wave function},\ }\href
  {https://doi.org/http://dx.doi.org/10.1063/1.4816627} {\bibfield  {journal}
  {\bibinfo  {journal} {The Journal of Chemical Physics}\ }\textbf {\bibinfo
  {volume} {139}},\ \bibinfo {eid} {044118} (\bibinfo {year}
  {2013})}\BibitemShut {NoStop}%
\bibitem [{\citenamefont {Friesecke}\ \emph {et~al.}(2023)\citenamefont
  {Friesecke}, \citenamefont {Barcza},\ and\ \citenamefont
  {Legeza}}]{Friesecke-2022b}%
  \BibitemOpen
  \bibfield  {author} {\bibinfo {author} {\bibfnamefont {G.}~\bibnamefont
  {Friesecke}}, \bibinfo {author} {\bibfnamefont {G.}~\bibnamefont {Barcza}},\
  and\ \bibinfo {author} {\bibfnamefont {O.}~\bibnamefont {Legeza}},\
  }\bibfield  {title} {\bibinfo {title} {Predicting the fci energy of large
  systems to chemical accuracy from restricted active space density matrix
  renormalization group calculations},\ }\href
  {https://doi.org/10.1021/acs.jctc.3c01001} {\bibfield  {journal} {\bibinfo
  {journal} {Journal of Chemical Theory and Computation}\ }\textbf {\bibinfo
  {volume} {20}},\ \bibinfo {pages} {87} (\bibinfo {year} {2023})}\BibitemShut
  {NoStop}%
\bibitem [{\citenamefont {Zgid}\ and\ \citenamefont
  {Nooijen}(2008)}]{Zgid-2008c}%
  \BibitemOpen
  \bibfield  {author} {\bibinfo {author} {\bibfnamefont {D.}~\bibnamefont
  {Zgid}}\ and\ \bibinfo {author} {\bibfnamefont {M.}~\bibnamefont {Nooijen}},\
  }\bibfield  {title} {\bibinfo {title} {The density matrix renormalization
  group self-consistent field method: Orbital optimization with the density
  matrix renormalization group method in the active space},\ }\href
  {https://doi.org/http://dx.doi.org/10.1063/1.2883981} {\bibfield  {journal}
  {\bibinfo  {journal} {The Journal of Chemical Physics}\ }\textbf {\bibinfo
  {volume} {128}},\ \bibinfo {eid} {144116} (\bibinfo {year}
  {2008})}\BibitemShut {NoStop}%
\bibitem [{\citenamefont {Neese}\ \emph {et~al.}(2020)\citenamefont {Neese},
  \citenamefont {Wennmohs}, \citenamefont {Becker},\ and\ \citenamefont
  {Riplinger}}]{neese_orca_2020}%
  \BibitemOpen
  \bibfield  {author} {\bibinfo {author} {\bibfnamefont {F.}~\bibnamefont
  {Neese}}, \bibinfo {author} {\bibfnamefont {F.}~\bibnamefont {Wennmohs}},
  \bibinfo {author} {\bibfnamefont {U.}~\bibnamefont {Becker}},\ and\ \bibinfo
  {author} {\bibfnamefont {C.}~\bibnamefont {Riplinger}},\ }\bibfield  {title}
  {\bibinfo {title} {The {ORCA} quantum chemistry program package},\ }\href
  {https://doi.org/10.1063/5.0004608} {\bibfield  {journal} {\bibinfo
  {journal} {The Journal of Chemical Physics}\ }\textbf {\bibinfo {volume}
  {152}},\ \bibinfo {pages} {224108} (\bibinfo {year} {2020})},\ \bibinfo
  {note} {\_eprint:
  https://pubs.aip.org/aip/jcp/article-pdf/doi/10.1063/5.0004608/16740678/224108\_1\_online.pdf}\BibitemShut
  {NoStop}%
\end{thebibliography}%

\end{document}